\documentclass[pra,twocolumn,groupedaddress,floatfix]{revtex4}

\usepackage{graphics}
\usepackage{graphicx}
\usepackage{bm}
\usepackage{amsmath}
\usepackage{amsfonts} 
\usepackage{amssymb}
\usepackage{latexsym}
\usepackage{color}
\usepackage{mathrsfs}
\usepackage{braket}
\usepackage{enumerate}
\usepackage[usenames,dvipsnames]{xcolor}
\usepackage[colorlinks=true,citecolor=blue,linkcolor=blue,urlcolor=blue,backref=true]{hyperref}
\usepackage{float}
\usepackage{mathtools}
\usepackage[raggedright,tight]{subfigure}  %side-by-side figures
\usepackage[inline]{enumitem}
\usepackage{upgreek}

\begin{document}

\title{%Gauge ambiguities in ultrastrong-coupling QED the Jaynes-Cummings model is as fundamental as the Rabi model \\ 
Gauge ambiguities imply Jaynes-Cummings physics remains valid in ultrastrong coupling QED}
\author{Adam Stokes}\email{adamstokes8@gmail.com}
\affiliation{School of Physics and Astronomy, University of Manchester, Oxford Road, Manchester M13 9PL, United Kingdom}
\author{Ahsan Nazir}\email{ahsan.nazir@manchester.ac.uk}
\affiliation{School of Physics and Astronomy, University of Manchester, Oxford Road, Manchester M13 9PL, United Kingdom}

\date{\today}

\begin{abstract}
Ultrastrong-coupling between two-level systems and radiation is important for both fundamental and applied quantum electrodynamics (QED). Such regimes are identified by the breakdown of the rotating-wave approximation, which applied to the quantum Rabi model (QRM) yields the apparently less fundamental Jaynes-Cummings model (JCM). We show that when truncating the material system to two levels, each gauge gives a different description whose predictions vary significantly for ultrastrong-coupling. QRMs are obtained through specific gauge choices, but so too is a JCM without needing the rotating-wave approximation. Analysing a circuit QED setup, we find that this JCM provides more accurate predictions than the QRM for the ground state, and often for the first excited state as well. Thus, Jaynes-Cummings physics is not restricted to light-matter coupling below the ultrastrong limit. Among the many implications is that the system's ground state is not necessarily highly entangled, which is usually considered a hallmark of ultrastrong-coupling. 
\end{abstract}

\maketitle
  
\section*{INTRODUCTION}

Progress in experimental cavity and circuit quantum electrodynamics has granted unprecedented access to the strong, ultrastrong, and deep-strong light-matter coupling regimes \cite{liu_optical_2005,manucharyan_fluxonium:_2009,bourassa_ultrastrong_2009,forn-diaz_observation_2010,niemczyk_circuit_2010,peropadre_switchable_2010,peropadre_nonequilibrium_2013,bell_spectroscopic_2016,baust_ultrastrong_2016,chen_single-photon-driven_2017,yoshihara_superconducting_2017,bosman_multi-mode_2017,yoshihara_characteristic_2017,yoshihara_inversion_2018}. Recently circuit QED experiments involving a single $LC$-oscillator mode with frequency $\omega$ coupled to a flux-qubit with transition frequency $\omega_{\rm m}$ have realised couplings $g$ as large as $g/\omega$ ranging from 0.72 to 1.34, with $g/\omega_{\rm m}\gg 1$ \cite{yoshihara_superconducting_2017}. Such regimes offer a new testing ground for the foundations of quantum theory, and offer opportunities for the development of quantum technologies.

Our interest is in material systems that possess anharmonic spectra, and which are commonly truncated to two levels (qubits). In this case conventional forms of light-matter interaction Hamiltonian yield the so-called quantum Rabi model (QRM), which consists of a linear interaction between the radiation mode and the qubit. Performing the rotating-wave approximation (RWA) then yields the celebrated Jaynes-Cummings model (JCM), which owing to its simple exact solution, has provided deep physical understanding in a wide range of contexts \cite{arroyo-correa_jaynes-cummings_1990,shore_Jaynes-cummings_1993,li_jaynes-cummings_1998,berman_collapse_2014}. In the ultrastrong-coupling regime $0.1<g/\omega<1$ the RWA is no longer valid \cite{bourassa_ultrastrong_2009,niemczyk_circuit_2010,yoshihara_superconducting_2017} and it is therefore widely believed that the Jaynes-Cummings model breaks down. For this reason the QRM is considered indispensible and has found myriad applications in condensed matter, quantum optics, and quantum information theory \cite{auffeves_strong_2013,nielsen_quantum_2000,romero_ultrafast_2012,kyaw_creation_2015,felicetti_dynamical_2014,rossatto_entangling_2016}. A disadvantage of the QRM when compared to the JCM is the lack of any simple solution, which makes its physical interpretation more difficult \cite{braak_integrability_2011}. Despite this difficulty, the QRM is known to possess some markedly different physical features compared to the JCM. For example, the JCM predicts that there is no atom-photon entanglement within the ground state, while the ground state of the QRM is highly entangled within the ultrastrong-coupling regime \cite{hepp_superradiant_1973}.

It was noted sometime ago in the context of scattering theory that retaining only a subset of states raises the prospect of gauge non-invariance \cite{lamb_fine_1952,yang_gauge_1982,kobe_question_1978,fried_vector_1973,bassani_choice_1977,forney_choice_1977,cohen-tannoudji_photons_1997,woolley_gauge_2000,stokes_gauge_2013}. Yet, when the coupling is weak it possible to elicit gauge-invariance through systematically accounting for the effects of the truncation \cite{stokes_master_2018}, and the choice of gauge has no practical implications for the qualitative physical conclusions. It has also been shown in the traditional setting of a single atom weakly-coupled to a (multimode) radiation reservior, that number-conserving (JCM-type) light-matter interaction Hamiltonians can be obtained without recourse to the RWA \cite{drummond_unifying_1987,baxter_gauge_1990,stokes_extending_2012}.

Very recently, the validity of two-level truncations performed in the Coulomb and multipolar gauges has been assessed \cite{de_bernardis_cavity_2018,de_bernardis_breakdown_2018}. The multipolar-gauge was found to offer a more accurate QRM than the Coulomb-gauge for the particular systems and regimes considered there. This was directly attributed to differences in the corresponding forms of coupling. Specifically, contributions of material levels above the first two were found to be suppressed for dipole moment matrix elements that feature in the multipolar-gauge coupling, but not for canonical momentum matrix elements that feature in the Coulomb-gauge coupling.

While Refs. \cite{de_bernardis_cavity_2018,de_bernardis_breakdown_2018} provide valuable comparisons of the Coulomb and multipolar gauges, we employ a more general approach whereby gauge-freedom is encoded into the value of a single real parameter. Our methods are applicable to arbitrary systems in QED including both cavity and circuit QED implementations. We show that corresponding to a given unique light-matter Hamiltonian there is a continuous infinity of non-equivalent two-level models, each of which corresponds to a different choice of gauge. We thereby obtain the most general possible Hermitian interaction operator that is bilinear in qubit and oscillator raising and lowering operators, and which is therefore more general than the JCM or QRM forms. We show that a specific choice of gauge, which we call the JC-gauge, yields a JCM without any need for the RWA. There are also two gauges that yield distinct QRMs. To understand the implications of our approach within the ultrastrong-coupling regime we consider in detail a fluxonium-$LC$ oscillator circuit QED system. We show that the breakdown of the RWA in strong and ultrastrong coupling regimes does not imply a breakdown of the JCM.

\section*{RESULTS}

Our key findings are as follows:
\begin{enumerate}[label=(\roman*)]
\item{
A finite-level truncation of the matter system ruins the gauge-invariance of the theory. In the ultrastrong-coupling regime, the predictions relating to the same physical observable are generally significantly different within any two distinct two-level models. However, it remains meaningful to ask which truncation produces the best approximation of the unique physics. We are able to determine the accuracy of approximate two-level models by benchmarking against the unique predictions of the non-truncated (exact and gauge-invariant) theory.}
\item{
Each two-level model admits a RWA, which yields a corresponding JCM. The only exception to this occurs in the case of the two-level model associated with the JC-gauge wherein the counter-rotating terms are automatically absent. This JCM is valid far beyond the regime of validity of the RWA as applied to the QRM. It follows that Jaynes-Cummings physics is not necessarily restricted to the weak-coupling regime. In particular, independent of the coupling strength the ground state is not entangled in the JC-gauge two-level model.}
\item{
When focusing on predictions that involve the lowest-lying energy eigenstates of the composite system, the JC-gauge two-level model nearly always outperforms the available QRMs within the regimes of interest. Thus, the JCM can and should be used in various situations previously thought to require use of the QRM.}
\end{enumerate}

\subsection*{Light-matter Hamiltonian}

We first present our approach within the context of cavity QED. We consider a material system with charge $e$ and mass $m$ described by position and velocity variables ${\bf r}$ and ${\dot {\bf r}}$ and with potential energy $V({\bf r})$. The material system interacts with an electromagnetic field described by the gauge-invariant transverse vector potential ${\bf A}$ and the associated transverse electric field $-{\dot {\bf A}}={\bf E}_{\rm T}$. The total vector potential is given by ${\bf A}_{\rm tot}={\bf A}+{\bf A}_{\rm L}$ where the longitudinal part ${\bf A}_{\rm L}$ determines the gauge. In the Coulomb-gauge ${\bf A}_{\rm L}={\bf 0}$ so ${\bf A}_{\rm tot}={\bf A}$. The scalar potential $A_0$ that then accompanies ${\bf A}$ is, upto a factor of $e$, the Coulomb potential. As is well-known, the Maxwell-Lorentz equations are invariant under a gauge transformation taking the form $A_0\to A_0-\partial \chi/\partial t$, ${\bf A}\to {\bf A}+\nabla\chi$ where ${\bf A}_{\rm L}=\nabla \chi$ and $\chi$ is an arbitrary function. Here we employ a formulation in which this gauge-freedom is contained within a single real parameter $\alpha$, which determines the gauge through the function $\chi_\alpha$. This function in turn defines a Lagrangian $L_\alpha$ (see Methods). The value $\alpha=0$ specifies the Coulomb gauge while the Poincar\'e (multipolar) gauge also commonly used in atomic physics is obtained by choosing $\alpha=1$.

Moving to the Hamiltonian description canonical momenta are defined in the usual way as ${\bf p}_\alpha = \partial L_\alpha /\partial {\dot {\bf r}}$ and ${\bf \Pi}_\alpha=\delta L_\alpha /\delta {\dot {\bf A}}$. Quantisation of the system is carried out using Dirac's method \cite{dirac_lectures_2003} full details of which are given in Supplementary Note 1. As in conventional derivations of the QRM and JCM we restrict our attention to a single cavity mode. Recently it was shown that the single-mode approximation can breakdown in the ultrastrong coupling regime and in particular that it eliminates the requisite spatio-temporal structure necessary to elicit causal signal propagation \cite{munoz_resolution_2018}. However, the single-mode approximation does not result in a breakdown of gauge-invariance because gauge transformations remain unitary in the single-mode theory. The generalisation to the multimode case is straightforward \cite{stokes_extending_2012,stokes_master_2018}, but is not necessary for understanding the implications of gauge-freedom within two-level models. Following conventional derivations we also make the electric dipole approximation, which similarly does not affect the gauge-invariance of the theory.

With these simplifications the $\alpha$-gauge canonical momenta ${\bf p}_\alpha,~{\bf \Pi}_\alpha$ are related to manifestly gauge-invariant observables by
\begin{align}
&m{\dot {\bf r}} = {\bf p}_\alpha + e(1-\alpha){\bf A},\label{gimomsa} \\  &{\bf E}_{\rm T} = -{\bf \Pi}_\alpha - {\alpha{\bm \varepsilon}({\hat {\bf d}}\cdot {\bm \varepsilon})\over v},\label{gimomsb}
\end{align}
where ${\hat {\bf d}}=-e{\bf r}$ is the material dipole moment, $v$ denotes the cavity volume, $\omega$ denotes the cavity frequency, and ${\bm \varepsilon}$ is a cavity unit polarisation vector. The Hamiltonian is the sum of material and cavity energies
\begin{align}\label{cav_ham}
H=E_{\rm matter}+E_{\rm cavity},
\end{align}
where $E_{\rm matter}= m{\dot {\bf r}}^2/2+V({\bf r})$ and $E_{\rm cavity} =v({\bf E}_{\rm T}^2+\omega^2{\bf A}^2)/2$. The Hamiltonian is expressible in terms of the $\alpha$-gauge canonical operators using Eqs.~(\ref{gimomsa}) and (\ref{gimomsb}), with the well-known Coulomb-gauge ($\alpha=0$) and Poincar\'e-gauge ($\alpha=1$) forms  obtained as specific examples. 

The energy is a particular example of a gauge-invariant observable, which in Eq.~(\ref{cav_ham}) has been expressed as a function of the elementary gauge-invariant observables ${\bf x}= \{{\bf r},{\dot {\bf r}},{\bf A},{\bf E}_{\rm T}\}$. More generally when written in terms of ${\bf x}$ any observable $O$ possesses a unique functional form $O\equiv O({\bf x})$. The theory is gauge-invariant in that the predictions concerning any gauge-invariant observable can be calculated using any gauge and these predictions are unique. The canonical momenta $\{ {\bf p}_\alpha,{\bf \Pi}_\alpha\}$ are however, manifestly gauge-dependent in that for each different $\alpha$ they constitute different functions of the gauge-invariant observables ${\bf x}$. When written in terms of canonical operators ${\bf y}_\alpha = \{{\bf r},{\bf p}_\alpha,{\bf A},{\bf \Pi}_\alpha\}$, an observable $O$ generally possesses an $\alpha$-dependent functional form $O=o^\alpha({\bf y}_\alpha)$. The canonical operators belonging to fixed gauges $\alpha$ and $\alpha'$ are related using the unitary gauge-fixing transformation $R_{\alpha\alpha'} = \exp[{\rm i}(\alpha-\alpha'){\hat{\bf d}}\cdot {\bf A}]$. This implies that distinct functional forms $o^\alpha$ and $o^{\alpha'}$ of the observable $O$ are related according to 
\begin{align}\label{funform}
O=o^\alpha({\bf y}_\alpha)=R_{\alpha\alpha'}o^{\alpha}({\bf y}_{\alpha'})R_{\alpha\alpha'}^{-1} \equiv o^{\alpha'}({\bf y}_{\alpha'}).
\end{align}
This equation expresses the uniqueness of physical observables independent of the chosen gauge.

The unitarity of the gauge transformation $R_{\alpha\alpha'}$ also ensures that in all gauges the canonical operators satisfy the canonical commutation relations $[r_i,p_{\alpha,j}]={\rm i}\delta_{ij}$, $[A_i,\Pi_{\alpha,j}]={\rm i}\varepsilon _i \varepsilon_j/v$ with all remaining commutators between canonical operators being zero. These relations allow us to decompose the state space ${\cal H}$ of the light-matter system into $\alpha$-dependent matter and cavity state spaces ${\cal H}_{\rm m}^\alpha$ and ${\cal H}_{\rm c}^\alpha$ such that ${\cal H}={\cal H}_{\rm m}^\alpha\otimes {\cal H}_{\rm c}^\alpha$. The eigenstates of the canonical operators ${\bf r},{\bf p}_\alpha$ provide a basis for the material space ${\cal H}_{\rm m}^\alpha$ while the eigenstates of the canonical operators ${\bf A},{\bf \Pi}_\alpha$ provide a basis for the cavity space ${\cal H}_{\rm c}^\alpha$. It is not possible to define gauge-invariant ($\alpha$-independent) light and matter quantum subsystem state spaces directly in terms of the gauge-invariant observables ${\bf x}$, because Eqs.~(\ref{gimomsa}) and (\ref{gimomsb}) along with the canonical commutation relations imply that $[m{\dot r}_i,E_{{\rm T},j}]=-{\rm i}e \varepsilon_i \varepsilon_j/v \neq 0$.

The present theory yields unique physical predictions despite the $\alpha$-dependence of the quantum subsystems. This is because the representation of an observable by operators is unique as expressed by Eq.~(\ref{funform}), which implies that the average of an observable $O$ in the state $\ket{\psi}$ is unambiguously $\bra{\psi}O\ket{\psi}$. The $\alpha$-dependence of the quantum subsystems is however an important feature of the theory, which is made transparent within our formulation. An approximation performed on one of the quantum subsystems will constitute a different approximation in each gauge, and may ruin the gauge-invariance of the theory.

\subsection*{Non-equivalent two-level models}

In conventional approaches a gauge is chosen at the outset and the Hamiltonian is partitioned into matter and cavity bare energies plus an interaction part. Here we follow this same procedure, but with the important exception that the gauge is left open rather than fixed. This is achieved through substitution of Eqs.~(\ref{gimomsa}) and (\ref{gimomsb}) into Eq.~(\ref{cav_ham}), which casts the total Hamiltonian in the form $H=H_{\rm m}^\alpha({\bf r},{\bf p}_\alpha)\otimes I_{\rm c}^\alpha+I_{\rm m}^\alpha\otimes H_{\rm c}^\alpha({\bf A},{\bf \Pi}_\alpha)+V^\alpha({\bf y}_\alpha)$. Here $I_{\rm m}^\alpha$ and $I_{\rm c}^\alpha$ are the identity operators in ${\cal H}_{\rm m}^\alpha$ and ${\cal H}_{\rm c}^\alpha$ respectively, $H_{\rm m}^\alpha$ and $H_{\rm c}^\alpha$ are material and cavity bare energies in ${\cal H}_{\rm m}^\alpha$ and ${\cal H}_{\rm c}^\alpha$ respectively, and $V^\alpha$ denotes the interaction Hamiltonian. The explicit forms of $H_{\rm m}^\alpha,~H_{\rm c}^\alpha$ and $V^\alpha$ are given in Eqs.~(\ref{hparts00}), (\ref{hparts0}) and (\ref{hparts}) in Methods.

One of the most useful and widespread approximations in light-matter theory is a two-level truncation of the material system whereby only the first two eigenstates $\ket{\epsilon^\alpha_0}, ~\ket{\epsilon^\alpha_1}$ of the material bare energy $H_{\rm m}^\alpha$ are retained. Our approach reveals that this procedure ruins the uniqueness of physical predictions that results from Eq.~(\ref{funform}). %We will show further that the ensuing ambiguity becomes significant when the light-matter coupling is sufficiently strong.
Using the projection $P^\alpha=\ket{\epsilon^\alpha_0}\bra{\epsilon^\alpha_0}+\ket{\epsilon^\alpha_1}\bra{\epsilon^\alpha_1}$ we obtain the $\alpha$-gauge two-level model Hamiltonian
\begin{align}\label{HPalph}
H_2^\alpha =& \, \omega_{\rm m}\sigma^+_\alpha \sigma^-_\alpha +\omega_\alpha\left(c_\alpha^\dagger c_\alpha +{1\over 2}\right)+\Delta_\alpha\nonumber \\ &+{\rm i}u_\alpha^-(\sigma^+_\alpha c_\alpha -\sigma^-_\alpha c_\alpha^\dagger) + {\rm i}u_\alpha^+(\sigma^+_\alpha c^\dagger_\alpha - \sigma^-_\alpha c_\alpha)
\end{align}
where $u_\alpha^\pm = \pm ({\bf d}\cdot {\bm \varepsilon})[\omega_\alpha\alpha \mp \omega_{\rm m}(1-\alpha)]/\sqrt{2\omega_\alpha v}$ and $\Delta_\alpha=\epsilon_0+\alpha^2({\bf d}\cdot {\bm \varepsilon})^2/2v$ is an $\alpha$-dependent zero-point shift. The transition dipole moment ${\bf d} = \bra{\epsilon^\alpha_1}-e{\bf r}\ket{\epsilon_0^\alpha}$, which is assumed to be real, is $\alpha$-independent, because ${\bf r}$ commutes with $R_{\alpha\alpha'}$. The material Hamiltonian's eigenvalues $\epsilon_0$ and $\epsilon_1=\omega_{\rm m}+\epsilon_0$ corresponding to material states $\ket{\epsilon_0^\alpha}$ and $\ket{\epsilon_1^\alpha}$ respectively, are also $\alpha$-independent because $H^\alpha_{\rm m}= R_{\alpha\alpha'}H_{\rm m}^{\alpha'} R_{\alpha\alpha'}^{-1}$. The complete derivation of Eq. (\ref{HPalph}) is given in Methods.

An important topic relating to two-level models and the choice of gauge in light-matter physics concerns the occurrence or otherwise of a super-radiant phase transition in the Dicke-model at strong-coupling  \cite{rzazewski_phase_1975,keeling_coulomb_2007,viehmann_superradiant_2011,nataf_no-go_2010,vukics_elimination_2014,de_bernardis_cavity_2018}. A precursor already occurs in the QRM whereby beyond a critical coupling point an exponential closure of the first transition energy occurs \cite{ashhab_qubit-oscillator_2010,ashhab_superradiance_2013,bamba_superradiant_2016}. We note that in Eq. (\ref{HPalph}) counter-rotating and number-conserving interactions generally have different coupling strengths, and a strict bound cannot be given for either coupling independent of the material potential, except if $\alpha=0$. It follows that the standard ``no-go theorem" concerning the ground state instability of a single-dipole, holds in general only in the Coulomb gauge \cite{rzazewski_phase_1975,keeling_coulomb_2007,viehmann_superradiant_2011,nataf_no-go_2010,de_bernardis_breakdown_2018}. An arbitrary-gauge analysis of this topic is important, but lies beyond the scope of this article and will be discussed elsewhere.
%%%%%%%%%%%%%%%%%%%%%%%%%%%%%%%%%%%%%%%%%%%%%%%%%%%%%%%%%%%%%%%%%%%%%%%%%%%%%%%%%%%%%%%%%%%%%
%%
%%	F I G U R E S  S T A R T
%%
%%%%%%%%%%%%%%%%%%%%%%%%%%%%%%%%%%%%%%%%%%%%%%%%%%%%%%%%%%%%%%%%%%%%%%%%%%%%%%%%%%%%%%%%%%%%%
%\onecolumngrid
%\widetext
\begin{figure}[t]
\vspace*{.2cm}
\includegraphics[scale=1.15]{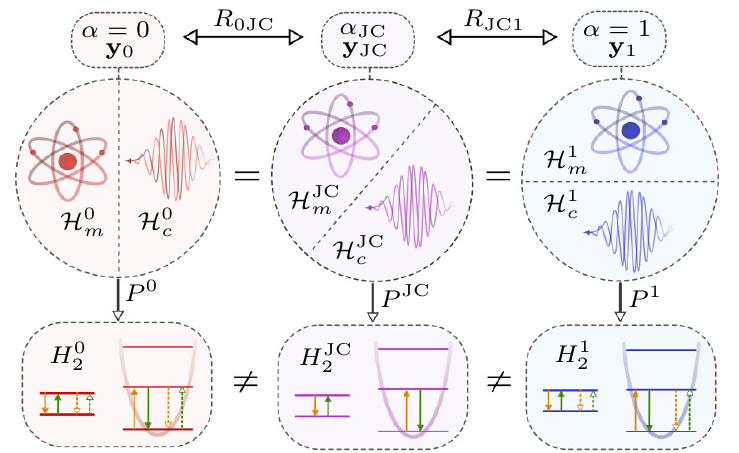}
\caption{\textbf{Three important gauges and their non-equivalent two-level models}. The $\alpha=0$, $\alpha=\alpha_{\rm JC}$ and $\alpha=1$ gauges, and their associated two-level truncations. The different gauges are associated with different unitarily related canonical operators ${\bf y}_0$, ${\bf y}_{\rm JC}$ and ${\bf y}_1$ respectively, which induce different subsystem decompositions of the light-matter Hilbert space. The composite Hilbert space and the Hamiltonian are unique, but a projection onto the first two levels of the material system results in distinct two-level models with Hamiltonians $H_2^0$, $H_2^{\rm JC}$ and $H_2^1$ respectively. The $\alpha=0$ and $\alpha=1$ gauge two-level model interaction Hamiltonians both have Rabi form and therefore describe real processes represented by the solid green and orange arrows, as well as counter-rotating processes represented by the dashed arrows. The $\alpha_{\rm JC}$-gauge two-level model interaction has Jaynes-Cummings form and therefore all processes it describes are real.}\label{pic}
\end{figure}
%\vspace*{-1cm}
%%%%%%%%%%%%%%%%%%%%%%%%%%%%%%%%%%%%%%%%%%%%%%%%%%%%%%%%%%%%%%%%%%%%%%%%%%%%%%%%%%%%%%%%%%%%%
%%
%%	F I G U R E S  E N D
%%
%%%%%%%%%%%%%%%%%%%%%%%%%%%%%%%%%%%%%%%%%%%%%%%%%%%%%%%%%%%%%%%%%%%%%%%%%%%%%%%%%%%%%%%%%%%%%\\ 
%\\ \\
%\twocolumngrid

We are concerned with the $\alpha$-dependence of predictions obtained when using the Hamiltonian in Eq. (\ref{HPalph}). This Hamiltonian has neither JC nor Rabi form, because $|u^+_\alpha| \neq |u_\alpha^-|$ and $u^+_\alpha\neq 0$ except when particular values of $\alpha$ are chosen. Specifically, two distinct QRMs are obtained for the choices $\alpha=0$ and $\alpha=1$, which are nothing but the Coulomb and Poincar\'e-gauge QRMs frequently encountered in quantum optics. On the other hand, by choosing $\alpha=\alpha_{\rm JC}$, which solves the coupled equations $\alpha_{\rm JC}(\omega_{\rm m}+\omega_{\rm JC}) = \omega_{\rm m}$ and $\omega^2_{\rm JC} = \omega^2 + {e^2}(1-\alpha_{\rm JC})^2/ m v$ we obtain $u^+_{\rm JC}\equiv 0$ and $u^-_{\rm JC}=-2({\bf d}\cdot {\bm \varepsilon})\omega_{\rm m} \sqrt{\omega_{\rm JC}}/[ \sqrt{2v}(\omega_{\rm JC}+ \omega_{\rm m})]$. This choice therefore yields a JC Hamiltonian without any need for the RWA. The JCM derived in this way possesses the same advantage of exact solvability as conventional JCMs obtained as RWAs of the Coulomb and Poincar\'e-gauge QRMs. However, the states $\ket{\epsilon_{\rm JC}}$, operators $\sigma^\pm_{\rm JC}$, and parameters $u^-_{\rm JC}$, $\omega_{\rm JC}$ are different to their counterparts within conventional JCMs. In particular, the renormalised cavity frequency $\omega_{\rm JC}$ together with the zero-point shift $\Delta_{\rm JC}$ yield a ground state energy that is a non-constant function of the Coulomb-gauge and multipolar-gauge QRM coupling parameters.

Having derived an expression for the energy, most properties of practical interest can now be calculated using the two-level model associated with any gauge. This includes atomic populations and coherences, as well as various cavity properties such as photon number. It is however possible to go further by defining the two-level representation of any additional observable of interest $O$ as $O_2^\alpha =P^\alpha O P^\alpha$. Restricting the state space ${\cal H}_{\rm m}^\alpha$ to the two-dimensional subspace spanned by the eigenstates $\ket{\epsilon^\alpha_0},~\ket{\epsilon^\alpha_1}$ then completes the construction of the two-level model.

Two-level models corresponding to distinct gauges $\alpha$ and $\alpha'$ must be distinguished, because when $\alpha\neq\alpha'$ the projection $P^\alpha$ involves all eigenstates of $H^{\alpha'}_{\rm m}$, and similarly $P^{\alpha'}$ involves all eigenstates of $H^{\alpha}_{\rm m}$. This is because the gauge transformation does not have product form; $R_{\alpha\alpha'}\neq R_{\rm m}\otimes R_{\rm c}$. A pictorial representation of the relationship between different gauges and their associated two-level models is given in Fig. \ref{pic}. After a two-level truncation the uniqueness of the representation of observables expressed by Eq.~(\ref{funform}) no longer holds, that is, $O_2^\alpha \neq  O^{\alpha'}_2$ when $\alpha\neq \alpha'$. Distinct two level-models will therefore give different predictions for the same physical quantity. %It is however possible to determine which two-level model is the most accurate through comparison of its predictions with those of the exact (non-truncated) theory.

An observable of particular importance is the energy represented by the Hamiltonian, which we focus on hereafter. There is generally no simple relation between distinct two-level model Hamiltonians $H_2^\alpha$ and $H^{\alpha'}_2$ when $\alpha\neq \alpha'$. In fact, it was noted some time ago that two-level models associated with different gauges can give different results even in the weak-coupling regime \cite{barton_frequency_1974}. %(see also supplementary note \ref{weak}). This stems from the breakdown of sum rules that rely for their proof on the CCR algebra, which cannot be supported in finite-dimensional Hilbert space.
However, provided that the two-level modification of the operator algebra is accounted for, it can be shown that certain two-level model predictions are gauge-invariant up to order $d^2$ \cite{stokes_master_2018}. This is discussed in more detail in Supplementary Note 2. Regardless, one expects predictions of two-level models corresponding to different gauges to be significantly different when the coupling is sufficiently strong. We show how a comparison of the predictions of different two-level models can be achieved for an arbitrary observable in Methods. We show further that if the material system is a harmonic oscillator, then it is possible to derive a JCM that is necessarily more accurate than any derivable QRM for finding ground state averages.

\subsection*{Application to ultrastrong coupling in circuit QED}

When considering less artificial systems than a material oscillator the relative accuracies of two-level models is more difficult to determine. We now consider an experimentally relevant circuit QED set-up consisting of a fluxonium atom coupled to an $LC$-oscillator. The fluxonium is described by the flux variables $\phi$, ${\dot \phi}$ and the external flux $\phi_{\rm ext}$, along with three energy parameters $E_{\rm c}$, $E_{\rm J}$ and $E_{\rm l}$ which are the capacitive energy, tunnelling Josephson energy and inductive energy respectively. The external flux $\phi_{\rm ext}={\rm \uppi}/2e$ specifies maximum frustration of the atom. The $LC$-oscillator is described by analogous flux variables $\theta,~{\dot \theta}$, with inductance $L$ and capacitance $C$ defining the oscillator frequency $\omega=1/\sqrt{LC}$.

In terms of ${\bf x}= \{\phi,\theta,{\dot \phi},{\dot \theta}\}$ the functional form of an observable $O$ is unique $O\equiv O({\bf x})$. On the other hand different canonical operators ${\bf y}_\alpha=\{\phi,\xi_\alpha,\theta_\alpha,\zeta \}$ are related by $\theta_\alpha = R^{-1}_{0\alpha}\theta_0 R_{0\alpha}$ and $\xi_\alpha = R^{-1}_{0\alpha}\xi_0 R_{0\alpha}$ where $R_{0\alpha} = {\rm e}^{{\rm i}\alpha \zeta \phi}$ is a unitary gauge transformation with $\alpha$ real and dimensionless. Here $\xi_\alpha$ and $\zeta$ are canonical momenta conjugate to $\phi$ and $\theta_\alpha$ respectively. The gauge choices $\alpha=0$ and $\alpha=1$ are called the charge-gauge and flux-gauge respectively \cite{manucharyan_resilience_2017}. The Hamiltonian $H$ describing the system is derived in Supplementary Note 3 and is given in Methods.

In exactly the same way as for the cavity QED Hamiltonian the projection $P_\alpha$ onto the first two eigenstates $\ket{\epsilon^\alpha_0},~\ket{\epsilon^\alpha_1}$ of the material bare energy $H_{\rm m}^\alpha$ can be used to obtain an $\alpha$-dependent two-level model Hamiltonian, which at maximal frustration reads
\begin{align}\label{HPbet}
H_2^\alpha =&\, \omega_{\rm m}\sigma^+_\alpha \sigma^-_\alpha +\omega_\alpha \left(c_\alpha^\dagger c_\alpha +{1\over 2}\right)+\Delta_\alpha\nonumber \\ &+u_\alpha^-(\sigma^+_\alpha c_\alpha +\sigma^-_\alpha c_\alpha^\dagger) + u_\alpha^+(\sigma^+_\alpha c^\dagger_\alpha + \sigma^-_\alpha c_\alpha),
\end{align}
where $u_\alpha^\pm=\varphi[\alpha\omega_\alpha \mp (1-\alpha)\omega_{\rm m}]/\sqrt{2\omega_\alpha L}$ and $\Delta_\alpha=\epsilon_0+\alpha^2\varphi^2/2L$, in which $\varphi=\bra{\epsilon_1^\alpha}\phi\ket{\epsilon_0^\alpha}=\varphi^*$ and $\epsilon_0$ denotes the ground energy of $H^\alpha_{\rm m}$. The two-level system parameters $\omega_{\rm m}$, $\varphi$ and $\epsilon_0$ depend implicitly on $E_{\rm c},~E_{\rm J},~E_{\rm l}$ and $\phi_{\rm ext}$. The renormalised cavity frequency is $\omega_\alpha =  \omega\sqrt{1+2E_{\rm c}(1-\alpha)^2C/e^2}$. Away from the maximal frustration point the flux $\phi$ possesses diagonal matrix elements in the basis $\{\ket{\epsilon_0^\alpha},\ket{\epsilon_1^\alpha}\}$, such that $\sigma_\alpha^+\sigma_\alpha^-$ and $\sigma_\alpha^-\sigma_\alpha^+$ are also linearly coupled to the mode operators $c_\alpha,~c_\alpha^\dagger$. In analogy to the cavity QED case the charge and flux-gauges yield distinct Rabi Hamiltonians, but there also exists a value $\alpha=\alpha_{\rm JC}=\omega_{\rm m}/(\omega_{\rm m}+\omega_{\rm JC})$ such that $u_\alpha^+\equiv 0$, which casts the Hamiltonian in JC form.

The ratio $\delta=\omega/\omega_{\rm m}$ in which $\omega_{\rm m}$ is taken as the qubit transition at maximal frustration $\phi_{\rm ext}={\rm \uppi}/2e$, specifies the relative qubit-oscillator detuning. To quantify the relative coupling strength we use the ratio $\eta =g/\omega$ where $g=\varphi \sqrt{\omega/2L}$. The parameters $g$ and $\omega$ are the coupling strength and cavity frequency of the flux-gauge QRM, but we note that the corresponding parameters associated with any other two-level model could also be used. For different $\alpha$ the $\alpha$-dependent two-level truncation yields different predicted behaviour of physical observables as functions of the model parameters $\delta$, $\eta$ and $\phi_{\rm ext}$. In contrast the exact predictions resulting from the non-truncated model are $\alpha$-independent (gauge-invariant).

\begin{figure}[t]
\begin{minipage}{\columnwidth}
\begin{center}
\hspace*{-0.6cm}\includegraphics[scale=0.73]{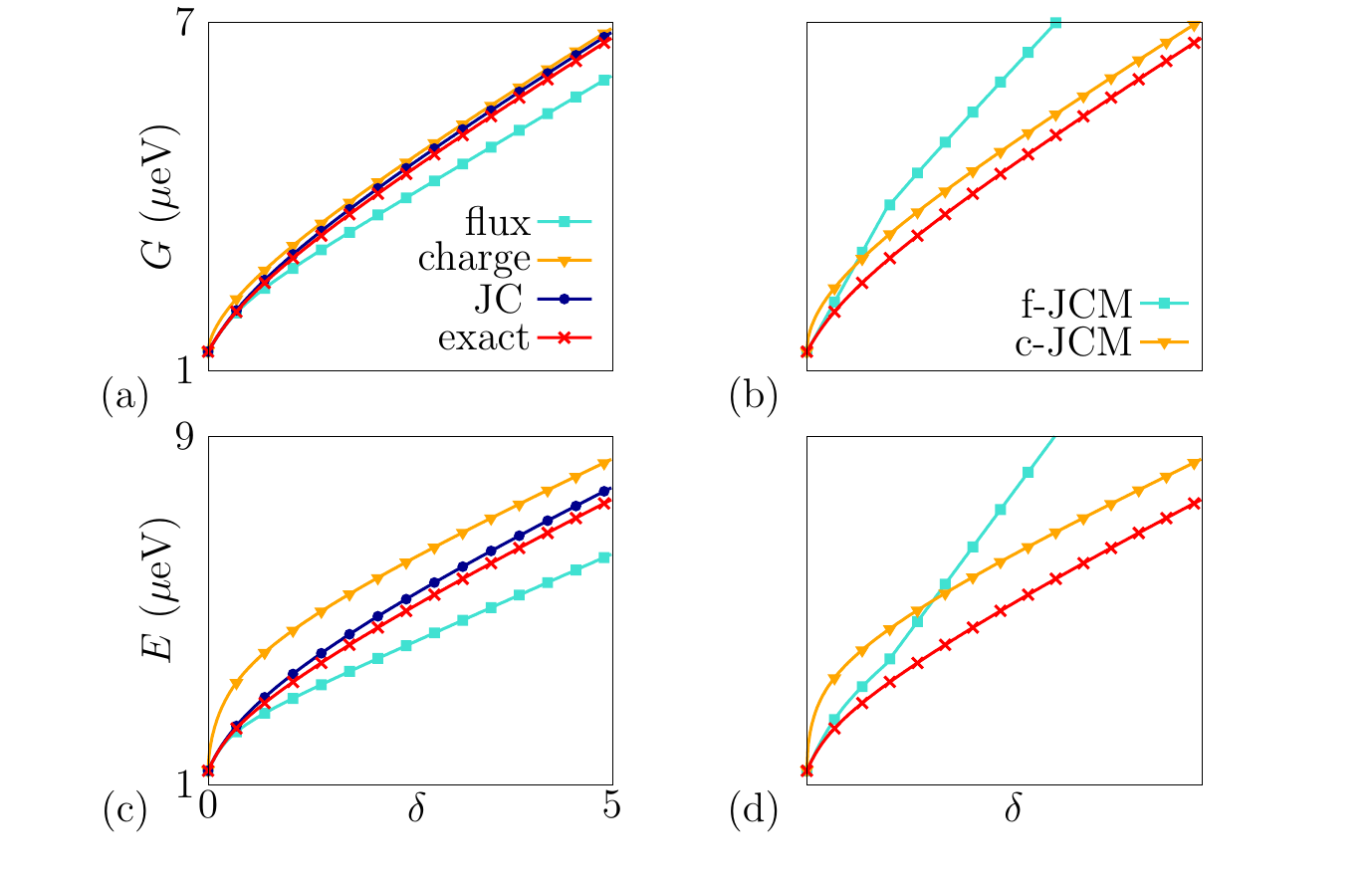}
\vspace*{-7mm}\caption{\textbf{Lowest energy levels as functions of detuning.} \textbf{(a)} $E_{\rm l}=0.33\mu$eV, $E_{\rm J}=10E_{\rm l}=E_{\rm c}$, $\phi_{\rm ext}={\rm \uppi}/2e$ and $\eta =1$. The ground energy is plotted with $\delta$ for the flux-gauge and charge-gauge QRMs, for the JC-gauge two-level model, and for the exact model. \textbf{(b)} Same as (a) for the charge and flux-gauge JCMs obtained as RWAs of the corresponding QRMs. \textbf{(c)} Same as (a) for the first excited energy. \textbf{(d)} Same as (c) for the charge and flux-gauge JCMs.}\label{en_det1}
\end{center}
\end{minipage}
\end{figure}

We begin by determining how the ground energy $G$ and first excited energy $E$ vary with the detuning $\delta$ at maximal frustration $\phi_{\rm ext}={\rm \uppi}/2e$ and fixed coupling $\eta =1$ (Fig.  \ref{en_det1}). Regimes with large $\delta$ are presently more experimentally relevant \cite{yoshihara_superconducting_2017,yoshihara_characteristic_2017,yoshihara_inversion_2018}, yet, unless $\delta$ is relatively small ($\delta<1$), we find that all two-level models become inaccurate in predicting eigenvalues $E_n>E$ of the non-truncated Hamiltonian. This can be traced to the occurence of resonances in energy shifts, that occur for large $\delta$ (see Supplementary Note 4). Indeed, deviations from the predictions of the QRM have been observed experimentally for such $E_n$ within the ultrastrong-coupling regime \cite{yoshihara_inversion_2018}.

%%%%%%%%%%%%%%%%%%%%%%%%%%%%%%%%%%%%%%%%%%%%%%%%%%%%%%%%%%%%%%%%%%%%%%%%%%%%%%%%%%%%%%%%%%%%%
%%
%%	F I G U R E S  S T A R T
%%
%%%%%%%%%%%%%%%%%%%%%%%%%%%%%%%%%%%%%%%%%%%%%%%%%%%%%%%%%%%%%%%%%%%%%%%%%%%%%%%%%%%%%%%%%%%%%
%%%%%%%%%%%%%%%%%%%%%%%%%%%%%%%%%%%%%%%%%%%%%%%%%%%%%%%%%%%%%%%%%%%%%%%%%%%%%%%%%%%%%%%%%%%%%
\begin{figure}[t]
\begin{minipage}{\columnwidth}
\begin{center}
\hspace*{-0.8cm}\includegraphics[scale=0.7]{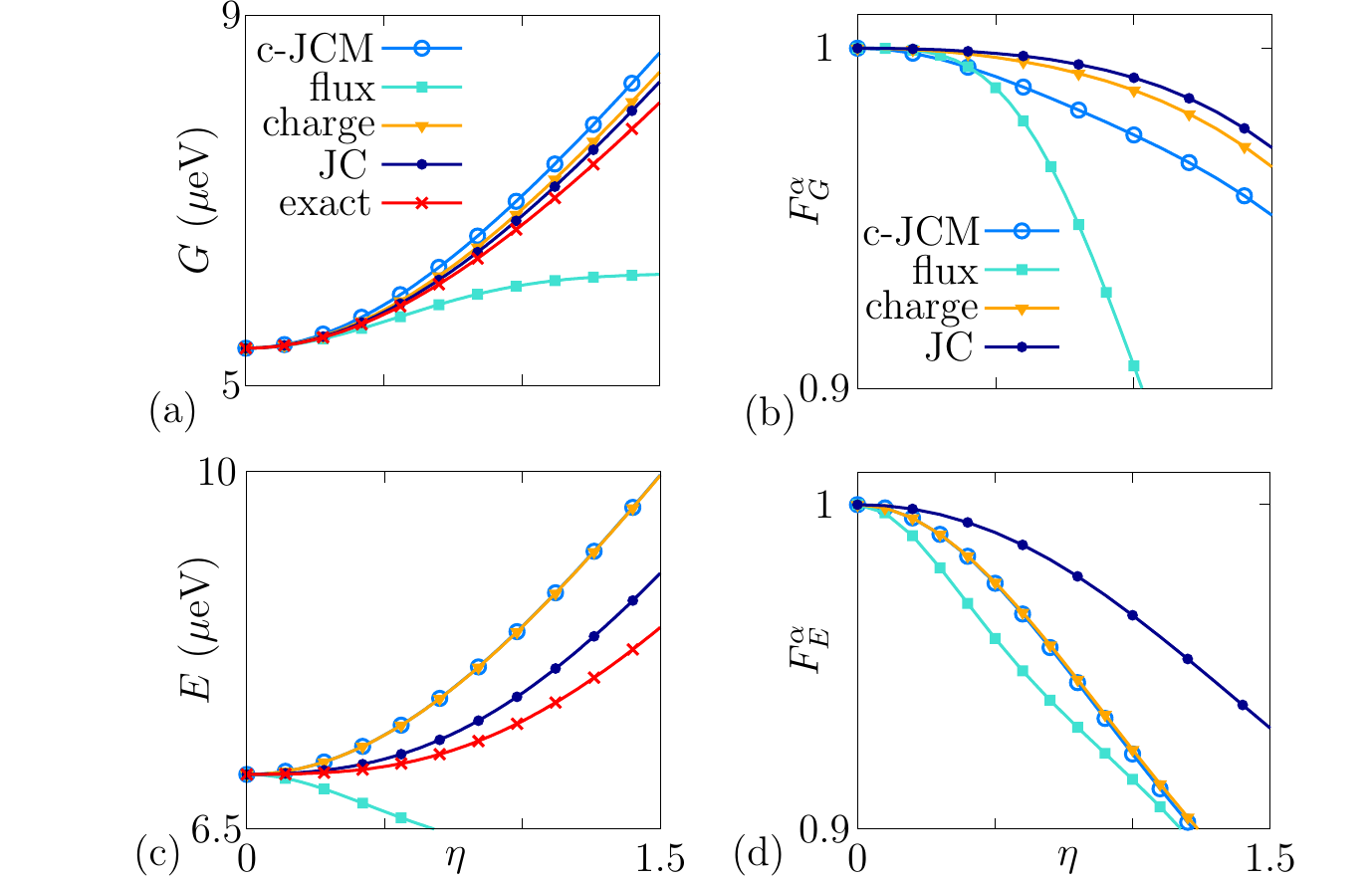}
\caption{\textbf{Lowest energies and eigenstate fidelities as functions of coupling strength.} In all plots $E_{\rm l}=0.33\mu$eV, $E_{\rm J}=10E_{\rm l}=E_{\rm c}$, $\delta=5$ and $\phi_{\rm ext}={\rm \uppi}/2e$. \textbf{(a)} The ground energy is plotted with $\eta$ for the flux-gauge and charge-gauge QRMs, for the JC-gauge two-level model, for the exact model, and for the charge-gauge JCM obtained via the RWA. The flux-gauge JCM is extremely inaccurate in this regime and is not shown.  \textbf{(b)} The ground state fidelity $F_G^\alpha$ is plotted with $\eta$ for the flux-gauge $\alpha=1$ and charge-gauge $\alpha=0$ QRMs, for the JC-gauge and for charge-gauge JCM (c-JCM) that is obtained as the RWA of the corresponding Rabi model. \textbf{(c)} Same as (a) for the first excited energy. \textbf{(d)} Same as (b) for the first excited state. For the excited state the RWA remains valid in the charge gauge.}\label{c}
\end{center}
\end{minipage}
\end{figure}
%%%%%%%%%%%%%%%%%%%%%%%%%%%%%%%%%%%%%%%%%%%%%%%%%%%%%%%%%%%%%%%%%%%%%%%%%%%%%%%%%%%%%%%%%%%%%
\begin{figure}[t]
\begin{minipage}{\columnwidth}
\begin{center}
\hspace*{-0.8cm}\includegraphics[scale=0.7]{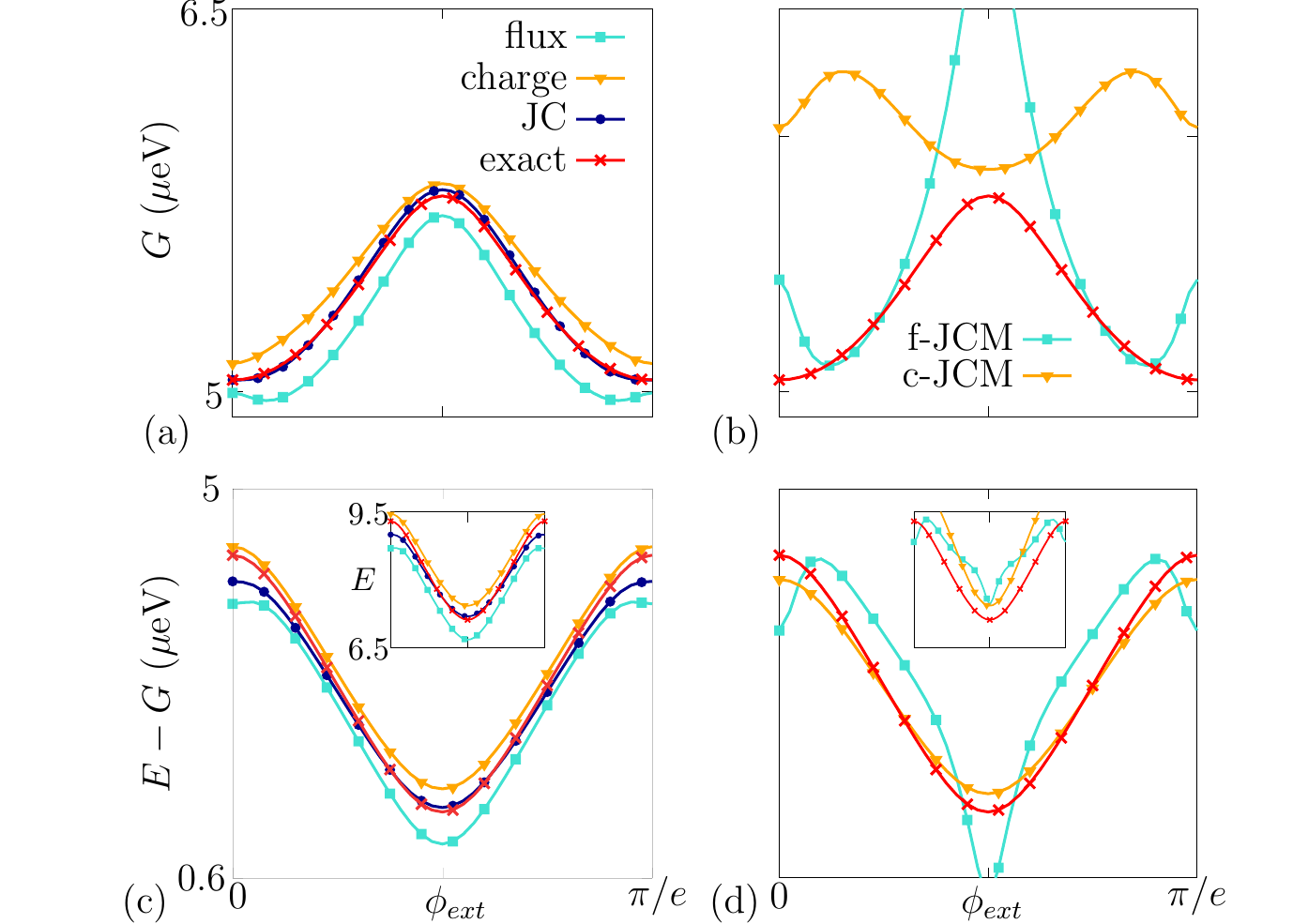}
\caption{\textbf{Lowest energies as functions of external flux}. In all plots $E_{\rm l}=0.33\mu$eV, $E_{\rm J}=10E_{\rm l}=E_{\rm c}$, $\delta=5$ and $\eta =0.5$. \textbf{(a)} The ground energy $G$ is plotted with $\phi_{\rm ext}$ for the flux-gauge and charge-gauge QRMs, for the JC-gauge two-level model and for the exact model.  \textbf{(b)} For the same range as (a) the ground energy is plotted for the flux-gauge and charge-gauge JCMs that are obtained as RWAs of the corresponding Rabi models, and for the exact model. \textbf{(c)} Same as (a) for the first transition energy $E-G$. The inset shows the corresponding first excited energies $E$. The JC-gauge is generally the most accurate two-level model, but because the charge gauge QRM overestimates both the ground and excited energy it becomes relatively accurate in the transition $E-G$ for fluxes away from the maximal frustration point. \textbf{(d)} Same as (b) for the first transition energy. The inset shows the corresponding excited energies over the same range as the inset in (c). We see that especially in the case of the charge-gauge, the JCM is inaccurate for the ground and excited energies, but it is by comparison more accurate for the transition.}\label{f}
\end{center}
\end{minipage}
\end{figure}
%%%%%%%%%%%%%%%%%%%%%%%%%%%%%%%%%%%%%%%%%%%%%%%%%%%%%%%%%%%%%%%%%%%%%%%%%%%%%%%%%%%%%%%%%%%%%
%%
%%	F I G U R E S  E N D
%%
%%%%%%%%%%%%%%%%%%%%%%%%%%%%%%%%%%%%%%%%%%%%%%%%%%%%%%%%%%%%%%%%%%%%%%%%%%%%%%%%%%%%%%%%%%%%%

We focus primarily on the experimentally relevant large $\delta$ regime by choosing $\delta=5$. Other detunings may also be considered and various results for the cases $\delta=1$ (resonance) and $\delta= 1/5$ are presented in Supplementary Note 5. In Figs. \ref{c} (a) and \ref{c} (b) we compare the ground and first excited energies found using various two-level models with the corresponding gauge-invariant energies of the exact theory. The ground and excited level-shifts are obtained by subtracting the corresponding (bare) eigenenergies of the non-interacting system. At maximal frustration the shift of the ground state can be identified as the Bloch-Siegert shift \cite{forn-diaz_observation_2010}.  The first transition shift is the difference between the ground and excited shifts and is commonly termed the Lamb shift by analogy with atomic hydrogen \cite{yoshihara_inversion_2018}. In the RWA the coupling-dependent zero-point contribution $\omega_\alpha/2 +\Delta_\alpha$ in Eq. (\ref{HPbet}) gives the ground energy. For $\alpha\neq \alpha_{\rm JC}$ this results in an incorrect expression for the Lamb shift even for weak-coupling \cite{allen_optical_1975,stokes_master_2018} (see also Supplementary Note 2). It is therefore unsurprising that the flux and charge-gauge JCMs are inaccurate in predicting the associated dressed energies within the ultrastrong-coupling regime, as illustrated in Figs. \ref{c} (a) and \ref{c} (b). In contrast, for the two-level model of the JC-gauge ($\alpha=\alpha_{\rm JC}$) the RWA is no longer an approximation. The ground energy $\omega_{\rm JC}/2+\Delta_{\rm JC}$, is different to the results of the RWA applied in the $\alpha=0$ and $\alpha=1$ gauges, and it does lead to the expected expression for the Lamb-shift within the weak-coupling regime \cite{stokes_master_2018} (see Supplementary Note 2). Thus, even though the Hamiltonian has Jaynes-Cummings form it is not evident that like the charge and flux-gauge JCMs the JC-gauge two-level model will necessarily be inaccurate in predicting dressed energies within the ultrastrong-coupling regime. Indeed,  Figs. \ref{c} (a) and \ref{c} (b) show that the JC-gauge two-level model is not only more accurate than the flux and charge-gauge JCMs it is also more accurate than the flux and charge-gauge QRMs.

To determine which two-level model yields the most accurate lowest energy eigenstates we compute the ground and first excited state fidelities $F_G^\alpha=|\braket{G^\alpha_2|G}|^2$ and $F_E^\alpha=|\braket{E^\alpha_2|E}|^2$ where $\ket{G}$ and $\ket{E}$ are the exact ground and first excited eigenstates of the non-truncated Hamiltonian $H$, while $\ket{G_2^\alpha}$ and $\ket{E_2^\alpha}$ are the corresponding eigenstates of $H_2^\alpha$. Figs. \ref{c} (c) and \ref{c} (d) show that the JC-gauge model is more accurate than both QRMs, and much more accurate than conventional JCMs, especially in the case of the ground state. Since the JC-gauge two-level model tends to produce a more accurate representation of the lowest two energy states of the system it is natural to suppose that it will generally be more accurate than the QRM in predicting observable averages in these states. This is verified for the cases of ground state photon number averages in Supplementary Note 6.

To link with recent experiments in which circuit properties are measured for varying external flux $\phi_{\rm ext}$, Fig. \ref{f} shows the behaviour with $\phi_{\rm ext}$ of the lowest dressed energies when $\eta =1/2$. The JC-gauge again yields the most accurate two-level model (Fig.~\ref{f} (a), (b)) despite the clear breakdown of the RWA (Fig.~\ref{f} (c), (d)).  It follows that Jaynes-Cummings physics is not synonymous with the RWA, and that a departure from Jaynes-Cummings physics is not implied within the ultrastrong-coupling regime. For larger $\eta$ two-level models become increasingly inaccurate, though the JC-gauge continues to give the best agreement with exact energies even within the deep-strong coupling regime (see Supplementary Note 5). 

\section{DISCUSSION}

The behaviour shown in Figs. \ref{en_det1}-\ref{f} can be understood by deriving an effective Hamiltonian valid sufficiently far from resonance (dispersive regime) \cite{zhu_circuit_2013}, details of which are given in Supplementary Note 4. In this context let us first consider the flux-gauge wherein the light and matter systems are coupled through the material position operator $\phi$. The matrix elements of this operator between material states $\ket{\epsilon_n^1}$ are largest between adjacent levels $n$, $n\pm1$ \cite{de_bernardis_breakdown_2018} (see Supplementary Note 4). Thus, provided higher material levels are sufficiently separated from the lowest two the coupling to them can be neglected, unless the light-matter coupling $\eta$ is very large, or $\delta$ is large enough that several material energies lie within the first oscillator band $\omega$. For such large $\delta$, contributions of energy denominators in the effective Hamiltonian become large due to the occurence of resonances $\epsilon_{ni}\sim \omega$, $\epsilon_{ni}=\epsilon_n-\epsilon_i,~i=0,1,~n>1$ (see Supplementary Note 4). The flux-gauge QRM is therefore qualitatively accurate if $\delta$ and $\eta$ are sufficiently small. This includes accurately predicting higher system energy levels $E_n>E$ as well as the first two levels $G$ and $E$ \cite{de_bernardis_breakdown_2018} (see Supplementary Note 5).

In the charge-gauge the light-matter coupling occurs via the material canonical momentum $\xi_0$, for which matrix elements involving higher levels are not suppressed (see Supplementary Note 4). Independent of $\delta$, when the coupling is sufficiently large they cannot generally be neglected even for highly anharmonic material spectra, so the charge-gauge QRM generally breaks down \cite{de_bernardis_breakdown_2018}. However, the ratio of the flux-gauge QRM coupling strength $g$ and the coupling strength ${\tilde g}_0$ of the charge-gauge QRM, increases as $\delta$ increases (see Supplementary Note 4). For large enough $\delta$ the charge-gauge coupling is significantly weaker than that of the flux-gauge to the extent that for sufficiently large $\delta$ and provided $\eta$ does not become too large, the charge-gauge QRM is qualitatively accurate for the ground level $G$, and occasionally for first level $E$ (Figs. \ref{en_det1}-\ref{f}).

In the general $\alpha$-gauge all flux-gauge coupling terms are weighted by $\alpha$ and all charge-gauge coupling terms by $1-\alpha$. By tuning $\alpha$ the $\alpha$-gauge two-level model smoothly interpolates between the two available QRMs. In particular the $\alpha_{\rm JC}$-gauge JCM is defined such that the counter-rotating terms that give the dominant contribution to deviations between the exact and two-level model ground states are eliminated (see Supplementary Note 4). This allows us to understand why the $\alpha_{\rm JC}$-gauge JCM accurately represents the ground state across all parameter regimes. As $\delta$ and $\eta$ increase the $\alpha_{\rm JC}$-gauge becomes predominantly charge-like (see Supplementary Note 4) and like the charge-gauge QRM becomes inaccurate for predicting levels $E_n>E$.

Quite generally two-level models remain most accurate in predicting the first two system levels $G$ and $E$. For the lowest such levels of certain circuit QED systems spectroscopic experimental data has been matched to the predictions of the QRM defined by the Hamiltonian $h=-(\Delta \sigma^z + \epsilon \sigma^x )/2 + \omega a^\dagger a + g'\sigma^x(a+a^\dagger)$ where $\Delta$ and $\epsilon$ are tunnelling and bias parameters respectively, and $g'$ denotes the coupling strength \cite{yoshihara_superconducting_2017,yoshihara_characteristic_2017,yoshihara_inversion_2018}. In Ref. \cite{yoshihara_characteristic_2017} for example, the parameters $\Delta,~g'$, and $ \omega$ are treated as constant fitting parameters while $\epsilon$ is externally variable. It is important to note however, that fitting transitions between eigenenergies of $h$ to experimental data does not preclude the possibility of fitting other models to experimental data. 

It is possible to rotate the flux-gauge QRM $H_2^1$ into the form of $h$, but upon doing so each of $\Delta$, $\epsilon$ and $g'$ are found to be non-trivial functions of $\phi_{\rm ext}$. In particular, for the fluxonium-$LC$ system we consider $g'$ and $\Delta$ do not remain constant while varying $\epsilon$ by varying $\phi_{\rm ext}$. Moreover, the $\alpha$-gauge two-level model cannot be uniquely specified in terms of the parameters of $h$. Whenever $\phi_{\rm ext}\neq {\rm \uppi}/2e$ these properties obstruct meaningful comparison between our results and experimental results of the kind found for example in Ref. \cite{yoshihara_characteristic_2017}.

More relevant experimental results for the system we consider are given in  \cite{manucharyan_fluxonium:_2009} where spectroscopic data was found to agree well with the non-truncated fluxonium-$LC$ Hamiltonian $H$ of Eq. (\ref{Hcircmain}). There the fluxonium energies $E_{\rm c}$, $E_{\rm l}$ and $E_{\rm J}$ were treated as fitting parameters. Our results show that using such a fitting procedure, the JC-gauge two-level model would offer better agreement with experimental data than the QRM, at least for the lowest two levels $G$ and $E$. This occurs over the full range of $\delta$ shown in Fig. \ref{en_det1} with only a few exceptions in the case of the excited state $E$ when $\delta$ is small (see Supplementary Note 5).\\

%We have derived arbitrary gauge descriptions of cavity and circuit QED systems comprised of an atomic system interacting with a single oscillator. Each gauge incurs a different approximate model via truncation of the atomic bare-energy eigenstates. The Rabi and Jaynes-Cummings models are special cases of a more general form of interaction. A fluxonium-oscillator system was considered, and the JCM was shown to be accurate beyond the regime of validity of the RWA. It was also found to be more accurate than the QRM in various situations involving the lowest two energy eigenstates of the system.

The results presented here open up multiple avenues for further investigation. For example, our more general form of two-level model in which the gauge is left open is capable (albeit fortuitously) of exactly predicting a given energy value, but it remains to be understood in more detail. A comprehensive comparison of different methods for deriving two-level model descriptions is also yet to be performed.

An investigation of the implications of the arbitrary gauge formalism for the occurence of phase transitions in multi-dipole systems  constitutes further important work. The dependence on arbitrary gauge parameters of weaker truncations such as three-level atomic models remains to be investigated as does the generalisation to multi-mode situations for structured photonic environments. We note that issues with the single-mode approximation have been recognised and discussed elsewhere \cite{niemczyk_circuit_2010,munoz_resolution_2018}, but that this approximation does not result in a breakdown of gauge-invariance and does not therefore affect the results reported here. Within exact (non-truncated) models determining the dependence on the gauge parameter of light-matter entanglement, as well as averages of local light and matter observables such as photon number, is of experimental relevance and is important for applications. This too will be investigated in further work.

\section*{METHODS}

\subsection*{Lagrangians in different gauges}

The Coulomb gauge Lagrangian is denoted $L_0$ and is given in Supplementary Note 1. More generally the $\alpha$-gauge Lagrangian yielding the same correct equations of motion as $L_0$ is $L_\alpha=L_0-d \chi_\alpha/d t$ where the function $\chi_\alpha$ is defined as
\begin{align}\label{chialphmain}
&\chi_\alpha(t)=\alpha \int d^3x\, {\bf A}\cdot {\bf P}_{\rm mult}, \\ &P_{{\rm mult},i}({\bf x}) = -e\int_0^1 d\lambda \, r_j\delta_{ij}^{\rm T}({\bf x}-\lambda{\bf r}).
\end{align}
Here ${\bf P}_{\rm mult}$ denotes the usual multipolar transverse polarisation field. Latin indices denote spatial components and repeated indices are summed. This $\chi_\alpha$ is the generator of the unitary Power-Zienau-Woolley transformation, multiplied by $\alpha$. The $\alpha$-dependence of the Lagrangian can be understood as the underlying cause of the $\alpha$-dependence of the canonical momenta ${\bf p}_\alpha =\partial L_\alpha/\partial {\dot {\bf r}}$ and ${\bf \Pi}_\alpha=\delta L_\alpha/\delta {\dot {\bf A}}$.

\subsection*{Derivation of cavity QED two-level model Hamiltonian}

Substituting Eqs. (\ref{gimomsa}) and (\ref{gimomsb}) into Eq.~(\ref{cav_ham}) yields the Hamiltonian written in terms of canonical operators ${\bf y}_\alpha$ as $H=H_{\rm m}^\alpha +H_{\rm c}^\alpha +V^\alpha$ where
\begin{align}
H_{\rm m}^\alpha =&{{\bf p}_\alpha^2\over 2m}+V({\bf r}),\label{hparts00}  \\
H_{\rm c}^\alpha = &{v\over 2}\left({\bf \Pi}_\alpha^2 +\omega^2{\bf A}^2\right),\label{hparts0} \\
V^\alpha =& {e\over m}(1-\alpha){\bf p}_\alpha\cdot {\bf A} +\alpha {\hat {\bf d}}\cdot {\bf \Pi}_\alpha\nonumber \\
&+{e^2\over 2m}(1-\alpha)^2{\bf A}^2 +{\alpha^2\over 2v}({\bm \varepsilon}\cdot {\hat {\bf d}})^2.\label{hparts}
\end{align}
The Hamiltonian has a hybrid form between the Coulomb and multipolar gauges. Coulomb gauge coupling terms are weighted by $1-\alpha$ while multipolar gauge coupling terms are weighted by $\alpha$. The interaction includes the quadratic ``${\bf A}^2$" and ``${\hat {\bf d}}^2$" self-energy terms in addition to the linear coupling terms ``${\bf p}_\alpha \cdot {\bf A}$" and ``${\hat {\bf d}}\cdot {\bf \Pi}_\alpha$". This approach is easily adapted to describe multi-mode fields and more than one dipole \cite{stokes_master_2018}.

The first two eigenstates of the material bare energy $H_{\rm m}^\alpha$ are denoted $\ket{\epsilon^\alpha_0}$ and $\ket{\epsilon_1^\alpha}$, and the projection onto this subspace is $P^\alpha = \ket{\epsilon^\alpha_0}\bra{\epsilon^\alpha_0}+\ket{\epsilon^\alpha_a}\bra{\epsilon^\alpha_1}$. The operator $H_{\rm m}^\alpha$ admits the two-level truncation $H^\alpha_{{\rm m},2}=P^\alpha H^\alpha_{\rm m} P^\alpha=\epsilon_0+\omega_{\rm m}^\alpha \sigma^+_\alpha\sigma^-_\alpha$ where $\omega_{\rm m}=\epsilon_1-\epsilon_0$, $\sigma^+_\alpha= \ket{\epsilon^\alpha_1}\bra{\epsilon^\alpha_0}$ and $\sigma^-_\alpha= \ket{\epsilon^\alpha_0}\bra{\epsilon^\alpha_1}$. The eigenvalues $\epsilon_0$ and $\epsilon_1=\omega_{\rm m}+\epsilon_0$ corresponding to $\ket{\epsilon_0^\alpha}$ and $\ket{\epsilon_1^\alpha}$ respectively, are $\alpha$-independent because $H^\alpha_{\rm m}= R_{\alpha\alpha'}H_{\rm m}^{\alpha'} R_{\alpha\alpha'}^{-1}$. In practice two-level model Hamiltonians are found by first defining the interaction Hamiltonian as $V_2^\alpha=V^\alpha(P^\alpha{\bf y}_\alpha P^\alpha)$ and then combining this interaction with the bare energies to obtain the total Hamiltonian 
\begin{align}\label{h2def}
H_2^\alpha = P^\alpha H_{\rm m}^\alpha P^\alpha +H_{\rm c}^\alpha +V^\alpha(P^\alpha {\bf y}_\alpha P^\alpha).
\end{align}
If the interaction Hamiltonian $V^\alpha$ is linear in ${\bf r}$ and ${\bf p}_\alpha$ then the two-level model Hamiltonian can also be written $H_2^\alpha = P^\alpha H P^\alpha$. This is not the case for $H$ in Eq. (\ref{hparts}) due to the ``${\hat {\bf d}}^2$" term, which demonstrates the availability of different methods for deriving truncated models. Here we adopt the approach most frequently encountered in the literature, and outline other methods in Supplementary Note 2.

We can now define an arbitrary-gauge two-level model associated with the Hamiltonian $H$ in Eq. (\ref{hparts}) by using the definition (\ref{h2def}). The projection $P^\alpha$ does not alter the ``${\bf A}^2$" and $H_{\rm c}^\alpha$ terms of Eq. (\ref{hparts}), because these terms depend on the cavity canonical operators only. Combining them gives the renormalised cavity energy $H_{\rm c}^\alpha+{e^2/2m}(1-\alpha)^2{\bf A}^2=\omega_\alpha(c_\alpha^\dagger c_\alpha +1/2)$ with renormalised cavity frequency $ \omega_\alpha = \omega\sqrt{1+e^2(1-\alpha)^2/ mv\omega^2}$. The $c_\alpha,~c_\alpha^\dagger$ are cavity ladder operators of the renormalised energy satisfying $[c_\alpha,c_\alpha^\dagger]=1$. In terms of these operators the Hamiltonian $H_2^\alpha$ defined by Eq. (\ref{h2def}) is given by Eq. (\ref{HPalph}).

\subsection*{Method for comparing two-level model predictions}

A comparison of the predictions that different two-level models yield for an arbitrary observable requires that we determine how a given physical state is represented within each two-level model. To this end consider an observable $A$ with the property that both the exact representation $A$ and the two-level model representation $A_2^\alpha$ possess non-degenerate discrete spectra. The eigenvalues $a_n$ of $A$ and $a_{2,n}^\alpha$ of $A_2^\alpha$ are in one-to-one correspondence such that the eigenstates $\ket{A_n}$ and $\ket{A^\alpha_{2,n}}$ can be assumed to represent the same physical state. An arbitrary physical state can then be constructed via linear combination; the physical state $\ket{\psi}=\sum_n \psi_n\ket{A_n},~\sum_n|\psi_n|^2=1$ within the exact theory, is represented within the $\alpha$-gauge two-level model by $\ket{\psi^\alpha_2} =\sum_n \psi_n \ket{A_{2,n}^\alpha}$. A natural choice of observable $A$ for the purpose of representing states is the energy $A=H$, which we consider in Results.

The most accurate two-level model for the purpose of predicting the average $\bra{\psi} O \ket{\psi}$ of an arbitrary observable $O$, which may or may not equal $A$, is found by selecting the gauge $\alpha$ for which the difference between the exact and two-level model prediction, $z^\alpha(O,\psi)=|\bra{\psi} O\ket{\psi}-\bra{\psi^\alpha_2} O_2^\alpha \ket{\psi^\alpha_2}|$, is minimised. Since two-level models are indispensable practical tools within cavity and circuit QED it is important to ascertain which two-level models yield the best approximations of physical averages that are of interest in applications. In Results the energy is considered, both to represent states ($A=H$) and as the observable of interest ($O=H$). The averages $\bra{A_n}O\ket{A_n}$ are then nothing but the eigenvalues $E_n$ of $H$. 

As an example illustrating how the relative accuracies of two-level models can be determined let us consider the quantities $z^\alpha(O,G)$ where $\ket{G}$ denotes the ground state of a composite cavity-charge system. The charge is assumed to be confined in all directions except the direction ${\bm \varepsilon}$ of the cavity mode polarisation. In this direction it oscillates harmonically with bare frequency $\omega_{\rm m}$. In the gauge specified by choosing $\alpha=\omega_{\rm m}/( \omega_{\rm m} +\omega)$, the matter oscillator can be described by ladder operators for which the interaction Hamiltonian takes number-conserving form \cite{drummond_unifying_1987}. The exact ground state $\ket{G}$ is then the vacuum state of these modes, and the projection $P^{\rm JC}$ onto the first two material levels in this gauge defines a two-level JCM with ground state $\ket{G^{\rm JC}_2}=P^{\rm JC}\ket{G}=\ket{G}$. It follows that $z^\alpha(O,G)=0$ for all $O$ with $O_2^\alpha=P^\alpha O P^\alpha$. Thus, if the material system is a harmonic oscillator, it is possible to derive a JCM that is necessarily more accurate than any derivable QRM for finding ground state averages.

\subsection*{Fluxonium-$LC$ two-level model Hamiltonian}

The derivation in Supplementary Note 3 yields the $\alpha$-gauge fluxonium-$LC$ Hamiltonian
\begin{align}\label{Hcircmain}
H=&{E_{\rm c}\over e^2}[\xi_\alpha+(1-\alpha)\zeta]^2+2e^2 E_{\rm l}\phi^2 \nonumber \\ &-E_{\rm J}\cos\left(2e [\phi-\phi_{\rm ext}] \right)+ {\zeta^2\over 2C} + {1\over 2L}[\theta_\alpha+\alpha\phi]^2.
\end{align}
The fluxonium bare-energy is defined as
\begin{align}
H_{\rm m}^\alpha={E_{\rm c}\over e^2}\xi_\alpha^2+2e^2 E_{\rm l}\phi^2 -E_{\rm J}\cos\left(2e [\phi-\phi_{\rm ext}] \right).
\end{align}
The projection onto the first two eigenstates of this operator is used along with $H$ in Eq. (\ref{Hcircmain}) to define a two-level model Hamiltonian in precisely the same way as in the cavity QED case. The final result is given in Eq. (\ref{HPbet}).\\

\noindent\textit{Data Availability}. The data that support the findings of this study are available from the corresponding authors upon reasonable request.\\

\noindent\textit{Author Contributions}. All authors contributed to all aspects of this work.\\

\noindent\textit{Acknowledgement}. This work was supported by the UK Engineering and Physical Sciences Research Council, grant no. EP/N008154/1. We thank Zach Blunden-Codd for useful discussions.\\

\noindent\textit{Competing interests}. The authors declare no competing interests.\\

\renewcommand{\figurename}{Supplementary Figure}

\clearpage

\onecolumngrid

\begin{center}
\textbf{Supplementary Information}
\end{center}

\section*{Supplementary Note 1: Arbitrary gauge quantisation of the matter-radiation system}\label{gi}

In this part our aim is to start from first principles with the Maxwell-Lorentz system of equations and derive a quantum Hamiltonian, that describes the interaction of a material system with a single-mode of radiation. Throughout the derivation we will carefully keep track of the gauge-freedom that is inherent within the electromagnetic potentials. In doing so we arrive at the final result given in the main text wherein the gauge freedom is parametrised by $\alpha\in {\mathbb R}$.

Throughout this section we will frequently use the Helmholtz decomposition of a vector field ${\bf V}$ into transverse and longitudinal parts ${\bf V}_{\rm T}$ and ${\bf V}_{\rm L}$ such that for all ${\bf x}$
\begin{align}
&{\bf V}={\bf V}_{\rm T}+{\bf V}_{\rm L},\\ &\nabla \cdot {\bf V}_{\rm T}({\bf x})=0,\\ &\nabla \times {\bf V}_{\rm L}({\bf x}) ={\bf 0}.
\end{align}
We assume that all vector fields vanish at the boundaries $|{\bf x}|\to \infty$, which allows free use of integration by parts such as
\begin{align}\label{intbpts}
\int d^3 x\, {\bf V}({\bf x})\cdot \nabla f({\bf x}) = -\int d^3 x \, f({\bf x})\nabla \cdot {\bf V}({\bf x}).
\end{align}
Recalling that $\nabla \times \nabla f({\bf x}) ={\bf 0}$ for any $f$ and for all ${\bf x}$, we have that for any longitudinal field ${\bf U}_{\rm L}$ there exists an $f$ such that  ${\bf U}_{\rm L}=\nabla f$. It follows from Supplementary Eq.~(\ref{intbpts}) that
\begin{align}\label{vu}
 \int d^3 x\, {\bf V}_{\rm T}({\bf x})\cdot {\bf U}_{\rm L}({\bf x}) =0
\end{align}
for any vector fields ${\bf V}$ and ${\bf U}$. These formulae will be frequently used in what follows.

For notational simplicity in the main text we used ${\bf A}$ to denote the transverse component of the vector potential. However, here we will deal with both the transverse and full vector potentials. We denote the total vector potential by ${\bf A}$ while its gauge-invariant transverse component is denoted ${\bf A}_{\rm T}$. The scalar potential is denoted $A_0$. A gauge transformation of the potentials
\begin{align}
&A_0\to A_0-{\dot \chi},\\ &{\bf A}\to {\bf A}+\nabla\chi
\end{align}
where $\chi$ is arbitrary, leaves the Maxwell-Lorentz equations invariant. By defining ${\bf B}=\nabla\times {\bf A}$ the non-dynamical Maxwell equation $\nabla\cdot {\bf B}=0$, which specifies the transverasailty of the magnetic field is satisfied identically, because the divergence of the curl is identically zero; $\nabla\cdot \nabla \times {\bf V}({\bf x})=0$ for any ${\bf V}$ and for all ${\bf x}$. On the other hand the non-dynamical Gauss law $\nabla\cdot {\bf E}=\rho$ where $\rho$ is the charge density, is a primary constraint, which generates gauge transformations, and which indicates redundancy within the unconstrained theory. The remaining constraint required to eliminate this redundancy is a gauge-fixing condition. As will be made precise in what follows, a convenient method of handling the gauge freedom uses the arbitrary transverse component ${\bf g}_{\rm T}={\bf g}-{\bf g}_{\rm L}$ of the green's function ${\bf g}$ for the divergence operator \cite{woolley_r._g._charged_1999,stokes_noncovariant_2012}. The green's function is defined by the equation
\begin{align}\label{pol}
\nabla \cdot {\bf g}({\bf x},{\bf x}') = \delta ({\bf x}-{\bf x}'),
\end{align}
but since $\nabla \cdot {\bf g}({\bf x},{\bf x'}) \equiv \nabla \cdot {\bf g}_{\rm L}({\bf x},{\bf x}')$, Supplementary Eq.~(\ref{pol}) only fixes ${\bf g}_{\rm L}={\bf g}-{\bf g}_{\rm T}$ uniquely as
\begin{align}\label{gl}
{\bf g}_{\rm L}({\bf x},{\bf x}') = -\nabla {1\over 4\uppi |{\bf x}-{\bf x}'|}.
\end{align}
Any field ${\bf g}_{\rm T}$ with $\nabla \cdot {\bf g}_{\rm T}({\bf x},{\bf x}')=0$, can be added to ${\bf g}_{\rm L}$ in Supplementary Eq.~(\ref{gl}) to obtain a ${\bf g}$ that satisfies Supplementary Eq.~(\ref{pol}).

We consider bound charges $-e$ and $+e$ where the charge $+e$ is stationary and fixed at the origin. For generality we include an additional external potential $V_{\rm ext}({\bf r})$ acting on the charge $-e$ at ${\bf r}$. We define the gauge-invariant non-relativistic Lagrangian as \cite{woolley_r._g._charged_1999}
\begin{align}\label{lag2}
L(t):= {1\over 2}m{\dot {\bf r}}^2 - V_{\rm ext}({\bf r}) - \int d^3x\, \left( \rho A_0- {\bf J}\cdot{\bf A} \right) + {1\over 2}\int d^3x\, \left({\bf E}^2-{\bf B}^2\right) -{d \over d t}\int d^3x\, {\bf A}\cdot {\bf P}
\end{align}
where $\rho({\bf x})=-e\delta({\bf x}-{\bf r})+e\delta({\bf x})$, ${\bf J}({\bf x}) = -e{\dot {\bf r}}\delta({\bf x}-{\bf r})$ and the polarisation field is defined by $-\nabla\cdot {\bf P}=\rho$. Using Supplementary Eq.~(\ref{pol}) we obtain
\begin{align}\label{Pintermsofg}
{\bf P}({\bf x}) := -\int d^3 x'\, {\bf g}({\bf x},{\bf x}')\rho({\bf x}').
\end{align}
Note that ${\bf P}_{\rm L}$ is fixed by Supplementary Eq.~(\ref{gl}), but ${\bf P}_{\rm T}$ is arbitrary. In Supplementary Eq.~(\ref{lag2}) ${\bf E}$ and ${\bf B}$ are electric and magnetic fields respectively. For each time $t$ the components of these vector fields belong to the real Hilbert space $L^2({\mathbb R}^3)$. The final term in Supplementary Eq.~(\ref{lag2}) is a total time derivative, so it does not affect the equations of motion. It does however ensure that the Lagrangian in Supplementary Eq.~(\ref{lag2}) is gauge-invariant. The gauge-invariance of $L(t)$ can easily be verified by making a gauge transformation of the potentials, followed by integration by parts, and then use of the continuity equation ${\dot \rho}=-\nabla \cdot {\bf J}$.

We identify two constraint functions, which are the Gauss law
\begin{align}
C_1:=\rho-\nabla\cdot{\bf E},
\end{align}
and a constraint on the form of ${\bf A}$
\begin{align}
C_2:=\int d^3 x' {\bf g}({\bf x}',{\bf x})\cdot {\bf A}({\bf x'}).
\end{align}
From $C_1=0$ it follows that ${\bf E}_{\rm L}=-{\bf P}_{\rm L}$, where ${\bf P}_{\rm L}$ is found using Eqs. (\ref{Pintermsofg}) and (\ref{gl}), while the constraint $C_2=0$ is chosen such that the final term in Supplementary Eq.~(\ref{lag2}), which is a total time derivative, vanishes. It is easily verified using Eqs. (\ref{intbpts}) and (\ref{pol}) that a set of vector potentials satisfying $C_2=0$ identically, are those such that
\begin{align}\label{arbvecpo1}
{\bf A}({\bf x})={\bf A}_{\rm T}({\bf x}) + \nabla \int d^3 x' \, {\bf g}({\bf x}',{\bf x})\cdot {\bf A}_{\rm T}({\bf x}') \equiv {\bf A}_{\rm T}({\bf x})+{\bf A}_{\rm L}({\bf x})
\end{align}
where in writing the second equality we have recalled that the gradient of a function is necessarily longitudinal. Supplementary Eq.~(\ref{arbvecpo1}) shows that we can use the components of the gauge-invariant transverse vector potential ${\bf A}_{\rm T}$ to specify any \textit{total} vector potential in the field configuration space that satisfies $C_2=0$. A particular choice of ${\bf g}_{\rm T}$ uniquely specifies the longitudinal vector potential as
\begin{align}\label{al}
{\bf A}_{\rm L}({\bf x}) = \nabla \int d^3 x' \, {\bf g}({\bf x}',{\bf x})\cdot {\bf A}_{\rm T}({\bf x}') \equiv \nabla \int d^3 x' \, {\bf g}_{\rm T}({\bf x}',{\bf x})\cdot {\bf A}_{\rm T}({\bf x}')
\end{align}
where we have used Supplementary Eq.~(\ref{vu}). It follows from Supplementary Eq.~(\ref{al}) that the longitudinal vector potential is completely independent of ${\bf g}_{\rm L}$, and is fully specified by choosing ${\bf g}_{\rm T}$. Note that throughout our approach ${\bf g}_{\rm T}$ has remained arbitrary. All of the above equations, as well as the constraints $C_1=0=C_2$ hold independently of the choice of ${\bf g}_{\rm T}$. Since ${\bf g}_{\rm T}$ uniquely specifies ${\bf A}_{\rm L}$ the freedom to choose ${\bf g}_{\rm T}$ must be interpreted as \textit{gauge} freedom. The only constraint on ${\bf g}_{\rm T}$ is transversality, and for reasons of practicality it may also be required that ${\bf g}_{\rm T}$ is suitably well-behaved, having for example, a well-defined Fourier transform.

The purpose of encoding gauge-freedom in this way, is that we can now view ${\bf A}_{\rm T}$ as the only dynamical coordinate of the electromagnetic field, that is, ${\bf A}_{\rm L}$ can be eliminated in favour of ${\bf A}_{\rm T}$ and ${\bf g}_{\rm T}$ according to Supplementary Eq.~(\ref{al}). In conventional approaches a choice of gauge is made at the outset and the theory is then quantised. In contrast within the present framework quantisation can be achieved while keeping the gauge ${\bf A}_{\rm L}$ arbitrary, because it is possible to quantise the theory via Dirac's method without committing to an explicit choice for ${\bf g}_{\rm T}$. As detailed below we therefore obtain a fully quantum framework, such that the gauge can be specified at the quantum level by choosing the $c$-number function ${\bf g}_{\rm T}$. Different gauges are then seen to be related by unitary transformations.

Since ${\bf A}_{\rm T}$ is the vector potential associated with the Coulomb gauge, we deduce that the scalar potential $A_0$ accompanying the vector potential ${\bf A}$ given in Supplementary Eq.~(\ref{arbvecpo1}) is
\begin{align}\label{arbscpo1}
A_0({\bf x}) = \phi_{\rm coul}({\bf x}) - {\partial \over \partial t}\int d^3 x' \, {\bf g}({\bf x}',{\bf x})\cdot {\bf A}_{\rm T}({\bf x}')
\end{align}
where $\phi_{\rm coul}$ is, upto a factor of $e$, the Coulomb potential associated with the charge density $\rho$. If we substitute Eqs (\ref{arbvecpo1}) and (\ref{arbscpo1}) into Supplementary Eq.~(\ref{lag2}) and use $C_1=0$ we obtain
\begin{align}\label{lag4}
L(t)=L_0(t)-{d \over d t}\chi(t)
\end{align}
where $L_0(t)$ is the Lagrangian associated with the Coulomb gauge and is given by \cite{cohen-tannoudji_photons_1997}
\begin{align}\label{cglag}
L_0(t)={1\over 2}m{\dot {\bf r}}^2 - V({\bf r}) +\int d^3x\, {\bf J}\cdot{\bf A}_{\rm T} + {1\over 2}\int d^3x\, \left({\bf E}_{\rm T}^2-{\bf B}^2\right).
\end{align}
The arbitrary function $\chi(t)$ appearing in Supplementary Eq.~(\ref{lag4}) is independent of ${\bf g}_{\rm L}$ and is determined entirely through a choice of gauge ${\bf g}_{\rm T}$. It is given by
\begin{align}\label{lag2b}
&\chi(t)=\int d^3x\, {\bf A}_{\rm T}\cdot {\bf P}\equiv \int d^3x\, {\bf A}_{\rm T}\cdot {\bf P}_{\rm T},\\ &{\bf P}_{\rm T}({\bf x}) := -\int d^3 x'\, {\bf g}_{\rm T}({\bf x},{\bf x}')\rho({\bf x}').
\end{align}
In Supplementary Eq.~(\ref{cglag}) $V({\bf r})=V_{\rm ext}({\bf r})+ V_{\rm coul}({\bf r})$ with
\begin{align}\label{Coulombself}
V_{\rm coul}({\bf r})={1\over 2}\int d^3 x \, \rho({\bf x})\phi_{\rm coul}({\bf x}) = {1\over 2} \int d^3 x \,{\bf E}_{\rm L}({\bf x})^2 ={1\over 2} \int d^3 x \,{\bf P}_{\rm L}({\bf x})^2.
\end{align}
This term includes the divergent Coulomb self-energy of each charge as well as the inter-charge Coulomb energy. Note that since $L_0(t)$ and $L(t)$ differ by a total time derivative they necessarily yield the same equations of motion.

We can now conveniently parametrise the choice of gauge by restricting our attention to functions ${\bf g}_{\rm T}$ that have the form
\begin{align}\label{galph}
g_{{\rm T},i}({\bf x},{\bf x}') := -\alpha\int_0^1 d\lambda \, x'_j \delta_{ij}^{\rm T}({\bf x}-\lambda{\bf x}')
\end{align}
where $\alpha \in {\mathbb R}$ is arbitrary. When $\alpha=0$ we have ${\bf A}={\bf A}_{\rm T}$, which specifies the Coulomb gauge. When $\alpha=1$ Eqs. (\ref{Pintermsofg}) and (\ref{galph}) yield the well-known multipolar transverse polarisation field in closed form \cite{cohen-tannoudji_photons_1997}
\begin{align}
P_{{\rm T},i}({\bf x})|_{\alpha=1}=P_{{\rm mult},i}({\bf x}) = -e\int_0^1 d\lambda \, r_j\delta_{ij}^{\rm T}({\bf x}-\lambda{\bf r}).
\end{align}
This polarisation field represents a continuum of infinitesimal dipoles each consisting of charges $+e$ and $-e$ that are stacked end-on-end, and which start at the charge $+e$ located at ${\bf 0}$ and end at the charge $-e$ located at ${\bf r}$. The vector potential corresponding to $\alpha=1$ is
\begin{align}
{\bf A}({\bf x})= {\bf A}_{\rm T}({\bf x}) - \nabla \int_0^1 d\lambda \, {\bf x}\cdot {\bf A}_{\rm T}(\lambda{\bf x}),
\end{align}
which satisfies ${\bf x}\cdot {\bf A}({\bf x})=0$. Denoting the Fourier transform of ${\bf A}$ by ${\tilde {\bf A}}$ we see that ${\bf x}\cdot {\bf A}({\bf x})=0$ is the position-space version of the condition ${\bf k}\cdot {\tilde {\bf A}}({\bf k})=0$ that defines the Coulomb gauge. The gauge defined by ${\bf x}\cdot {\bf A}({\bf x})=0$ is called the \textit{Poincar\'e} or multipolar gauge \cite{cohen-tannoudji_photons_1997}.

With the restriction given by Supplementary Eq.~(\ref{galph}) the Lagrangian in Supplementary Eq.~(\ref{lag4}) becomes
\begin{align}
L(t)\equiv L_\alpha(t) = L_0(t)-{d \over d t}\chi_\alpha(t)
\end{align}
where
\begin{align}\label{chialph}
\chi_\alpha(t)=\alpha \int d^3x\, {\bf A}_{\rm T}\cdot {\bf P}_{\rm mult}.
\end{align}
Using $L_\alpha(t)$ we can define the following canonical momenta
\begin{align}
&{\bf p}_\alpha = {\partial L_\alpha \over \partial {\dot {\bf r}}} =  m{\dot {\bf r}}-e{\bf A}_{\rm T}({\bf r}) + e\alpha \nabla \int_0^1 d\lambda \, {\bf r}\cdot {\bf A}_{\rm T}(\lambda{\bf r}),\label{conalphs}\\ &{\bf \Pi}_{{\rm T},\alpha} ={\delta L_\alpha \over \delta {\dot {\bf A}}_{\rm T}} = {\dot {\bf A}}_{\rm T}-\alpha {\bf P}_{\rm mult} = -{\bf E}_{\rm T}-\alpha {\bf P}_{\rm mult} \label{conalphsb}
\end{align}
where in finding the expression for ${\bf p}_\alpha$ we have used
\begin{align}
-{\partial \over \partial {\dot {\bf r}}}{d\chi_\alpha\over dt} = e\alpha \int_0^1 d\lambda \, \left[{\bf A}_{\rm T}(\lambda{\bf r})+r_i \nabla A_{{\rm T},i}(\lambda{\bf r})\right] = e\alpha \nabla \int_0^1 d\lambda \, {\bf r}\cdot {\bf A}_{\rm T}(\lambda{\bf r}).
\end{align}
Here the repeated index is summed and the first equality follows from Supplementary Eq.~(\ref{chialph}) and the chain rule
\begin{align}
{d\over dt}A_{{\rm T},i}(\lambda{\bf r}(t),t)={\dot A}_{{\rm T},i}(\lambda{\bf r}(t),t)+{\dot {\bf r}}(t)\cdot \nabla  A_{{\rm T},i}(\lambda{\bf r}(t),t).
\end{align}

Although we have been able to exhibit expressions for the canonical momenta in an arbitrary gauge $\alpha$, in order to pass to the canonical formalism we need to determine the algebraic properties of the canonical momenta and the position variables ${\bf r}$ and ${\bf A}_{\rm T}$. The Lie algebra of these variables must be consistent with the constraints, and must also suffice to obtain the correct equations of motion once we have obtained the Hamiltonian. Before we commit to the specific form of ${\bf g}_{\rm T}$ given in Supplementary Eq.~(\ref{galph}), we will quantise the classical description while keeping ${\bf g}_{\rm T}$ completely arbitrary. This is achieved using Dirac's method \cite{dirac_lectures_2003}, which yields the Hamiltonian
\begin{align}\label{Harbg}
H = & {1 \over 2m}\left({\bf p}+e\left[{\bf A}_{\rm T}({\bf r}) + \nabla \int d^3 x\, {\bf g}({\bf x},{\bf r})\cdot {\bf A}_{\rm T}({\bf x})\right]\right)^2 + V({\bf r})\nonumber \\ &+ {1\over 2} \int d^3 x \, \left[ \left( {\bf \Pi}_{\rm T} - \int d^3 x' {\bf g}_{\rm T}({\bf x},{\bf x}')\rho({\bf x}')\right)^2+ (\nabla \times {\bf A}_{\rm T})^2\right]
\end{align}
where the canonical variables $\{{\bf r}, {\bf p}, {\bf A}_{\rm T}, {\bm \Pi}_{\rm T}\}$ are fully specified by the commutation relations
\begin{align}
&[r_i,p_j] ={\rm i}\delta_{ij},\label{db3} \\  &[{\rm A}_{\rm T,i}({\bf x}),\Pi_{{\rm T},j}({\bf x}')] ={\rm i}\delta_{ij}^{\rm T}({\bf x}-{\bf x}'). \label{db32}
\end{align}
All other commutators between elements of $\{{\bf r}, {\bf p}, {\bf A}_{\rm T}, {\bm \Pi}_{\rm T}\}$ vanish identically. All observables are expressed as functions of these operators and Eqs.~(\ref{db3}) and (\ref{db32}), and the Hamiltonian in Supplementary Eq.~(\ref{Harbg}) provide all that is needed to obtain the time evolution of a given observable. In particular it is straightforward to verify that the Hamiltonian in Supplementary Eq.~(\ref{Harbg}) yields the correct Maxwell-Lorentz equations. The gauge-invariant vector potential ${\bf A}_{\rm T}$ appearing in Supplementary Eq.~(\ref{Harbg}) belongs to the Coulomb gauge in the sense that ${\bf A} \equiv {\bf A}_{\rm T}$ in this gauge, but the Hamiltonian itself has been expressed in an arbitrary gauge $g$, which is determined by ${\bf g}_{\rm T}$. Using the Heisenberg equation we see that the arbitrary $g$-gauge canonical momenta ${\bf p}$ and ${\bf \Pi}_{\rm T}$ can be identified in terms of the gauge-invariant observables $\{{\bf r}, {\bf A}_{\rm T}, {\dot {\bf r}}, {\dot {\bf A}}_{\rm T} = -{\bf E}_{\rm T}\}$, and the gauge dependent function ${\bf g}_{\rm T}$, as
\begin{align}
{\bf p}&= m{\dot {\bf r}} - e\left({\bf A}_{\rm T}({\bf r}) + \nabla \int d^3 x \,{\bf g}_{\rm T}({\bf x},{\bf r})\cdot {\bf A}_{\rm T}({\bf x})\right)\equiv m{\dot {\bf r}} - e{\bf A}({\bf r}),\label{conmomingaugeg} \\ {\bf \Pi}_{\rm T}({\bf x})&=-{\bf E}_{\rm T}({\bf x})-{\bf P}_{\rm T}({\bf x})\label{conmomingaugegb}
\end{align}
where ${\bf P}_{\rm T}$ is determined by ${\bf g}_{\rm T}$ as in Supplementary Eq.~(\ref{lag2b}). This shows clearly that the canonical momenta ${\bf p}$ and ${\bf \Pi}_{\rm T}$ are manifestly gauge-dependent. Upon restricting ourselves to the specific form of ${\bf g}_{\rm T}$ given in Supplementary Eq.~(\ref{galph}) the canonical momenta in Eqs.~(\ref{conmomingaugeg}) and (\ref{conmomingaugegb}) are seen to coincide with those given in Eqs.~(\ref{conalphs}) and (\ref{conalphsb}). Using Eqs.~(\ref{conmomingaugeg}) and (\ref{conmomingaugegb}) we see that in \textit{any} gauge $H$ can be written entirely in terms of gauge-invariant observables as the sum of material and field energies;
\begin{align}\label{giham}
H&= H_{\rm matter}+H_{\rm field},\\  \qquad H_{\rm matter}&:={1\over 2}m{\dot {\bf r}}^2 + V({\bf r}),\\ H_{\rm field} &:= {1\over 2} \int d^3 x\, \left({\bf E}_{\rm T}^2 + {\bf B}^2\right).
\end{align}
A unitary gauge-fixing transformation between gauges ${\bf g}_{\rm T}$ and ${\bf g}'_{\rm T}$ can be defined as
\begin{align}\label{fix}
R_{gg'} := \exp \left[{\rm i}\int d^3 {\bf x} \, [{\bf P}_{{\rm T},g'}({\bf x})-{\bf P}_{{\rm T},g}({\bf x})] \cdot {\bf A}_{\rm T}({\bf x})\right]
\end{align}
where ${\bf P}_{{\rm T},g}$ and ${\bf P}_{{\rm T},g'}$ are defined as in Supplementary Eq.~(\ref{lag2b}) in terms of ${\bf g}_{\rm T}$ and ${\bf g}'_{\rm T}$ respectively. When used to transform the canonical momenta in Eqs.~(\ref{conmomingaugeg}) and (\ref{conmomingaugegb}) $R_{gg'}$ replaces the function ${\bf g}_{\rm T}$ with the alternative choice ${\bf g}_{\rm T}'$.

Let us now return to the specific form of ${\bf g}_{\rm T}$ given in Supplementary Eq.~(\ref{galph}), wherein the freedom to choose a gauge reduces to the freedom to choose the value of the real parameter $\alpha$. In terms of this form of ${\bf g}_{\rm T}$ the canonical momenta and Hamiltonian are found using Eqs.~(\ref{conmomingaugeg}) and (\ref{conmomingaugegb}), and Supplementary Eq.~(\ref{Harbg}). As noted previously, in this case the canonical momenta are seen to coincide with those given in Supplementary Eq.~(\ref{conalphs}). The unitary gauge-fixing transformation between different gauges $\alpha$ and $\alpha'$ takes the form of a generalised Power-Zienau-Woolley transformation;
\begin{align}\label{pzw}
R_{\alpha\alpha'} := \exp \left[{\rm i}(\alpha-\alpha')\int d^3 {\bf x} \, {\bf P}_{\rm mult}({\bf x}) \cdot {\bf A}_{\rm T}({\bf x})\right].
\end{align}
The usual Power-Zienau-Woolley transformation, which is used to relate the Coulomb and Poincar\'e gauges is obtained if $\alpha-\alpha'=1$. The $\alpha$-gauge polarisation field is $\alpha {\bf P}_{\rm mult}$. A multipole expansion of the polarisation field allows one to perform the electric-dipole approximation as $P_{{\rm mult},i}^{\rm EDA}=-er_i\delta_{ij}^{\rm T}({\bf x})$. Equivalently, the dipole approximation can be realised via
\begin{align}
g_{{\rm T},i}^{\rm EDA}({\bf x},{\bf x}') = -\alpha x'_j\delta_{ij}^{\rm T}({\bf x}).
\end{align}
The dipole approximated $\alpha$-gauge canonical momenta can then be read-off from Eqs.~(\ref{conmomingaugeg}) and (\ref{conmomingaugegb}) as
\begin{align}
{\bf p}_\alpha &= m{\dot {\bf r}}-e(1-\alpha){\bf A}_{\rm T}({\bf 0}),\label{pedas} \\ \Pi_{{\rm T}.\alpha,i}({\bf x}) &= - E_{{\rm T},i}({\bf x})-\alpha d_j \delta_{ij}^{\rm T}({\bf x})\label{pedasb}
\end{align}
where ${\bf d}=-e{\bf r}$. The unitary gauge-fixing transformation becomes
\begin{align}\label{dipR}
R_{\alpha\alpha'} := \exp \left[{\rm i}(\alpha-\alpha'){\bf d}\cdot {\bf A}_{\rm T}({\bf 0})\right].
\end{align}
Since the gauge-fixing transformation remains unitary the dipole approximation does not destroy the gauge-invariance of the theory. The dipole approximated Hamiltonian is
\begin{align}\label{Heda}
H=&  {1\over 2m}[{\bf p}_\alpha + e(1-\alpha){\bf A}_{\rm T}({\bf 0})]^2 +V({\bf r})+ {1\over 2} \int d^3 x \left(\left[\Pi_{{\rm T},\alpha,i}({\bf x}) + \alpha d_j\delta_{ij}^{\rm T}({\bf x}) \right]^2 + [\nabla\times{\bf A}_{\rm T}({\bf x})]^2\right) \nonumber \\ =&  {1\over 2m}[{\bf p}_\alpha + e(1-\alpha){\bf A}_{\rm T}({\bf 0})]^2 +V({\bf r}) + {1\over 2} \int d^3 k \left(\bigg|{\tilde {\bf \Pi}}_{{\rm T},\alpha}({\bf k}) +{\alpha \over (2\uppi)^3}\sum_\lambda {\bm \varepsilon}_\lambda({\bf k})[{\bf d}\cdot{\bm \varepsilon}_\lambda({\bf k})] \bigg|^2 + |{\rm i}{\bf k}\times{\tilde {\bf A}}_{\rm T}({\bf k})|^2\right) 
\end{align}
where ${\bm \varepsilon}_\lambda({\bf k}),~\lambda=1,2$ are mutually orthogonal unit polarisation vectors that are both orthogonal to ${\bf k}$, and ${\tilde f}$ denotes the Fourier transform of $f$. We have also used
\begin{align}\label{pmulteda}
P_{{\rm mult},i}^{\rm EDA}({\bf x}) = d_j\delta_{ij}^{\rm T}({\bf x})=\int {d^3 k\over (2\uppi)^3} \sum_\lambda \varepsilon_{\lambda,i}({\bf k})[{\bf d}\cdot {\bm \varepsilon}_\lambda({\bf k})]{\rm e}^{{\rm i}{\bf k}\cdot {\bf x}}.
\end{align}
The above expressions are applicable for general field operators ${\bf A}_{\rm T}$ and ${\bf \Pi}_{{\rm T},\alpha}$. We define the operator
\begin{align}
a_{\alpha,\lambda}({\bf k}) := \sqrt{1\over 2\omega}\bigg(\omega {\tilde A}_{{\rm T},\lambda}({\bf k})+{\rm i}{\tilde \Pi}_{{\rm T},\alpha,\lambda}({\bf k})\bigg)
\end{align}
where ${\tilde A}_{{\rm T},\lambda}({\bf k})={\bm \varepsilon}_\lambda({\bf k})\cdot {\tilde {\bf A}}_{{\rm T}}({\bf k})$ and ${\tilde \Pi}'_{{\rm T},\alpha,\lambda}({\bf k})={\bm \varepsilon}_\lambda({\bf k})\cdot {\tilde {\bf \Pi}}'_{{\rm T},\alpha}({\bf k})$. From the transverse canonical commutation relation
\begin{align}
[A_{{\rm T},i}({\bf x}),\Pi_{{\rm T},\alpha,j}({\bf x}')]={\rm i}\delta_{ij}^{\rm T}({\bf x}-{\bf x}')
\end{align}
it follows that
\begin{align}\label{boscom2}
[a_{\alpha,\lambda}({\bf k}),a_{\alpha,\lambda'}^\dagger({\bf k}')]=\delta_{\lambda\lambda'}\delta({\bf k}-{\bf k}').
\end{align}
The operators $a_{\alpha,\lambda}({\bf k})$ and $a_{\alpha,\lambda}^\dagger({\bf k})$ are recognisable as annihilation and creation operators for a photon with momentum ${\bf k}$ and polarisation $\lambda$. In terms of these operators the canonical fields support the Fourier representations
\begin{align}\label{modeex}
&{\bf A}_{\rm T}({\bf x}) = \int d^3 k \sum_\lambda g{\bm \varepsilon}_\lambda({\bf k})\left(a^\dagger_{\alpha,\lambda}({\bf k}){\rm e}^{-{\rm i}{\bf k}\cdot {\bf x}} + a_{\alpha,\lambda}({\bf k}){\rm e}^{{\rm i}{\bf k}\cdot {\bf x}}\right),\nonumber \\ &
{\bm \Pi}_{{\rm T},\alpha}({\bf x}) = {\rm i}\int d^3 k \sum_\lambda \omega g{\bm \varepsilon}_\lambda({\bf k})\left(a^\dagger_{\alpha,\lambda}({\bf k}){\rm e}^{-{\rm i}{\bf k}\cdot {\bf x}} - a_{\alpha,\lambda}({\bf k}){\rm e}^{i{\bf k}\cdot {\bf x}}\right)
\end{align}
where $\omega=|{\bf k}|$ and $g:={1 / \sqrt{2\omega (2\uppi)^3}}$. 

If we assume an implicit cavity with volume $v$ that satisfies periodic boundary conditions, the continuous label ${\bf k}$ becomes discrete. The pair ${\bf k}\lambda$ then labels a radiation mode. As a less realistic, but simpler model for the cavity we may restrict our attention to a single mode, in which case the field operators become
\begin{align}\label{consm}
&{\bf A}_{\rm T} =  g {\bm \varepsilon} \left(a_\alpha^\dagger+a_\alpha\right),\\ &{\bf \Pi}_{\rm T,\alpha} ={\rm i} \omega g {\bm \varepsilon} \left(a_\alpha^\dagger-a_\alpha\right),
\end{align}
where $[a_\alpha,a_\alpha^\dagger]=1$ and $g=1/\sqrt{2\omega v}$. Eqs. (\ref{consm}) imply that the cavity canonical operators now satisfy the commutation relation 
\begin{align}
[A_{{\rm T},i},\Pi_{{\rm T},\alpha,j}]={{\rm i}\varepsilon_i \varepsilon_j \over v}
\end{align}
as specified in the main text. For consistency with Eqs.~(\ref{conmomingaugeg}), (\ref{conmomingaugegb}), (\ref{pedas}), and (\ref{pedasb}), within the single-mode approximation we must also restrict the Fourier transform of the polarisation field ${\bf P}_{\rm mult}$ in Supplementary Eq.~(\ref{pmulteda}) to a single mode such that the transverse electric field satisfies ${\bf E}_{\rm T} = -{\dot {\bf A}}_{\rm T}=-{\bf \Pi}_{{\rm T},\alpha}-\alpha{\bf P}_{\rm mult}$. If in the single-mode approximation we write the Hamiltonian in Supplementary Eq.~(\ref{Heda}) as
\begin{align}\label{H}
H=& \, {1 \over 2m}({\bf p}_\alpha + e(1-\alpha){\bf A}_{\rm T})^2 +V({\bf r})+ {v\over 2}\left(\left[{\bf \Pi}_{{\rm T},\alpha} + {\alpha{\bm \varepsilon}({\bf d}\cdot {\bm \varepsilon})\over v}\right]^2 + \omega^2{\bf A}_{\rm T}^2\right)
\end{align} 
where we have restricted the polarisation field to a single polarisation as $\alpha {\bf P}_{\rm mult} = \alpha{\bm \varepsilon}({\bf d}\cdot {\bm \varepsilon})/v$, we obtain
\begin{align}
{\bf E}_{\rm T}=-{\dot {\bf A}}_{\rm T}= -g {\bm \varepsilon} \left({\dot a}_\alpha^\dagger+{\dot a}_\alpha\right)=-{\rm i} \omega g {\bm \varepsilon} \left(a_\alpha^\dagger-a_\alpha\right)-{\alpha{\bm \varepsilon}({\bf d}\cdot {\bm \varepsilon})\over v}=-{\bf \Pi}_{\rm T}-\alpha {\bf P}_{\rm mult}
\end{align}
as required. We therefore obtain a consistent single-mode theory with Hamiltonian given by Supplementary Eq.~(\ref{H}), and cavity canonical operators ${\bf A}_{\rm T}$ and ${\bf \Pi}_{{\rm T},\alpha}$ fully specified by Eqs.~(\ref{consm}). Like the dipole approximation the single-mode approximation preserves the gauge-invariance of the theory, because it does not alter the unitary property of the gauge-fixing transformation $R_{\alpha\alpha'}$, which retains the form given in Supplementary Eq.~(\ref{dipR}) but with ${\bf A}_{\rm T}$ specifying the single-mode vector potential from Supplementary Eq.~(\ref{consm}). The Hamiltonian and the Heisenberg equation yield
\begin{align}
&m{\dot {\bf r}} = {\bf p}_\alpha + e(1-\alpha){\bf A}_{\rm T},\label{gimomsap}\\ &{\bf E}_{\rm T} = -{\bf \Pi}_{{\rm T},\alpha} - {\alpha{\bm \varepsilon}({\bf d}\cdot {\bm \varepsilon})\over v}\label{gimomsapb}
\end{align}
which are the single-mode versions of Eqs.~(\ref{pedas}) and  (\ref{pedasb}). Eqs.~(\ref{gimomsap}) and (\ref{gimomsapb}) allow us to write the Hamiltonian as  $H= E_{\rm matter}+E_{\rm cavity}$ where $E_{\rm matter}= m{\dot {\bf r}}^2/2+V({\bf r})$ and $E_{\rm cavity} =v({\bf E}_{\rm T}^2+\omega^2{\bf A}^2_{\rm T})/2$. This is merely the dipole-approximated single-mode version of Supplementary Eq.~(\ref{giham}).

In summary, the restriction to functions ${\bf g}_{\rm T}$ of the form given in Supplementary Eq.~(\ref{galph}), together with the electric-dipole approximation, and the restriction to a single-mode of radiation yield the expressions given in the main text.  For simplicity, in the main text we use the notation ${\bf A}$ for ${\bf A}_{\rm T}$ and ${\bf \Pi}_\alpha$ for ${\bf \Pi}_{\rm T,\alpha}$. The gauge is completely determined by $\alpha$. The theory is gauge-invariant in the sense that the predictions concerning any gauge-invariant observable can be calculated using any gauge and these predictions are unique. Choosing a specific gauge is merely a matter of convenience for performing calculations. As explained in the main text this is no longer the case within two-level models for the material system.

\section*{Supplementary Note 2: Can gauge-invariant predictions be obtained from two-level models in the weak-coupling regime?}\label{weak}

A well-known drawback of two-level models is the breakdown of sum-rules involving matrix elements of operators which satisfy the CCR algebra. This occurs because the CCR algebra cannot be supported by a finite-dimensional Hilbert space. A well-known example is given by the Thomas-Reiche-Kuhn (TRK) sum rule \cite{barton_frequency_1974}
\begin{align}\label{id}
\sum_{r=0}^\infty \epsilon_{rs}d_{rs}^id_{sr}^j = {e^2\over 2m}\delta_{ij}
\end{align}
where $d_{rs}^i = \bra{\epsilon^r_\alpha}-er_i\ket{\epsilon^s_\alpha}$ and $\epsilon_{rs}=\epsilon_r-\epsilon_s$. In the full (infinite-dimensional) atomic Hilbert space the right-hand-side of this identity is independent of the dipole level $s$. Yet, when considering a two-level dipole the values of the indices $r$ and $s$ on the right-hand-side of Supplementary Eq.~(\ref{id}) must be $0$ or $1$. For the ground state $\sigma^-_\alpha\sigma^+_\alpha$ with $s=0$ Supplementary Eq.~(\ref{id}) becomes $\omega_{\rm m}({\bf d}\cdot {\bm \varepsilon})^2 = e^2/ 2m$ while for the excited state $\sigma^+_\alpha \sigma^-_\alpha$ with $s=1$ Supplementary Eq.~(\ref{id}) becomes $\omega_{\rm m}({\bf d}\cdot {\bm \varepsilon})^2 = -e^2/ 2m$. These relations cannot be simultaneously satisfied. Furthermore, the second relation implies that $m<0$.

In conventional atomic physics it is necessary to use the TRK sum rule (\ref{id}) in order to show invariance, between the Coulomb and Poincar\'e gauges, of the Lamb shift derived using stationary second order perurbation theory \cite{craig_molecular_1998,stokes_master_2018}. Thus, one should already anticipate difficulties in the maintenance of gauge-invariance in two-level models even within the conventional weak-coupling regime. We note that the TRK sum rule has also been used extensively in the strong and ultrastrong light-matter physics literature, but in a different context. There it is applied on the level of the infinite-dimensional atom with the aim of deriving inequalities for atomic transitions involving the lowest two levels.

Our motivation here is different; we are concerned with the question of whether it is possible to establish gauge-invariance of predictions using non-equivalent two-level models. We therefore consider whether or not it is possible to elicit gauge-invariance of level-shifts through any systematic application of the TRK sum rule, \textit{after} the two-level approximation has been made in the arbitrary $\alpha$-gauge. We show that provided the TRK sum rule is applied judiciously, the precise meaning of which will be specified below, then gauge-invariance ($\alpha$-independence) can be elicited for the energy levels calculated in different two-level models, but only upto second order in $e$. Thus, the main conclusion of this Supplementary Note is that even with a somewhat ad hoc application of the TRK sum rule, in any two-level model one can at best expect to obtain $\alpha$-independent predictions upto order $d^2$ only. At the end of this section we also briefly discuss alternative definitions of two-level models.

We consider the two-level model Hamiltonian $H_2^\alpha$ given in Eq.~(5) of the main text, second order perturbation theory and judicious use of the relations $\omega_{\rm m}({\bf d}\cdot {\bm \varepsilon})^2 = e^2/ 2m$ and $\omega_{\rm m}({\bf d}\cdot {\bm \varepsilon})^2 = -e^2/ 2m$ yield $\alpha$-independent expressions for the ground and first excited energies. The ground energy of $H^\alpha_2$ found using second order perturbation theory in the interaction Hamiltonian is 
\begin{align}
\bra{G^\alpha_2} H^\alpha_2 \ket{G^\alpha_2} \approx {\omega_\alpha\over 2}+\Delta_\alpha - {{u_\alpha^+}^2\over \omega_{\rm m}+\omega_\alpha}.
\end{align}
Upon use of $\omega_{\rm m}({\bf d}\cdot {\bm \varepsilon})^2 = e^2/ 2m$ appropriate for the ground state we obtain to order $d^2$ the $\alpha$-independent result
\begin{align}
\bra{G_2^\alpha} H_2^\alpha \ket{G_2^\alpha} \approx \epsilon_0+{\omega\over  2} + [g {\bf d}\cdot {\bm \varepsilon}]^2 {\omega_{\rm m} \omega \over \omega_{\rm m}+\omega}.
\end{align}
Similarly, the first excited energy of $H^\alpha_2$ is to order $d^2$ given by 
\begin{align}
\bra{E_2^\alpha} H_2^\alpha \ket{E_2^\alpha} \approx {\omega_\alpha\over 2}+\Delta_\alpha + {{u_\alpha^-}^2\over \omega_{\rm m}-\omega_\alpha} \approx \epsilon_0+{\omega\over  2} + [g {\bf d}\cdot {\bm \varepsilon}]^2 {\omega_{\rm m} \omega \over \omega_{\rm m}-\omega}.
\end{align}
where in writing the second equality we have used $\omega_{\rm m}({\bf d}\cdot {\bm \varepsilon})^2 = -e^2/ 2m$ appropriate for the excited state. As in the case of an infinite-dimensional atom, the above exhibition of gauge-invariance relies upon the elimination of the mass $m$ in favour of the dipole moment ${\bf d}$ and other parameters \cite{stokes_master_2018,craig_molecular_1998}. The difference in the two-level model case is only that the separate relations $\omega_{\rm m}({\bf d}\cdot {\bm \varepsilon})^2 = e^2/ 2m$ and $\omega_{\rm m}({\bf d}\cdot {\bm \varepsilon})^2 = -e^2/ 2m$ must be used for the ground and excited states respectively, which essentially accounts for the modification of the operator algebra incurred by the two-level truncation \cite{stokes_master_2018}.

Motivated by the above derivations we now show that the bare mass can similarly be eliminated on the level of the Hamiltonian, and this yields another possible form of two-level model. More precisely, if within the two-level truncation we apply the appropriate relations $\omega_{\rm m}({\bf d}\cdot {\bm \varepsilon})^2 = e^2/ 2m$ for the ground state and $\omega_{\rm m}({\bf d}\cdot {\bm \varepsilon})^2 = -e^2/ 2m$ for the excited state in the self-energy term $e^2(1-\alpha)^2 {\bf A}^2/2m$, then this term becomes $-\omega_{\rm m}({\bf d}\cdot {\bm \varepsilon})^2 (1-\alpha)^2 \sigma^z_\alpha {\bf A}^2$ where $\sigma^z_\alpha = [\sigma^+_\alpha,\sigma^-_\alpha]$. All dependence on the bare mass $m$ has now been eliminated and we obtain a well-defined two-level model Hamiltonian given by
\begin{align}\label{HPalph2}
{\bar H}_2^\alpha= \omega_{\rm m}\sigma^+_\alpha\sigma^-_\alpha + \Delta_\alpha  + \omega_{\rm m}(1-\alpha)({\bf d}\cdot {\bf A})\sigma^y_\alpha +\alpha({\bf d}\cdot {\bf \Pi}_\alpha)\sigma^x_\alpha - \omega_{\rm m}({\bf d}\cdot {\bm \varepsilon})^2 (1-\alpha)^2 \sigma^z_\alpha {\bf A}^2 +\omega\left(a^\dagger_\alpha a_\alpha +{1\over 2}\right)
\end{align}
where $\sigma^y_\alpha = {\rm i}(\sigma^-_\alpha-\sigma^+_\alpha)$ and $\sigma^x_\alpha = \sigma^+_\alpha+\sigma^-_\alpha$.

In contrast to $H_2^\alpha$ a simple approximate relation can be given between the average energy found using distinct two-level model Hamiltonians ${\bar H}_2^\alpha$ and ${\bar H}_2^{\alpha'}$ when $\alpha \neq \alpha'$. To see this we note that the gauge transformation $R_{0\alpha}={\rm e}^{{\rm i}e\alpha {\bf r}\cdot {\bf A}}$ is a function of the canonical variables ${\bf r},~{\bf A}\in {\bf y}_0$. We express this functional dependence as $R_{0\alpha}\equiv R_{0\alpha}({\bf y})$ where ${\bf y}\equiv {\bf y}_0$, and we define the unitary operator $U_\alpha = R_{0\alpha}(P^0{\bf y} P^0) = \cos(\alpha{\bf d}\cdot{\bf A})-{\rm i}\sigma^x \sin(\alpha{\bf d}\cdot {\bf A})$ where $\sigma^k\equiv \sigma^k_0,~k=\pm,x,y,z$. Noting further that ${\bar H}_2^0$ is a function of the Coulomb-gauge Rabi model raising and lowering operators, expressed as ${\bar H}_2^0(\sigma^\pm)$, it is straightforward to show using this notation that $U_\alpha {\bar H}_2^0(\sigma^\pm) U_\alpha^{-1} \approx {\bar H}_2^\alpha(\sigma^\pm)$ where the approximate equality means that equality holds upto second order in $\eta$. If in the Coulomb gauge two-level model with Hamiltonian ${\bar H}_2^0$ we represent an arbitrary state ${\cal S}$ by $\ket{\psi^0}  =\sum_n \psi_n\ket{{\bar E}_{2,n}^0}$, then in the $\alpha$-gauge two-level model with Hamiltonian ${\bar H}_2^\alpha$, ${\cal S}$ is represented by $\ket{\psi^\alpha} =\sum_n \psi_n \ket{{\bar E}_{2,n}^\alpha}\approx U_\alpha\ket{\psi^0}$ and we therefore obtain $\bra{\psi^0}{\bar H}_2^0\ket{\psi^0} \approx \bra{\psi^\alpha}{\bar H}_2^\alpha\ket{\psi^\alpha}$.

Finally we remark on the possibility of yet another form of two-level model. In the main text the two-level model Hamiltonian is found by projecting the canonical operators ${\bf y}_\alpha$ as $P^\alpha {\bf y}_\alpha P^\alpha$ and substituting the projected operators into the interaction Hamiltonian. This is only equivalent to a projection of the Hamiltonian itself $P^\alpha H P^\alpha$ if the interaction is a linear function of the material operators within ${\bf y}_\alpha$. In the non-truncated Hamiltonian of Eqs. (9), (10) and (11) of the main text, however there is a non-linear term $\alpha^2 /2v ({\bm \varepsilon} \cdot {\bf d})^2$. The lowest two energy levels obtained for two-level models of the form $P^\alpha H P^\alpha$ are therefore quantitatively different in some cases to those found in the main text. However the main conclusions remain unchanged. As stated within the main text a comprehensive comparison of distinct types of two-level model is beyond the scope of this article and will be given elsewhere.

\section*{Supplementary Note 3: Fluxonium $LC$-oscillator Hamiltonian}\label{cqedham}

Here we derive the full Hamiltonian describing a fluxonium-$LC$ oscillator circuit. The fluxonium is described by flux operator $\phi$ with conjugate momentum $\xi$ such that $[\phi,\xi]={\rm i}$. The Hamiltonian is \cite{manucharyan_resilience_2017}
\begin{align}\label{fluxo}
H_{\rm m}={E_{\rm c} \over e^2}\xi^2-E_{\rm J}\cos\left(2e[\phi-\phi_{\rm ext}]\right)+2e^2 E_{\rm l}\phi^2 
\end{align}
where $E_{\rm c},~E_{\rm J}$ and $E_{\rm l}$ are the capacitive, Josephson and inductive energies respectively, and $\phi_{\rm ext}$ is the applied external flux. The  $LC$-oscillator is described by flux operator $\theta$ with conjugate momentum $\zeta$ such that $[\theta,\zeta]={\rm i}$. Its Hamiltonian is
\begin{align}\label{lco}
H_{LC}={\zeta^2 \over 2C}+{\theta^2\over 2L}
\end{align}
where $C$ and $L$ are the capacitance and inductance respectively.

There is considerable freedom in describing the coupling between the fluxonium and the oscillator. Capacitively coupling the systems is achieved through the replacement $\xi\to \xi+\zeta$. This can be viewed as analogous to the replacement ${\bf p}\to {\bf p}+e{\bf A}$, which results in the Coulomb-gauge coupling between an atom and a cavity. Inductively coupling the fluxonium and oscillator is achieved through the replacement $\theta\to \theta+\phi$, which can be viewed as analogous to the replacement ${\bf \Pi}\to {\bf \Pi}+{\bm \epsilon}({\bm \epsilon}\cdot {\bf d})/v$ that gives the Poincar\'e-gauge coupling between an atom and a cavity. We have already seen in the context of an atom-cavity system that this freedom in the description of the coupling is a gauge-freedom, and that the different descriptions are unitarily related. Analogously we call the capacitive coupling the charge-gauge description, and we call the inductive coupling the flux-gauge description. Making the replacement $\xi\to \xi+\zeta$ in the Hamiltonian in Supplementary Eq.~(\ref{fluxo}) and adding the bare oscillator Hamiltonian in Supplementary Eq.~(\ref{lco}) yields the Hamiltonian expressed in the charge-gauge;
\begin{align}
H={E_{\rm c} \over e^2}(\xi+\zeta)^2-E_{\rm J}\cos\left(2e[\phi-\phi_{\rm ext}]\right)+2e^2 E_{\rm l}\phi^2 +{\zeta^2 \over 2C}+{\theta^2\over 2L}.
\end{align}
If we define the unitary gauge transformation $R_{01}={\rm e}^{{\rm i}\phi\zeta}$ then we can define new canonical operators $\xi_1=R_{01}^{-1}\xi R_{01}$ and $\theta_1=R_{01}^{-1}\theta R_{01}$ in terms of which the Hamiltonian is expressed in the flux-gauge;
\begin{align}
H={E_{\rm c} \over e^2}\xi_1^2-E_{\rm J}\cos\left(2e[\phi-\phi_{\rm ext}]\right)+2e^2 E_{\rm l}\phi^2 +{\zeta^2 \over 2C}+{1\over 2L}(\theta_1+\phi)^2.
\end{align}
More generally, we define the unitary gauge transformation $R_{0\alpha}={\rm e}^{i\alpha\phi\zeta}$ and associated $\alpha$-gauge canonical operators by $\xi_\alpha=R_{0\alpha}^{-1}\xi R_{0\alpha}$ and $\theta_\alpha=R_{0\alpha}^{-1}\theta R_{0\alpha}$. The Hamiltonian expressed in terms of the $\alpha$-gauge canonical operators ${\bf y}_\alpha=\{\phi,~\xi_\alpha,~\theta_\alpha,~\zeta\}$ is
\begin{align}\label{Hcirc}
H={E_{\rm c}\over e^2}[\xi_\alpha+(1-\alpha)\zeta]^2+2e^2 E_{\rm l}\phi^2 -E_{\rm J}\cos\left(2e [\phi-\phi_{\rm ext}] \right) + {\zeta^2\over 2C} + {1\over 2L}[\theta_\alpha+\alpha\phi]^2.
\end{align}
In all gauges the canonical operators satisfy the canonical commutation relations due to the unitarity of the gauge transformation. The Hamiltonian $H$ has $\alpha$-independent (gauge-invariant) spectrum. More generally, the predictions for any observable $O=o^\alpha({\bf y}_\alpha)=o^{\alpha'}({\bf y}_{\alpha'})$ can be calculated using any gauge and these predictions are unique. The charge and flux-gauge descriptions are obtained by choosing $\alpha=0$ and $\alpha=1$ respectively. The $\alpha$-gauge two-level model is obtained in exact analogy with the atom-cavity formalism by using the projection $P^\alpha$ onto the first two eigenstates of the $\alpha$-gauge bare fluxonium Hamiltonian
\begin{align}\label{fluxo2}
H_{\rm m}^\alpha={E_{\rm c} \over e^2}\xi_\alpha^2-E_{\rm J}\cos\left(2e[\phi-\phi_{\rm ext}]\right)+2e^2 E_{\rm l}\phi^2.
\end{align}

\section*{Supplementary Note 4: Further anaysis via an effective Hamiltonian in the dispersive regime}\label{effs}

Here we provide more detailed analysis of predictions in different gauges with the aim of understanding which two-level models will be more accurate in which regimes. Our results are derived using Schrieffer-Wolff perturbation theory (also known as Van Vleck perturbation theory), and our presentation is similar to the one in Ref. \cite{zhu_circuit_2013}. An effective Hamiltonian describing a linearly coupled material-oscillator system is derived, which is valid within the dispersive regime $|\epsilon_{nm}-\omega| \gg |g_{nm}|\sqrt{N+1}$ where $g_{nm}$ are the coupling constants of the linear interaction, $\epsilon_{nm}$ are the material transition frequencies and $N$ denotes photon population. For the fluxonium-$LC$ system we consider, this regime requires detunings $\delta$ sufficiently far from resonance in comparison to the coupling strengths $\eta$ even if $N=0$. Although this is not always the case in the regimes we consider, the Schrieffer-Wolff perturbation method can be used to gain physical insight into the regimes of large and small $\delta$ for various coupling strengths, and can thereby reveal why particular two-level models become more accurate in particular regimes. The method uses an appropriate unitary transformation ${\rm e}^{{\rm i}S}$ and perturbation theory to derive a diagonal Hamiltonian describing the system.

As in Ref. \cite{zhu_circuit_2013} let us consider the general linear-coupling of a multi-level material system and a single oscillator. Partitioned into unperturbed and interaction parts the most general such Hamiltonian is
\begin{align}\label{lingen}
H = H_0+V \equiv \left[\sum_n \epsilon_n \ket{\epsilon_n}\bra{\epsilon_n} + \Omega a^\dagger a\right] + \left[\sum_{nm} g_{nm} \ket{\epsilon_n}\bra{\epsilon_m}a+{\rm H.c.} \right].
\end{align}
If in the fluxonium-$LC$ Hamiltonian of Eq.~(6) of the main text we ignore the flux self-energy contribution $\alpha^2 \phi^2/2L$, which is not important for our analysis, then the Hamiltonian in Eq.~(6) of the main text is of the form of $H$ in Supplementary Eq.~(\ref{lingen}) upto a c-number offset term.

Using an appropriate unitary transformation ${\rm e}^{{\rm i}S}$ and second order perturbation theory, in the dispersive regime an effective Hamiltonian can be derived from Supplementary Eq.~(\ref{lingen}) as \cite{zhu_circuit_2013}
\begin{align}\label{psh}
H_{\rm eff} = \omega {\rm a}^\dagger {\rm a} + \sum_n (\epsilon_n +\kappa_n)\ket{n}\bra{n} + \sum_n \chi_n {\rm a}^\dagger {\rm a} \ket{n}\bra{n}
\end{align}
where $\ket{n}\bra{n}={\rm e}^{{\rm i}S}\ket{\epsilon_n}\bra{\epsilon_n}{\rm e}^{-{\rm i}S}$ and ${\rm a}={\rm e}^{{\rm i}S}a{\rm e}^{-{\rm i}S}$ are dressed operators. For details of the method of derivation of $H_{\rm eff}$ we refer to \cite{zhu_circuit_2013}. The additional energy coefficients $\kappa_n$ and $\chi_n$ are given by ordinary stationary second order perturbation theory and are found to be \cite{zhu_circuit_2013}
\begin{align}\label{shifts}
\kappa_n = \sum_{m} {|g_{nm}|^2 \over \epsilon_{nm}-\Omega},\qquad \chi_n =2 \sum_m  |g_{nm}|^2 {\epsilon_{nm} \over \epsilon_{nm}^2-\Omega^2}.
\end{align}
The $\kappa_n$ are material level shifts (Lamb-shifts) while the $\chi_n$ are ac-Stark shifts, which can be understood as material level-dependent oscillator shifts $\omega \to \omega + \chi_n$, or as oscillator-dependent material shifts; $\epsilon_n\to \epsilon_n+ \chi_n a^\dagger a$. It is important to note that for the fluxonium-$LC$ Hamiltonian in Eq.~(6) of the main text the unperturbed Hamiltonian $H_0$ as well as the interaction $V$ are different in each gauge. More precisely, for $H$ in Supplementary Eq.~(\ref{lingen}) to match the fluxonium-$LC$ Hamiltonian in Eq.~(6) of the main text the oscillator frequency $\Omega$ must be identified as $\omega_\alpha$ and the coupling constants $g_{nm}$ must be identified as
\begin{align}\label{mulcoup}
g_{nm} = {\phi_{nm} \over \sqrt{2\omega_\alpha L}}(\epsilon_{nm}(1-\alpha)+\omega_\alpha \alpha).
\end{align}
In the general $\alpha$-gauge we obtain $\kappa_n$ and $\chi_n$ through substitution of $\Omega=\omega_\alpha$ and Supplementary Eq.~(\ref{mulcoup}) into Supplementary Eq.~(\ref{shifts}). The expressions obtained are $\alpha$-dependent. Flux-gauge coupling is weighted in Supplementary Eq.~(\ref{mulcoup}) by $\alpha$ and depends on the matrix elements $\phi_{nm}$, which are shown for the first few fluxonium levels in Supplementary Fig.~\ref{dipsmoms} (a). Charge gauge coupling is weighted by $1-\alpha$ and depends on the matrix elements $\omega_{nm} \phi_{nm}$ shown in Supplementary Fig.~\ref{dipsmoms} (b). The matrix elements $\phi_{nm}$ are largest between adjacent states, so their higher level contributions are suppressed for a sufficiently anharmonic fluxonium system \cite{de_bernardis_breakdown_2018}. The same is not true for the matrix elements $\epsilon_{nm}\phi_{nm}$ and this is largely why the charge-gauge QRM is inaccurate even in the regime of small $\delta$ where the flux-gauge QRM is relatively accurate in predicting transition energies \cite{de_bernardis_breakdown_2018} (Supplementary Note 5).
%%%%%%%%%%%%%%%%%%%%%%%%%%%%%%%%%%%%%%%%%%%%%%%%%%%%%%%%%%%%%%%%%%%%%%%%%%%%%%%%%%%%%%%%%%%%%%%
%%	F I G U R E S  S T A R T
%%
%%%%%%%%%%%%%%%%%%%%%%%%%%%%%%%%%%%%%%%%%%%%%%%%%%%%%%%%%%%%%%%%%%%%%%%%%%%%%%%%%%%%%%%%%%%%%
\begin{figure}[t]
(a)  \includegraphics[width=.4\linewidth]{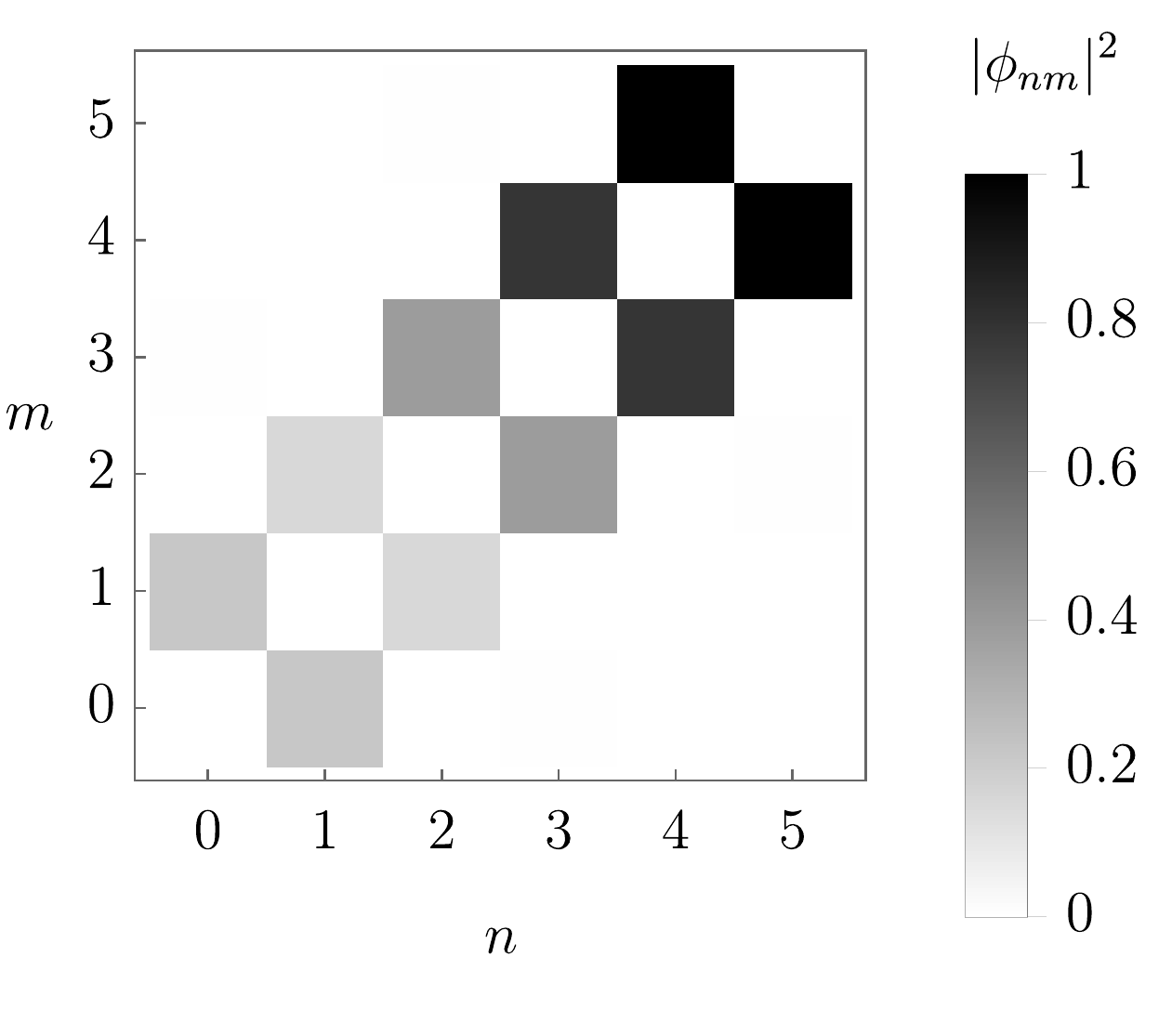}\hspace*{1.5cm}
(b) \includegraphics[width=.4\linewidth]{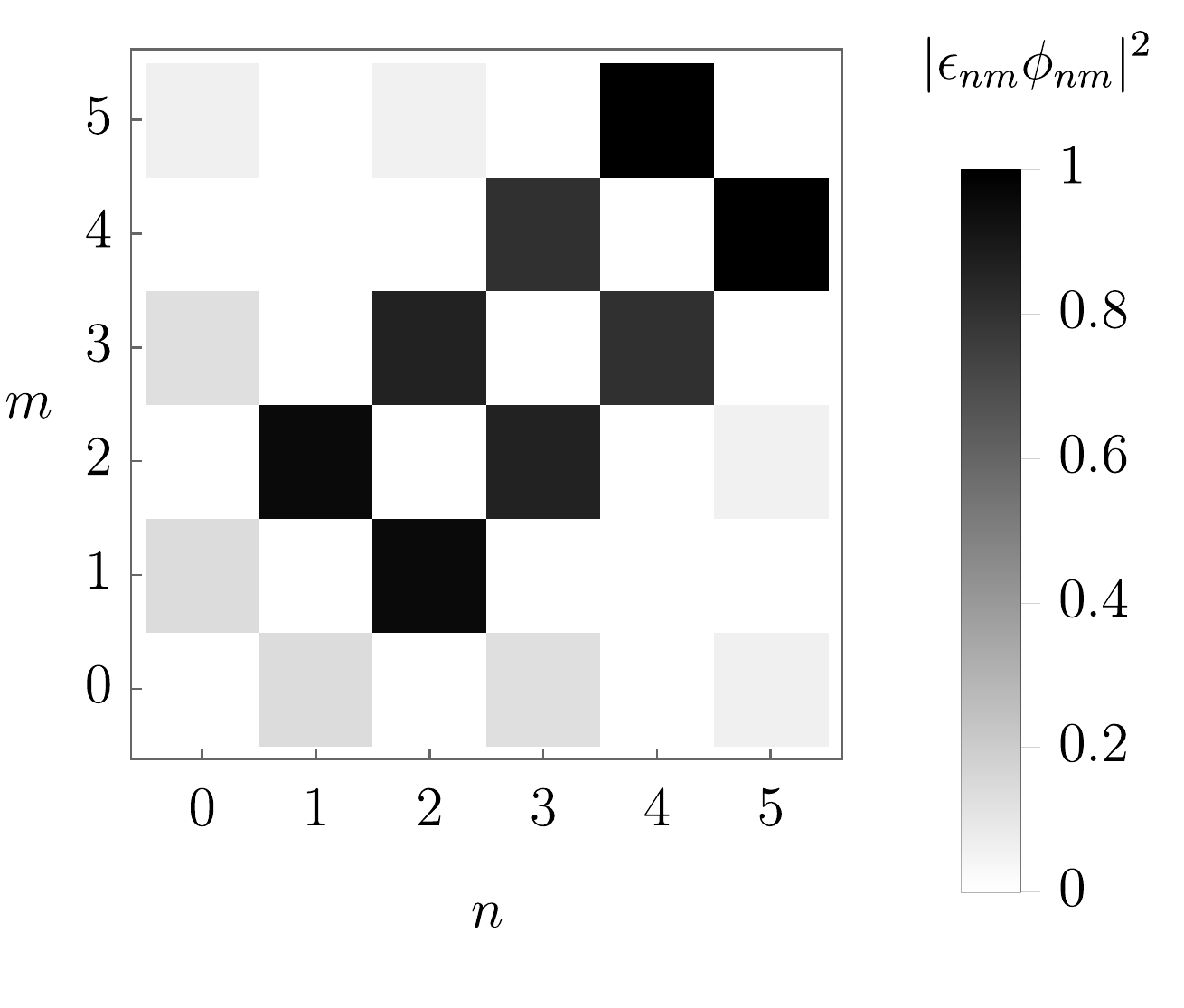}
\caption{\textbf{Matrix elements of material canonical operators}. In all plots $E_{\rm l}=0.33\mu$eV, $E_{\rm J}=10E_{\rm l}=E_{\rm c}$, and $\phi_{\rm ext}=\uppi/2e$.  \textbf{(a)} Magnitude of dipole matrix elements $|\phi_{nm}|^2$ normalised by the largest element in the array. For sufficiently anharmonic material spectra the contributions of levels $n,~m>1$ are negligible \textbf{(b)} Magnitude of effective canonical momentum matrix elements $|\epsilon_{nm}\phi_{nm}|^2$ normalised by the largest element in the array. Even for highly anharmonic spectra the contributions of levels $n,~m>1$ are non-negligible in general.}
\label{dipsmoms}
\end{figure}
%%%%%%%%%%%%%%%%%%%%%%%%%%%%%%%%%%%%%%%%%%%%%%%%%%%%%%%%%%%%%%%%%%%%%%%%%%%%%%%%%%%%%%%%%%%%
%%
%%	F I G U R E S  E N D
%%
%%%%%%%%%%%%%%%%%%%%%%%%%%%%%%%%%%%%%%%%%%%%%%%%%%%%%%%%%%%%%%%%%%%%%%%%%%%%%%%%%%%%%%%%%%%%

Let us now consider the three gauges of most importance for our purposes, namely, the flux-gauge, charge gauge and JC-gauge. We begin with the flux-gauge, for which we obtain
\begin{align}\label{shiftsflux}
&\kappa_n = g^2 \sum_{m} \left|{\phi_{nm}\over \varphi}\right|^2 {1 \over \epsilon_{nm}-\omega},\\ &\chi_n =2 g^2 \sum_m  \left|{\phi_{nm}\over \varphi}\right|^2 {\epsilon_{nm} \over \epsilon_{nm}^2-\omega^2}.
\end{align}
For the shifts $\kappa_n$ contributions from higher material levels $m>1$ are limited, because the ratios $|\phi_{nm}/\varphi|^2,~n=0,1,~m>1$ are generally small [Supplementary Fig.~\ref{dipsmoms} (a)]. Although the $\chi_n$ also depend on $|\phi_{nm}/\varphi|^2$ the remaining energy denominator $\epsilon_{nm}/(\epsilon_{nm}^2-\omega^2)$ is large for levels $m>n=0,~1$ whenever there is a resonance $\epsilon_{mn} \sim \omega$. If $\delta$ is large such that there exist higher levels $m>n=0,~1$ for which $\epsilon_m \in [0,\omega]$ it is clear that such resonances can occur, and that they will render higher material levels non-negligible despite the material anharmonicity. The flux-gauge QRM will therefore generally only be accurate for sufficiently small $\delta$ and $\eta$. We note that this accuracy often includes qualitatively accurate predictions of higher system levels $E_n>E$ for small enough $\delta$ and $\eta$ \cite{de_bernardis_breakdown_2018} (see Supplementary Note 5).

In the charge-gauge we obtain
\begin{align}
&\kappa_n = g^2 \sum_{m} \left|{\phi_{nm}\over \varphi}\right|^2{1\over \omega \omega_0} {\epsilon_{nm}^2 \over \epsilon_{nm}-\omega_0},\\ &\chi_n =2 g^2 \sum_m  \left|{\phi_{nm}\over \varphi}\right|^2 {1\over \omega \omega_0} {\epsilon_{nm}^3 \over \epsilon_{nm}^2-\omega_0^2}.
\end{align}
Like in the flux-gauge resonances in $\chi_n$ involving higher material levels can occur and will generally be non-negligible for large $\delta$. The charge-gauge oscillator renormalisation only exacerbates this situation by increasing the oscillator frequency $\omega_0\geq \omega$. Unlike in the flux-gauge an additional contribution from higher material levels also occurs via the matrix elements $\epsilon_{nm}\phi_{nm}$, For these matrix elements the contribution from higher material levels can be much bigger than that of the lowest two \cite{de_bernardis_breakdown_2018} [Supplementary Fig.~\ref{dipsmoms} (b)]. As a result the charge-gauge QRM can be expected to break down even when $\delta$ is small and the coupling strength is modest. In fact as explained below the charge-gauge becomes increasingly inaccurate as $\delta$ decreases. On the other hand, for large $\delta$ the charge-gauge linear coupling term $2E_{\rm c} \xi_0 \zeta /e^2$ (ignoring the oscillator renormalisation) has strength $g_0 = g/\delta$ within the two-level truncation. Thus, if $\delta$ is large, e.g., if $\delta=5$ as considered in the main text and further below (see Supplementary Note 5), then the linear charge-gauge light-matter coupling is much weaker than the corresponding flux-gauge coupling. A more relevant comparison must however account for the renormalisation of the oscillator frequency in the charge-gauge, which is $\omega_0 =\omega \mu_0$ with
\begin{align}
\mu_0^2 = 1 + {4 E_{\rm c} \over \omega_{\rm m} e^2\varphi^2 }\left[\eta^2 \over \delta \right].
\end{align}
This renormalisation in turn alters the charge-gauge QRM coupling strength, which becomes ${\tilde g}_0= g/(\delta \sqrt{\mu_0})$. The ratio ${\tilde g}_0/g$ is shown in Supplementary Fig.~\ref{relcoup}. Like $g_0$ the coupling ${\tilde g}_0$ remains relatively weak compared to $g$ for large $\delta$ [Supplementary Fig.~\ref{relcoup}]. Results in the main text and in Supplementary Note 5 confirm that the relative weakness of ${\tilde g}_0$ compared with $g$ allows the charge gauge QRM to (qualitatively) accurately predict the lowest two levels $G$ and $E$ for sufficiently large $\delta$, unless $\eta$ is very large. In contrast, for large $\eta$ and small $\delta<1$ the charge-gauge coupling becomes relatively large compared to the flux-gauge coupling. Combined with the generally non-negligible higher material levels already discussed above, two-level models with $\alpha \sim 0$ can be expected to become rather inaccurate for large $\eta$ and small $\delta$. Indeed, in agreement with Ref. \cite{de_bernardis_breakdown_2018} our results (see Supplementary Note 5) confirm that the performance of the charge-gauge QRM diminishes rapidly with decreasing $\delta$ and increasing $\eta$ even for the first two levels $G$ and $E$.
%%%%%%%%%%%%%%%%%%%%%%%%%%%%%%%%%%%%%%%%%%%%%%%%%%%%%%%%%%%%%%%%%%%%%%%%%%%%%%%%%%%%%%%%%%%%%%%
%%	F I G U R E S  S T A R T
%%
%%%%%%%%%%%%%%%%%%%%%%%%%%%%%%%%%%%%%%%%%%%%%%%%%%%%%%%%%%%%%%%%%%%%%%%%%%%%%%%%%%%%%%%%%%%%%
\begin{figure}[t]
\begin{minipage}{\columnwidth}
\begin{center}
\hspace*{-0.2cm}\includegraphics[scale=1]{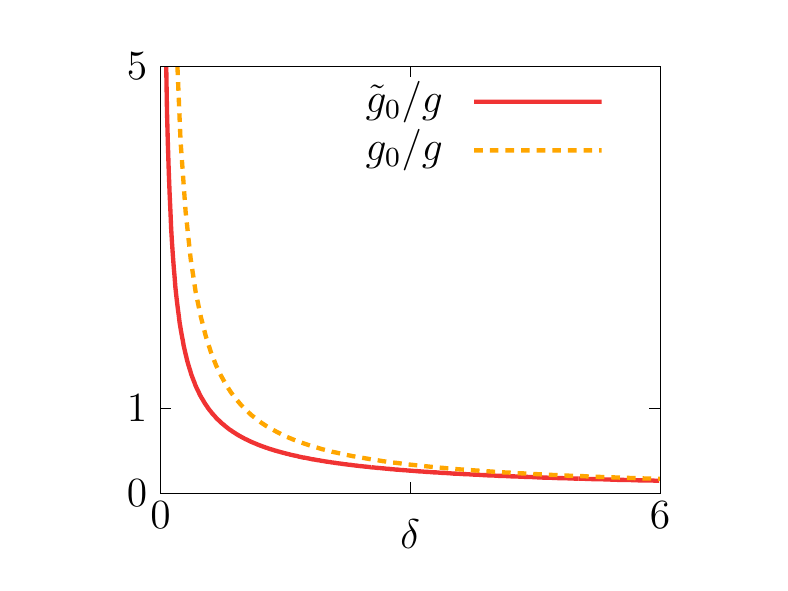}
\caption{\textbf{The ratio of charge-gauge QRM and flux-gauge QRM coupling strengths}. In all plots $E_{\rm l}=0.33\mu$eV, $E_{\rm J}=10E_{\rm l}=E_{\rm c}$, $\phi_{\rm ext}=\uppi/2e$. The ratio ${\tilde g}_0/g$ of the (renormalised) charge-gauge QRM and flux-gauge QRM coupling strengths is plotted as a function of $\delta$ with $\eta=1$. The effect of the charge-gauge renormalisation of the oscillator frequency is shown by comparison with $g_0/g=1/\delta$.}\label{relcoup}
\end{center}
\end{minipage}
\end{figure}
%%%%%%%%%%%%%%%%%%%%%%%%%%%%%%%%%%%%%%%%%%%%%%%%%%%%%%%%%%%%%%%%%%%%%%%%%%%%%%%%%%%%%%%%%%%%
%%
%%	F I G U R E S  E N D
%%
%%%%%%%%%%%%%%%%%%%%%%%%%%%%%%%%%%%%%%%%%%%%%%%%%%%%%%%%%%%%%%%%%%%%%%%%%%%%%%%%%%%%%%%%%%%%

Finally we remark on the $\alpha_{\rm JC}$-gauge JCM, which symmetrically mixes the flux and charge-gauge QRM couplings. In the JC-gauge we obtain
\begin{align}
&\kappa_n = g^2 \sum_{i} \left|{\phi_{ni}\over \varphi}\right|^2{\omega_{\rm JC}\over \omega } {(\omega_{\rm m}+\epsilon_{ni})^2 \over (\epsilon_{ni}-\omega_{\rm JC})(\omega_{\rm m}+\omega_{\rm JC})^2},\label{shiftsjc} \\ &\chi_n =2g^2 \sum_{i} \left|{\phi_{ni}\over \varphi}\right|^2{\omega_{\rm JC}\over \omega }  {(\omega_{\rm m}+\epsilon_{ni})^2\epsilon_{ni} \over (\epsilon_{ni}^2-\omega_{\rm JC}^2)(\omega_{\rm m}+\omega_{\rm JC})^2}
\end{align}
where $\omega_{\rm m} = \epsilon_{10}$ is the first material transition frequency. By construction, the contributions of the first material level $i=1$ to the ground shifts $\kappa_0$ and $\chi_0$, are zero due to the factor $\omega_{\rm m}+\epsilon_{ni}$ in the numerator of both shifts. Thus, in the JC-gauge the bare ground state is only coupled to other levels via counter-rotating contributions involving matrix elements of position and momentum between $\ket{\epsilon_0^{\rm JC}}$ and $\ket{\epsilon_i^{\rm JC}},~i>1$. These contributions make $\kappa_0$ in Supplementary Eq.~(\ref{shiftsjc}) non-zero in the non-truncated theory. Since $\kappa_0$ is the shift of the state $\ket{\epsilon^{\rm JC}_n,0^{\rm JC}}$, which is also the ground state of the JC-gauge two-level model, it quantifies the deviation between the ground state of the JC-gauge two-level model and the true ground state $\ket{G}$ of the non-truncated Hamiltonian. In the case of both position and momentum matrix elements the material ground state $\ket{\epsilon_0^{\rm JC}}$ is predominantly linked to the first level $\ket{\epsilon_1^{\rm JC}}$ (Supplementary Fig.~\ref{dipsmoms}). Matrix elements involving $\ket{\epsilon_0^{\rm JC}}$ and higher levels are smaller especially in the case of the flux operator $\phi$. As a result $\kappa_0$ is small in the JC-gauge, which explains why the JC-gauge two-level model gives a good representation of the ground state. Indeed our results confirm that the JC gauge two-level model actually always outperforms the available QRMs in predicting the ground state and its energy.
%%%%%%%%%%%%%%%%%%%%%%%%%%%%%%%%%%%%%%%%%%%%%%%%%%%%%%%%%%%%%%%%%%%%%%%%%%%%%%%%%%%%%%%%%%%%%%%
%%	F I G U R E S  S T A R T
%%
%%%%%%%%%%%%%%%%%%%%%%%%%%%%%%%%%%%%%%%%%%%%%%%%%%%%%%%%%%%%%%%%%%%%%%%%%%%%%%%%%%%%%%%%%%%%%
\begin{figure}[t]
\begin{minipage}{\columnwidth}
\begin{center}
\hspace*{-0.2cm}\includegraphics[scale=1]{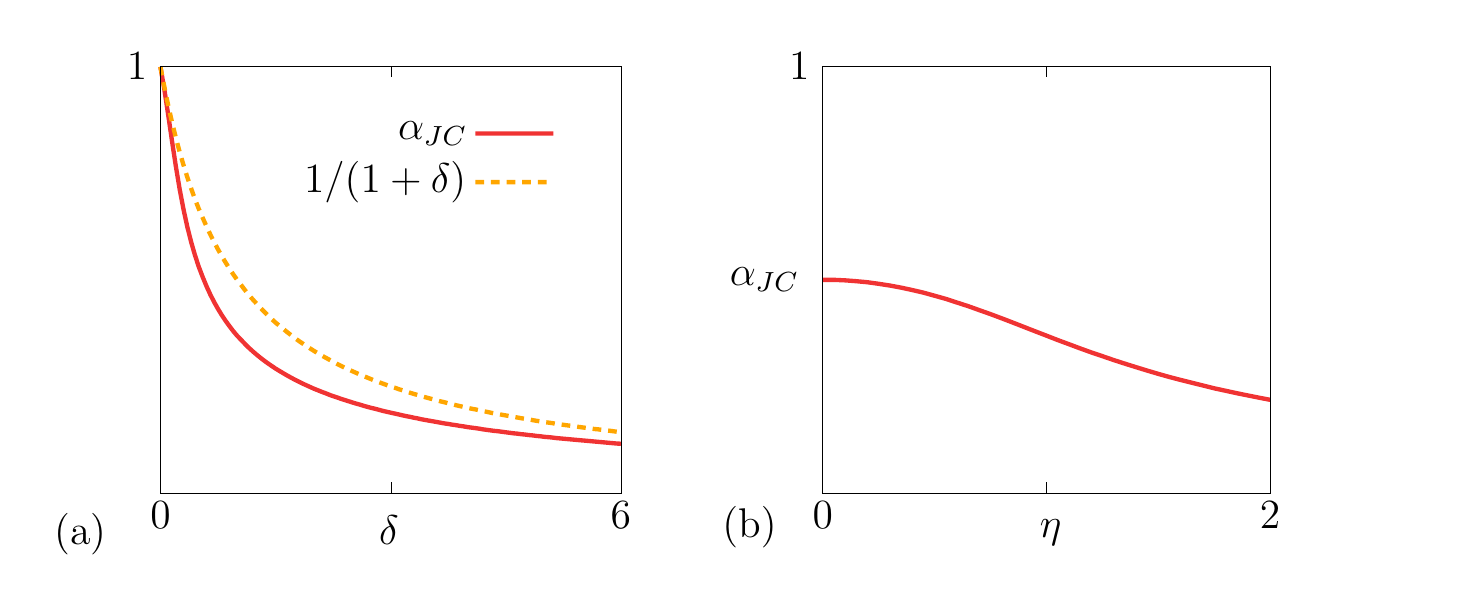}
\caption{\textbf{The $\alpha_{\rm JC}$-gauge parameter as a function of coupling and detuning}. In all plots $E_{\rm l}=0.33\mu$eV, $E_{\rm J}=10E_{\rm l}=E_{\rm c}$, $\phi_{\rm ext}=\uppi/2e$. \textbf{(a)} $\alpha_{\rm JC}$ and $1/(1+\delta)$ as a function of $\delta$ with $\eta=1$. \textbf{(b)} $\alpha_{\rm JC}$ as a function of $\eta$ with $\delta=1$.}\label{betajc}
\end{center}
\end{minipage}
\end{figure}
%%%%%%%%%%%%%%%%%%%%%%%%%%%%%%%%%%%%%%%%%%%%%%%%%%%%%%%%%%%%%%%%%%%%%%%%%%%%%%%%%%%%%%%%%%%%
%%
%%	F I G U R E S  E N D
%%
%%%%%%%%%%%%%%%%%%%%%%%%%%%%%%%%%%%%%%%%%%%%%%%%%%%%%%%%%%%%%%%%%%%%%%%%%%%%%%%%%%%%%%%%%%%%

To understand the accuracy of the JC-gauge two-level model more generally we consider how $\alpha_{\rm JC}$ varies with $\delta$ and $\eta$. This is shown in Supplementary Fig.~\ref{betajc}. The effect of the renormalisation of the oscillator frequency is shown by comparison with $\omega_{\rm m}/(\omega+\omega_{\rm m}) = 1/(1+\delta)$ which equals $\alpha_{\rm JC}$ if $\omega_{\rm JC}$ is approximated as $\omega$. As $\delta$ increases $\alpha_{\rm JC}$ decays quite rapidly from $1$ towards $0$ indicating that the JC-gauge becomes predominantly charge-like for even fairly small $\delta$. For large $\delta$ both the flux and charge-gauge QRMs are inaccurate for levels $E_n>E$, and the JC-gauge is therefore also inaccurate for these levels. More surprising is the inaccuracy of the JC-gauge for levels $E_n>E$ when $\delta$ is small. For example, when $\delta=1/5$ the flux-gauge QRM is relatively accurate (see Supplementary Note 5) and for this value of $\delta$ we obtain $\alpha_{\rm JC}\sim 0.8$ when $\eta=1$ indicating a predominantly flux-like coupling within the JC-gauge. For $\delta=1/5$ the JC-gauge two-level model is nevertheless relatively inaccurate (compared with the flux-gauge QRM) in predicting levels $E_n>E$ for sufficiently large couplings (see Supplementary Note 5). This can only be attributed to the quite severe breakdown of the charge-gauge QRM for small $\delta$ combined with the decrease in $\alpha_{\rm JC}$ towards the charge-gauge value $0$, as the coupling $\eta$ increases [Supplementary Fig.~\ref{betajc} (b)].\\

\section*{Supplementary Note 5: Further anaysis; alternative parameter regimes}\label{alt}

Here we provide further analysis of two-level models via comparison with exact predictions. Our findings are consistent with the analysis of Supplementary Note 4 above. We divide this part into three sections corresponding to three values of the detuning $\delta=1/5,~1,~5$. For each value of $\delta$ we consider variations in the remaining parameters $\eta,~\phi_{\rm ext},~\alpha$. For large detuning $\delta=5$ two-level models tend to be inaccurate in predicting dressed energies $E_n>E$ with $E$ the first excited level. The charge gauge QRM is more accurate than the flux gauge QRM for the first two levels while the JC-gauge is the most accurate two-level model. As $\delta$ decreases the charge gauge QRM becomes less accurate and the flux-gauge QRM more accurate. The JC-gauge two-level model typically remains the most accurate for the first two system levels, but unlike the charge-gauge QRM and JC-gauge two-level model the flux gauge QRM is able to give qualitative agreement with the non-truncated model for levels $E_n>E$ provided $\delta$ and $\eta$ are sufficiently small.\\

\subsubsection{$\delta=5$}\label{five}

We begin with the most experimentally relevant regime presently, $\delta=5$. We show in Supplementary Fig.~\ref{ext2} how the first two dressed energies behave as $\phi_{\rm ext}$ is varied within the deep-strong coupling regime $\eta =1.5$. The JC-gauge two-level model provides reasonable qualitative agreement with the exact energies while the flux and charge-gauge QRMs are significantly less accurate. The flux and charge-gauge JCMs are also inaccurate as expected.

%%%%%%%%%%%%%%%%%%%%%%%%%%%%%%%%%%%%%%%%%%%%%%%%%%%%%%%%%%%%%%%%%%%%%%%%%%%%%%%%%%%%%%%%%%%%%%%
%%	F I G U R E S  S T A R T
%%
%%%%%%%%%%%%%%%%%%%%%%%%%%%%%%%%%%%%%%%%%%%%%%%%%%%%%%%%%%%%%%%%%%%%%%%%%%%%%%%%%%%%%%%%%%%%%
\begin{figure}[t]
\begin{minipage}{\columnwidth}
\begin{center}
\hspace*{-1cm}\includegraphics[scale=0.9]{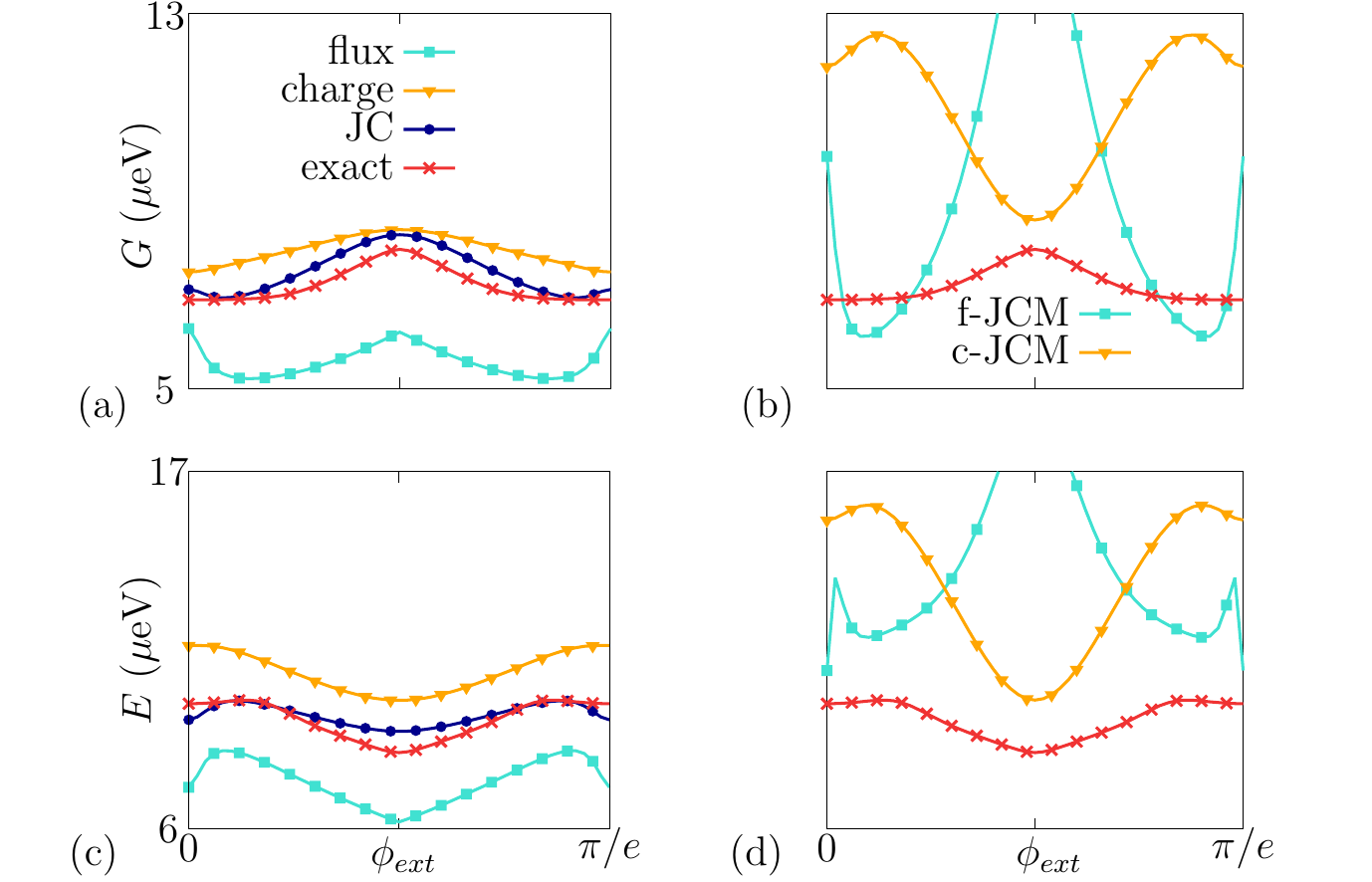}
\caption{\textbf{Lowest energy levels as functions of external flux}. In all plots $E_{\rm l}=0.33\mu$eV, $E_{\rm J}=10E_{\rm l}=E_{\rm c}$, $\delta=5$ and $\eta =1.5$. \textbf{(a)} The ground energy is plotted with $\phi_{\rm ext}$ for the flux-gauge and charge-gauge QRMs, for the JC-gauge two-level model and for the exact model (solid red). \textbf{(b)} For the same range as (a) the ground energy is plotted with $\phi_{\rm ext}$ for the flux-gauge and charge-gauge JCMs, and for the exact model. \textbf{(c)} Same as (a) for the first excited energy. \textbf{(d)} Same as (b) for the first excited energy.}\label{ext2}
\end{center}
\end{minipage}
\end{figure}
%%%%%%%%%%%%%%%%%%%%%%%%%%%%%%%%%%%%%%%%%%%%%%%%%%%%%%%%%%%%%%%%%%%%%%%%%%%%%%%%%%%%%%%%%%%%%
\begin{figure}[H]
\begin{minipage}{\columnwidth}
\begin{center}
\hspace*{-1cm}\includegraphics[scale=0.9]{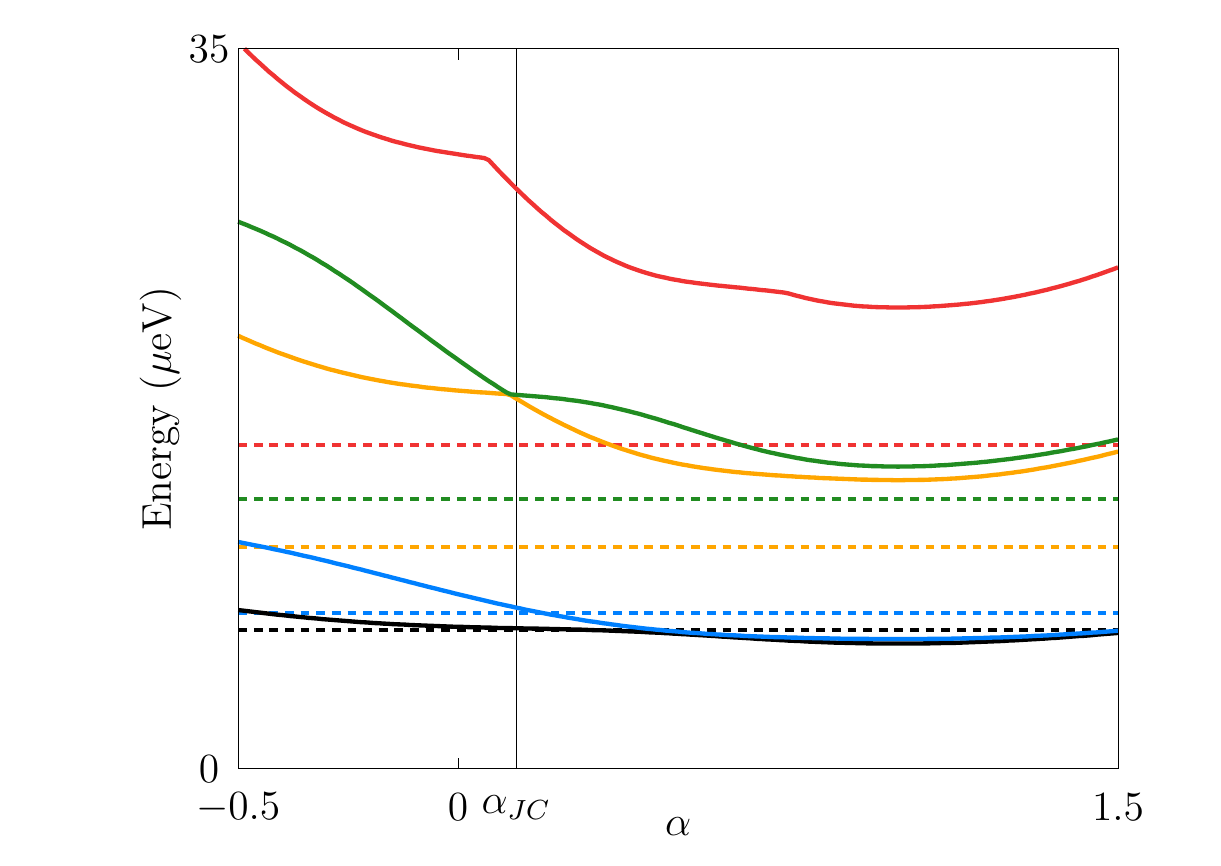}
\caption{\textbf{Lowest energy levels as functions of the gauge parameter}. In all plots $E_{\rm l}=0.33\mu$eV, $E_{\rm J}=10E_{\rm l}=E_{\rm c}$, $\delta=5$, $\phi_{\rm ext}=\uppi/2e$ and $\eta =1$. The first five Hamiltonian energies are plotted as functions of $\alpha$. The dashed lines give the $\alpha$-independent exact energies, while the solid curves give the corresponding energies found within the two-level truncation. The vertical line specifies the value $\alpha_{\rm JC}$.}\label{en1}
\end{center}
\end{minipage}
\end{figure}
%%%%%%%%%%%%%%%%%%%%%%%%%%%%%%%%%%%%%%%%%%%%%%%%%%%%%%%%%%%%%%%%%%%%
Next we restrict our attention to the maximal frustration $\phi_{\rm ext}=\uppi/2e$ point and consider how predictions vary with $\alpha$ while other parameters are held fixed. Supplementary Fig.~\ref{en1} shows how the dressed energies of the general $\alpha$-gauge two-level model varies with $\alpha$ when $\eta =1$. All two-level models become inaccurate for dressed levels $E_n>E$. For the specified parameters two-level models with $\alpha$ near to $\alpha_{\rm JC}=0.132$ are accurate in predicting the first two energy values. 
%%%%%%%%%%%%%%%%%%%%%%%%%
\begin{figure}[H]
\begin{minipage}{\columnwidth}
\begin{center}
\hspace*{-1cm}\includegraphics[scale=0.9]{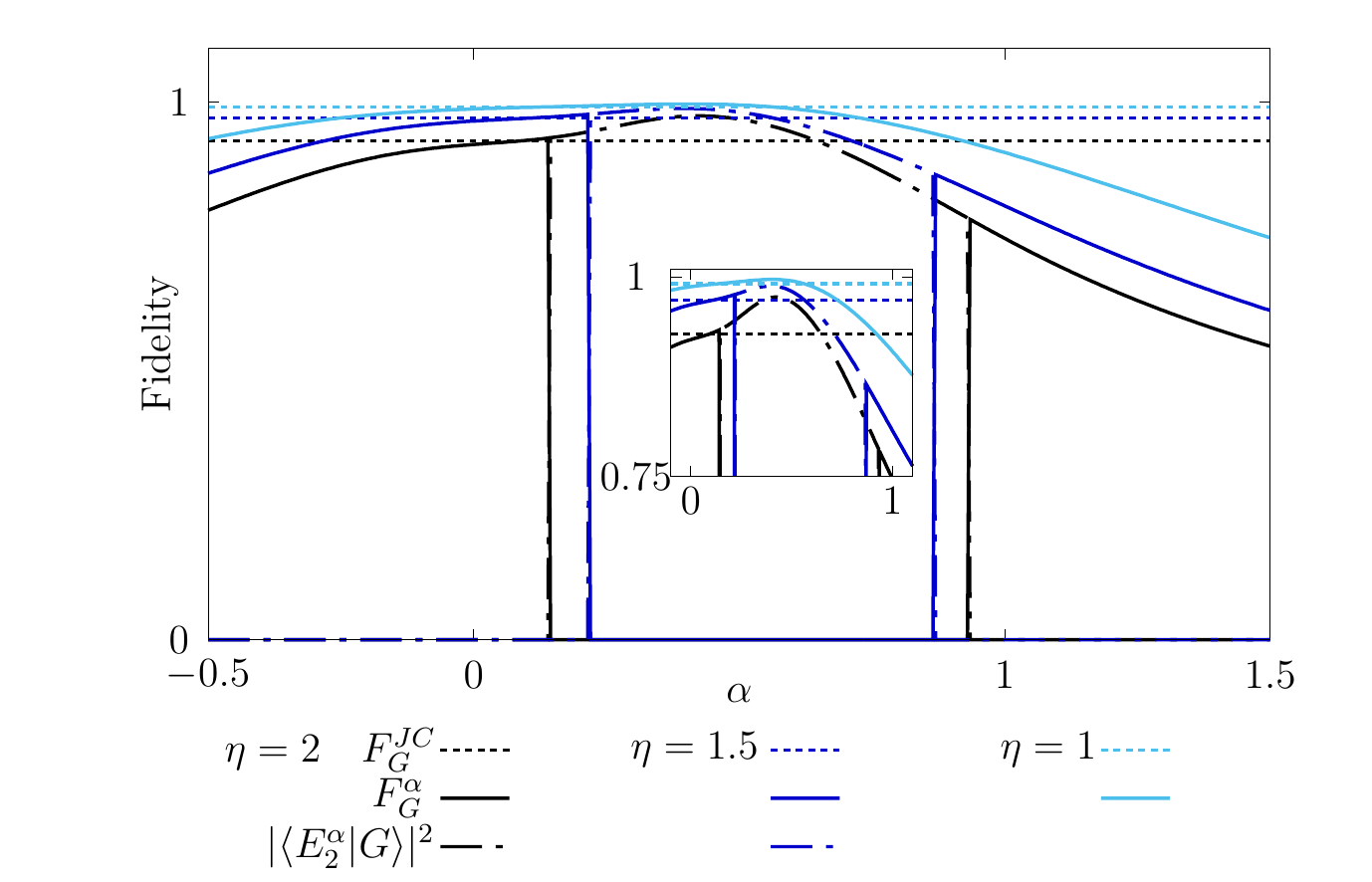}
\caption{\textbf{Fidelities of the lowest eigenstates as functions of the gauge parameter}. $E_{\rm l}=0.33\mu$eV, $E_{\rm J}=10E_{\rm l}=E_{\rm c}$, $\delta=5$ and $\phi_{\rm ext}=\uppi/2e$. The ground state fidelity of two-level model states $\ket{G_2^\alpha}$ and $\ket{E_2^\alpha}$ in the ground state $\ket{G}$ are plotted as functions of $\alpha$ for $\eta =1,~1.5,~2$ (solid lines). The straight dashed lines show the values of $F_G^{\rm JC}$ for the same three coupling strengths. The dash-dotted lines show the overlap $|\langle E_2^{\alpha} | G\rangle|^2$ as a function of $\alpha$ for $\eta =1.5,~2$.}\label{fid1}
\end{center}
\end{minipage}
\end{figure}
%%%%%%%%%%%%%%%%%%%%%%%%%%%%%%%%%%%%%%%%%%%%%%%%%%%%%%%%%%%%%%%%%%%%%%%%%%%%%%%%%%%%%%%%%%%%
%%
%%	F I G U R E S  E N D
%%
%%%%%%%%%%%%%%%%%%%%%%%%%%%%%%%%%%%%%%%%%%%%%%%%%%%%%%%%%%%%%%%%%%%%%%%%%%%%%%%%%%%%%%%%%%%%

Supplementary Fig.~\ref{fid1} shows how the ground and excited state state fidelities $F^\alpha_{g,e}$ vary with $\alpha$ for various couplings. For sufficiently small $\eta$ the JC-gauge two-level model is always close to the optimal two-level model for representing the ground state. For larger $\eta$ degeneracy points of the first two-levels occur for certain values of $\alpha$. For such $\alpha$ a transfer of population within the exact ground state $\ket{G}$ from the ground state $\ket{G_2^\alpha}$ into the excited state $\ket{E_2^\alpha}$ occurs, such that the state $\ket{E^\alpha_2}$ becomes a better representation of the true ground state $\ket{G}$. In fact for large enough $\eta$ there exists a value $\alpha_{\rm opt}$ that through $\ket{E_2^{\alpha_{\rm opt}}}$ gives the best possible representation of $\ket{G}$, i.e., is such that $|\langle E_2^{\alpha_{\rm opt}} | G\rangle|^2 \geq F_g^\alpha,~\forall \alpha$.

\subsubsection{$\delta=1$}\label{res}

\begin{figure}[H]
\vspace*{-1.2cm}
\begin{minipage}{\columnwidth}
\begin{center}
\hspace*{-0.7cm}\includegraphics[scale=0.9]{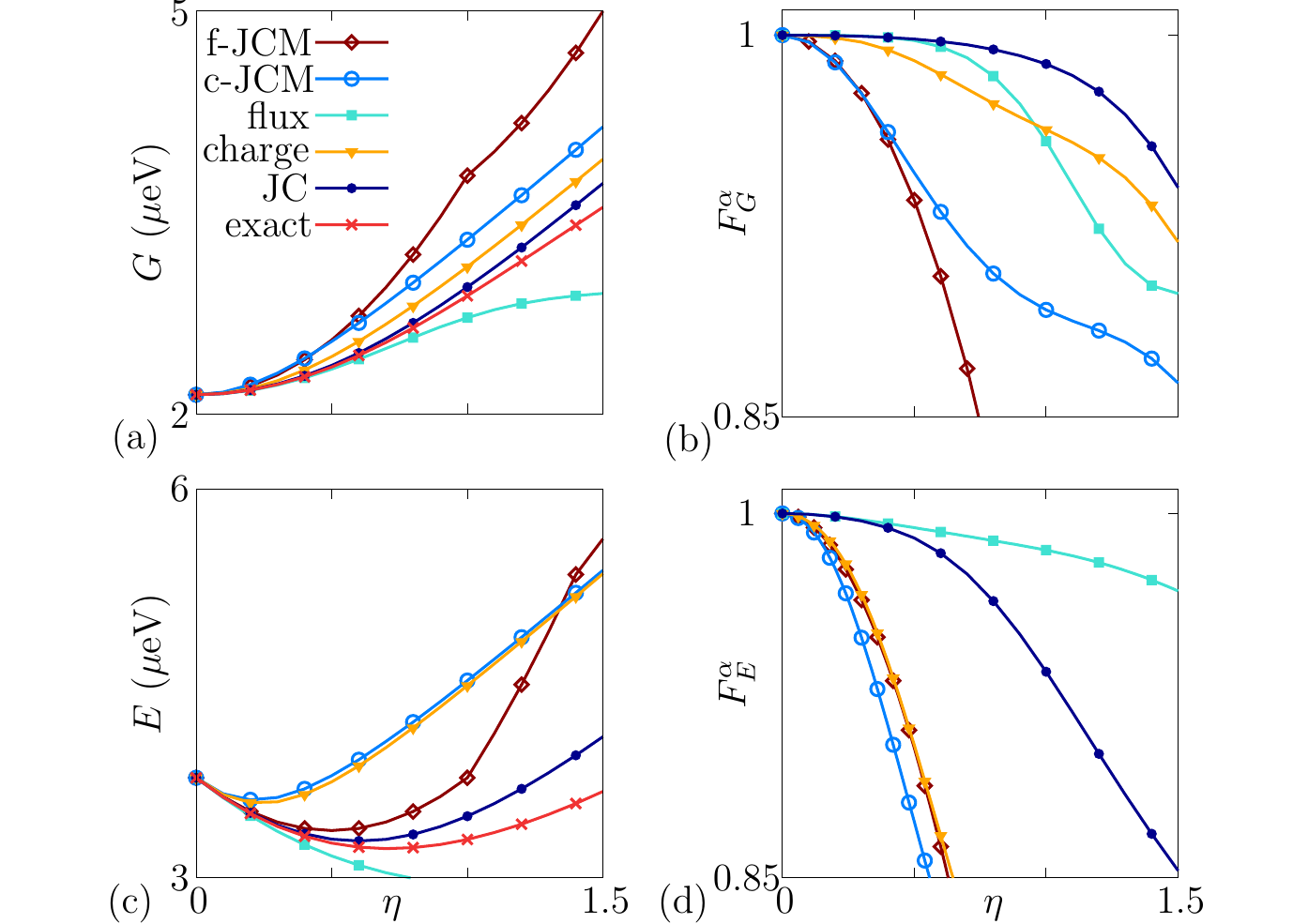}
\caption{\textbf{Lowest energies as functions of coupling strength}. In all plots $E_{\rm l}=0.33\mu$eV, $E_{\rm J}=10E_{\rm l}=E_{\rm c}$, $\delta=1$ and $\phi_{\rm ext}=\uppi/2e$. \textbf{(a)} The ground energy is plotted with $\eta$ for the flux-gauge and charge-gauge QRMs, for the JC-gauge two-level model, for the exact model, and for the flux and charge-gauge JCMs obtained via the RWA. \textbf{(b)} The ground state fidelity $F_G^\alpha$ is plotted with $\eta$ for the flux and charge-gauge QRMs, for the JC-gauge two-level model and for the flux and charge-gauge JCMs. \textbf{(c)} Same as (a) for the first excited energy. \textbf{(d)} Same as (b) for the first excited state.}\label{cres}
\end{center}
\end{minipage}
\end{figure}

\begin{figure}[H]
\begin{minipage}{\columnwidth}
\begin{center}
\hspace*{-2cm}\includegraphics[scale=0.9]{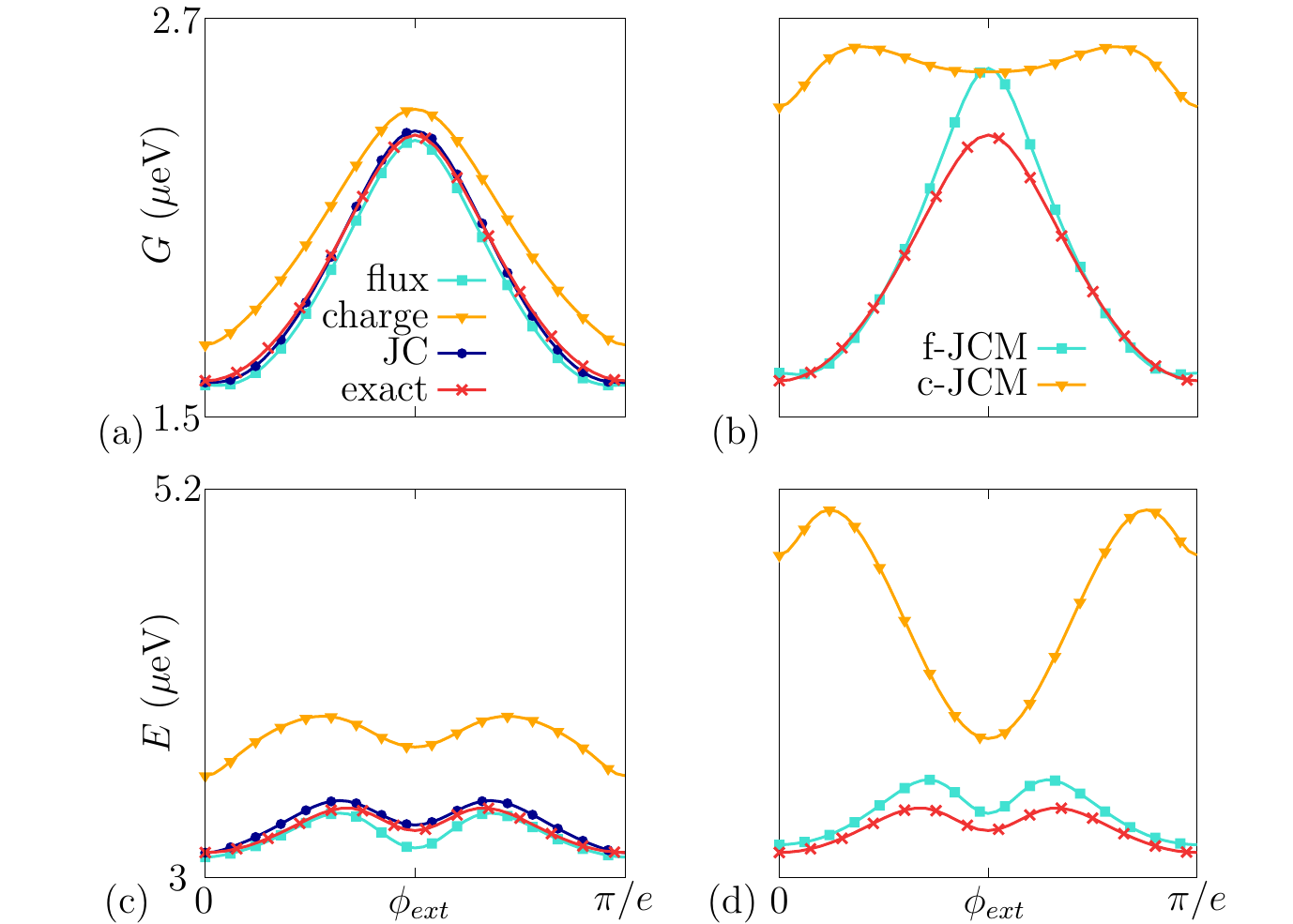}
\caption{\textbf{Lowest energies as functions of external flux}. In all plots $E_{\rm l}=0.33\mu$eV, $E_{\rm J}=10E_{\rm l}=E_{\rm c}$, $\delta=1$ and $\eta =1/2$. \textbf{(a)} The ground energy is plotted with $\phi_{\rm ext}$ for the flux-gauge and charge-gauge QRMs, for the JC-gauge two-level model and for the exact model. \textbf{(b)} For the same range as (a) the ground energy is plotted with $\phi_{\rm ext}$ for the flux-gauge and charge-gauge JCMs, and for the exact model. \textbf{(c)} Same as (a) for the first excited energy. \textbf{(d)} Same as (b) for the first excited energy.}\label{ext_res}
\end{center}
\end{minipage}
\end{figure}

Next we consider the case of resonance $\delta=1$. Here for the first two levels the JC-gauge two-level model again provides energies closest to the exact energies. The JC-gauge again also provides the best representation of the ground state (Supplementary Fig.~\ref{cres}). The flux-gauge QRM becomes more accurate and is typically more accurate than the charge-gauge QRM (Supplementary Fig.~\ref{ext_res}). For $\eta>0.4$ the excited state of the flux-gauge QRM has larger overlap with the exact excited state than the JC-gauge JCM. Two-level models remain largely inaccurate for levels $E_n>E$ in this regime of detuning, although the flux-gauge QRM accurately predicts certain energies for levels $E_n>E$ e.g. $n=3,~5,~8$ as shown in Supplementary Fig.~\ref{en_res} for $\eta =1$.
%%%%%%%%%%%%%%%%%%%%%%%%%%%%%%%%%%%%%%%%%%%%%%%%%%%%%%%%%%%%%%%%%%%%%%%%%%%%%%%%%%%%%%%%%%%%%%%
%%	F I G U R E S  S T A R T
%%
%%%%%%%%%%%%%%%%%%%%%%%%%%%%%%%%%%%%%%%%%%%%%%%%%%%%%%%%%%%%%%%%%%%%%%%%%%%%%%%%%%%%%%%%%%%%%
%%%%%%%%%%%%%%%%%%%%%%%%%%%%%%%%%%%%%%%%%%%%%%%%%%%%%%%%%%%%%%%%%%%%%%%%%%%%%%%%%%%%%%%%%%%%%
%%%%%%%%%%%%%%%%%%%%%%%%%%%%%%%%%%%%%%%%%%%%%%%%%%%%%%%%%%%%%%%%%%%%
%%%%%%%%%%%%%%%%%%%%%%%%%
\begin{figure}[H]
\begin{minipage}{\columnwidth}
\begin{center}
\hspace*{-1cm}\includegraphics[scale=0.9]{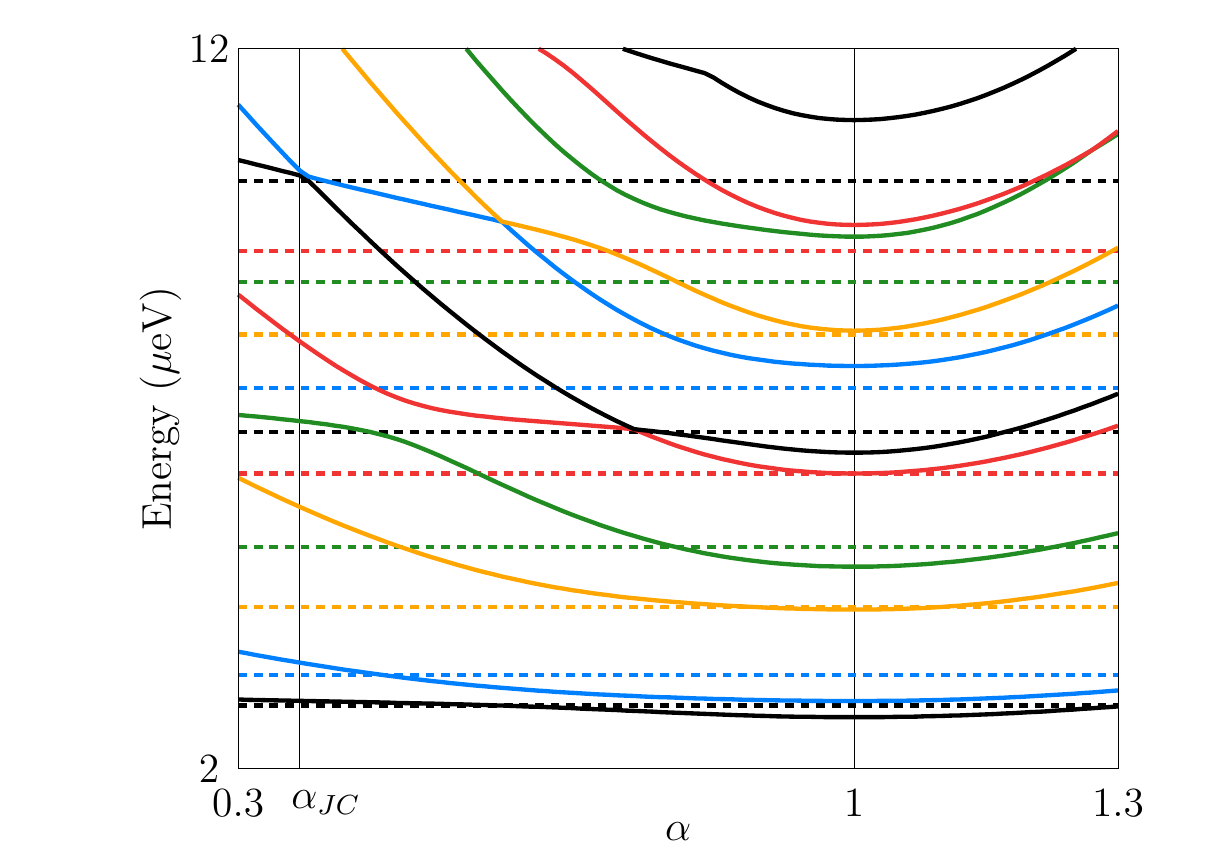}\vspace*{-0.2cm}
\caption{\textbf{Lowest energies as functions of the gauge parameter}. $E_{\rm l}=0.33\mu$eV, $E_{\rm J}=10E_{\rm l}=E_{\rm c}$, $\delta=1$, $\phi_{\rm ext}=\uppi/2e$ and $\eta =1$. The first 11 Hamiltonian energies are plotted as functions of $\alpha$. The dashed lines give the $\alpha$-independent exact energies, while the solid curves give the corresponding energies found within the two-level truncation. The vertical lines specifiy the value $\alpha_{\rm JC}$ and the flux-gauge $\alpha=1$ respectively.}\label{en_res}
\end{center}
\end{minipage}
\vspace*{-0.2cm}
\end{figure}
%%%%%%%%%%%%%%%%%%%%%%%%%%%%%%%%%%%%%%%%%%%%%%%%%%%%%%%%%%%%%%%%%%%%%%%%%%%%%%%%%%%%%%%%%%%%
%%
%%	F I G U R E S  E N D
%%
%%%%%%%%%%%%%%%%%%%%%%%%%%%%%%%%%%%%%%%%%%%%%%%%%%%%%%%%%%%%%%%%%%%%%%%%%%%%%%%%%%%%%%%%%%%%

\subsubsection{$\delta=1/5$}\label{fifth}

\begin{figure}[H]
\begin{minipage}{\columnwidth}
\begin{center}
\hspace*{-0.8cm}\includegraphics[scale=0.9]{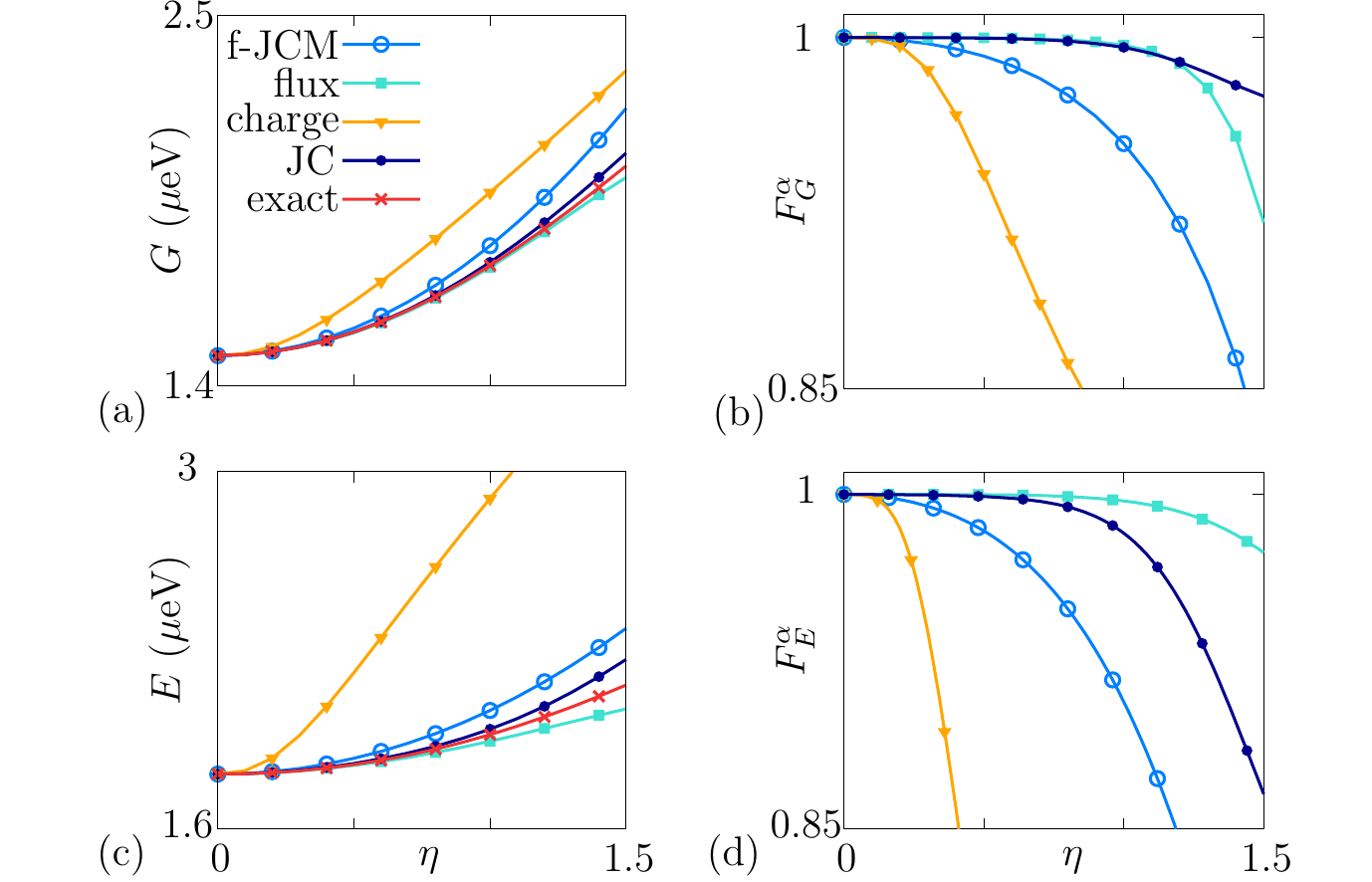}
\caption{\textbf{Lowest energies as functions of coupling strength}. In all plots $E_{\rm l}=0.33\mu$eV, $E_{\rm J}=10E_{\rm l}=E_{\rm c}$, $\delta=1/5$ and $\phi_{\rm ext}=\uppi/2e$. \textbf{(a)} The ground energy is plotted with $\eta$ for the flux-gauge and charge-gauge QRMs, for the JC-gauge two-level model, for the exact model, and for the charge-gauge JCM obtained via the RWA. \textbf{(b)} The ground state fidelity $F_G^\alpha$ is plotted with $\eta$ for the flux-gauge $\alpha=1$ and charge-gauge $\alpha=0$ QRMs, for the JC-gauge and for charge-gauge JCM.  \textbf{(c)} Same as (a) for the first excited energy. \textbf{(d)} Same as (b) for the first excited state.}\label{cf}
\end{center}
\end{minipage}
\end{figure}

\begin{figure}[H]
\begin{minipage}{\columnwidth}
\begin{center}
\includegraphics[scale=0.9]{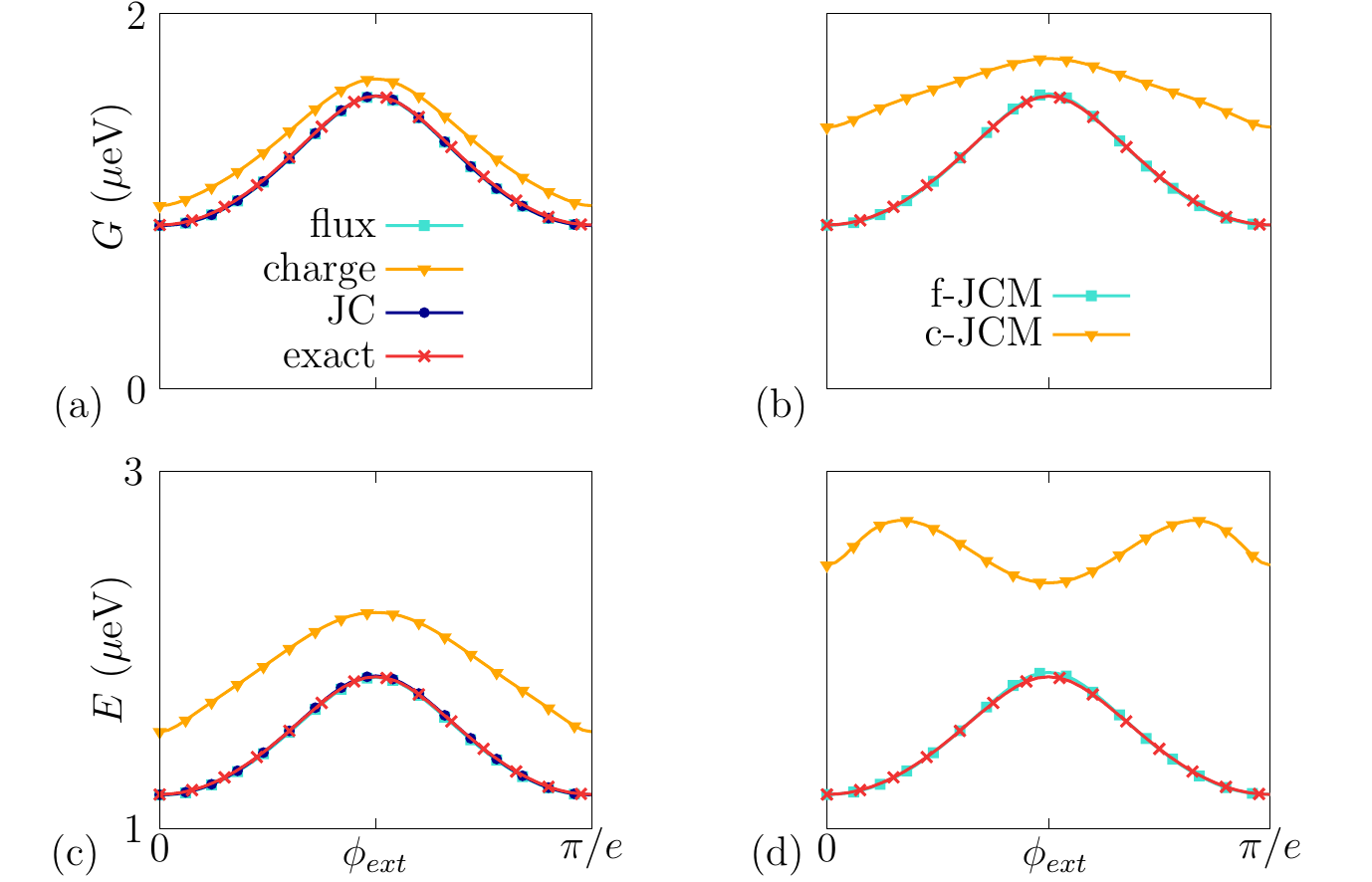}
\caption{\textbf{Lowest energies as functions of external flux}. In all plots $E_{\rm l}=0.33\mu$eV, $E_{\rm J}=10E_{\rm l}=E_{\rm c}$, $\delta=1/5$ and $\eta =1/2$. \textbf{(a)} The ground energy is plotted with $\phi_{\rm ext}$ for the flux-gauge and charge-gauge QRMs, for the JC-gauge two-level model and for the exact model (solid red). \textbf{(b)} For the same range as (a) the ground energy is plotted with $\phi_{\rm ext}$ for the flux-gauge and charge-gauge JCMs, and for the exact model. \textbf{(c)} Same as (a) for the first excited energy. \textbf{(d)} Same as (b) for the first excited energy.}\label{ext_f}
\end{center}
\end{minipage}
\end{figure}

Next we consider the regime $\omega_0>\omega$ by letting $\delta=1/5$. Here the JC-gauge again provides the best representation of the ground state and for the first two levels again provides energies closest to the exact energies (Supplementary Fig.~\ref{cf}). The flux-gauge QRM becomes yet more accurate while the charge-gauge QRM becomes yet more inaccurate (Supplementary Fig.~\ref{ext_f}). As in the case of the charge-gauge QRM in the regime $\delta=5$ the RWA is seen to incur very little error when applied to the flux-gauge QRM in the regime $\delta=1/5$. The flux-gauge JCM therefore occasionally outperforms the charge-gauge QRM (Supplementary Fig.~\ref{ext_f}). The flux-gauge QRM becomes much more accurate at predicting energy values for levels $E_n>E$ as shown in Supplementary Fig.~\ref{en_f}. Again all two-level models breakdown for these levels when the coupling is sufficiently large.

%%%%%%%%%%%%%%%%%%%%%%%%%%%%%%%%%%%%%%%%%%%%%%%%%%%%%%%%%%%%%%%%%%%%%%%%%%%%%%%%%%%%%%%%%%%%%%%
%%	F I G U R E S  S T A R T
%%
%%%%%%%%%%%%%%%%%%%%%%%%%%%%%%%%%%%%%%%%%%%%%%%%%%%%%%%%%%%%%%%%%%%%%%%%%%%%%%%%%%%%%%%%%%%%%
%%%%%%%%%%%%%%%%%%%%%%%%%%%%%%%%%%%%%%%%%%%%%%%%%%%%%%%%%%%%%%%%%%%%%%%%%%%%%%%%%%%%%%%%%%%%%
%%%%%%%%%%%%%%%%%%%%%%%%%%%%%%%%%%%%%%%%%%%%%%%%%%%%%%%%%%%%%%%%%%%%%%%%%%%%%%%%%%%%%%%%%%%%%
\begin{figure}[H]
\begin{minipage}{\columnwidth}
\begin{center}
\hspace*{-1cm}\includegraphics[scale=0.9]{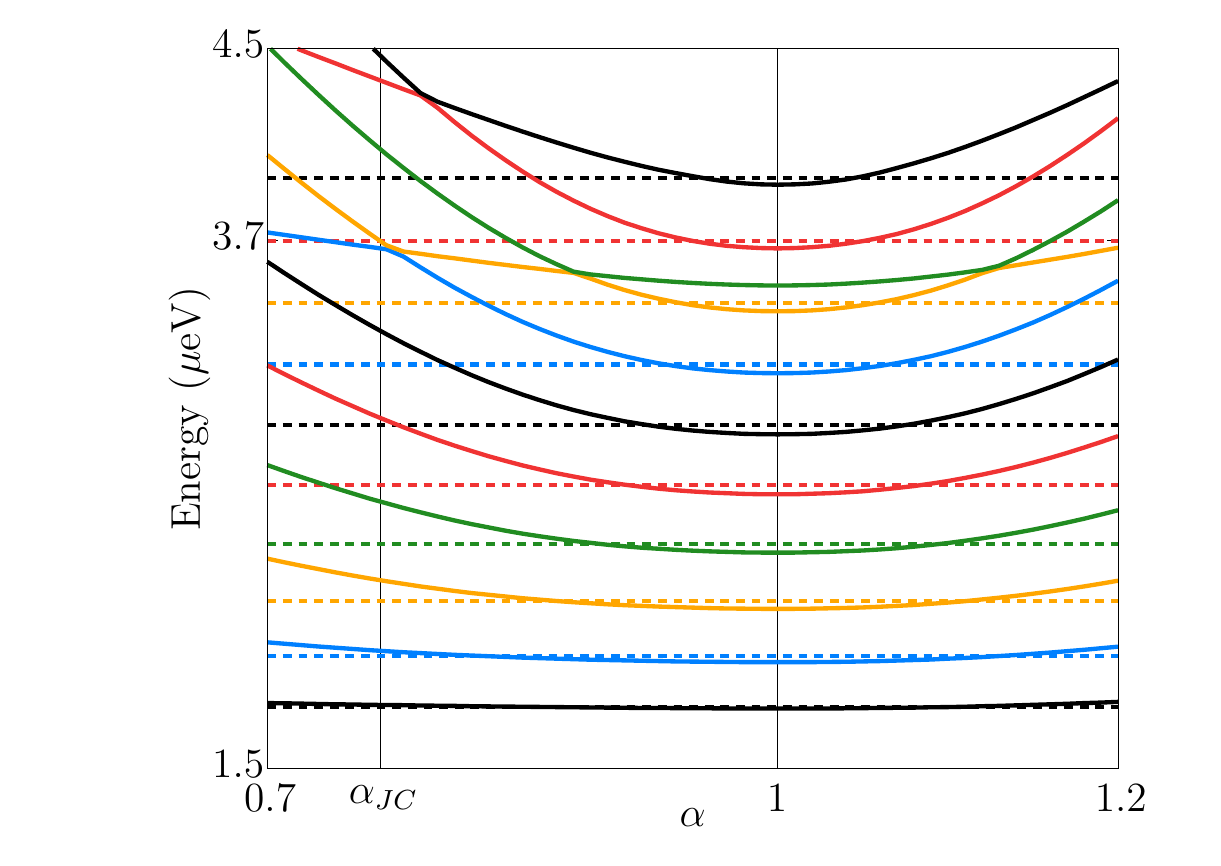}
\caption{\textbf{Lowest energies as functions of the gauge parameter}. $E_{\rm l}=0.33\mu$eV, $E_{\rm J}=10E_{\rm l}=E_{\rm c}$, $\delta=1/5$, $\phi_{\rm ext}=\uppi/2e$ and $\eta =1$. The first 11 Hamiltonian energies are plotted as functions of $\alpha$. The dashed lines give the $\alpha$-independent exact energies, while the solid curves give the corresponding energies found within the two-level truncation. The vertical lines specifiy the value $\alpha_{\rm JC}$ and the flux-gauge $\alpha=1$ respectively. The exact energies for levels $9$ and $10$ are nearly equal at $3.6995\mu$eV and $3.6997\mu$eV respectively.}\label{en_f}
\end{center}
\end{minipage}
\end{figure}
%%%%%%%%%%%%%%%%%%%%%%%%%%%%%%%%%%%%%%%%%%%%%%%%%%%%%%%%%%%%%%%%%%%%%%%%%%%%%%%%%%%%%%%%%%%%
%%
%%	F I G U R E S  E N D
%%
%%%%%%%%%%%%%%%%%%%%%%%%%%%%%%%%%%%%%%%%%%%%%%%%%%%%%%%%%%%%%%%%%%%%%%%%%%%%%%%%%%%%%%%%%%%%

\section*{Supplementary Note 6: Photon number averages}\label{photon}

As an example of an observable different from the energy we consider here photon number. Each different gauge $\alpha$ has two associated photon number operators, $a_\alpha^\dagger a_\alpha$ and $c_\alpha^\dagger c_\alpha$. The number operator $a_\alpha^\dagger a_\alpha$ has associated energy $H_{\rm c}^\alpha = \omega(a_\alpha^\dagger a_\alpha +1/2)$, whereas the number operator $c_\alpha^\dagger c_\alpha$ implicitly includes the oscillator self-energy contribution $E_{\rm c}(1-\alpha)^2 \zeta^2/e^2$ such that the associated energy is $H_{\rm c}^\alpha+E_{\rm c}(1-\alpha)^2 \zeta^2/e^2= \omega_\alpha(c_\alpha^\dagger c_\alpha +1/2)$ where $\omega_\alpha^2 =\omega^2+E_{\rm c}(1-\alpha)^2 C \zeta^2/e^2$. If and only if $\alpha=1$ does one obtain $c_\alpha^\dagger c_\alpha =a_\alpha^\dagger a_\alpha$. We focus on the renormalised number operator $c_\alpha^\dagger c_\alpha$, which can be expressed in terms of operators associated with any other gauge $\alpha'\neq \alpha$ by using the unitary relation between $c_\alpha$ and $a_\alpha$ and then using the unitary relation between $a_\alpha$ and $a_{\alpha'}$. Explicitly we have
\begin{align}
n_\alpha &= c_\alpha^\dagger c_\alpha = {1\over 2}\left({\omega\over \omega_\alpha}+{\omega_\alpha\over \omega}\right)a_\alpha^\dagger a_\alpha +{1\over 4}\left({\omega\over \omega_\alpha}-{\omega_\alpha\over \omega}\right)({a_\alpha^\dagger}^2+a_\alpha^2)+{1\over 4}\left({\omega\over \omega_\alpha}+{\omega_\alpha\over \omega}-2\right),\label{nalph}
\end{align}
and
\begin{align}
a_\alpha &= a_{\alpha'} + {\eta \over \varphi}(\alpha'-\alpha)\phi.\label{aalph}
\end{align}
Upon substitution of Supplementary Eq.~(\ref{aalph}) into the right-hand-side of Supplementary Eq.~(\ref{nalph}) one obtains $n_\alpha = n_\alpha({\bf y}_{\alpha'})$ expressed as a function of $\alpha'$-gauge ladder operators and $\phi$. The expression includes terms quadratic in $\phi$ implying that, like the Hamiltonian, there are at least two non-equivalent ways of defining $n_\alpha$ within the $\alpha'$-gauge two-level model, because $P^{\alpha'}\phi^2 P^{\alpha'} \neq (P^{\alpha'}\phi P^{\alpha'})^2$ (see Supplementary Note 2). The two possible definitions of $n_\alpha$ in the $\alpha'$-gauge two-level model are given by the left and right-hand-sides of the inequality 
\begin{align}\label{ineq}
P^{\alpha'}n_\alpha({\bf y}_{\alpha'}) P^{\alpha'} \neq n_\alpha(P^{\alpha'}{\bf y}_{\alpha'}P^{\alpha'}).
\end{align}

\begin{figure}[H]
\begin{minipage}{\columnwidth}
\begin{center}
\vspace*{-4cm}
\hspace*{-0.4cm}\includegraphics[scale=1]{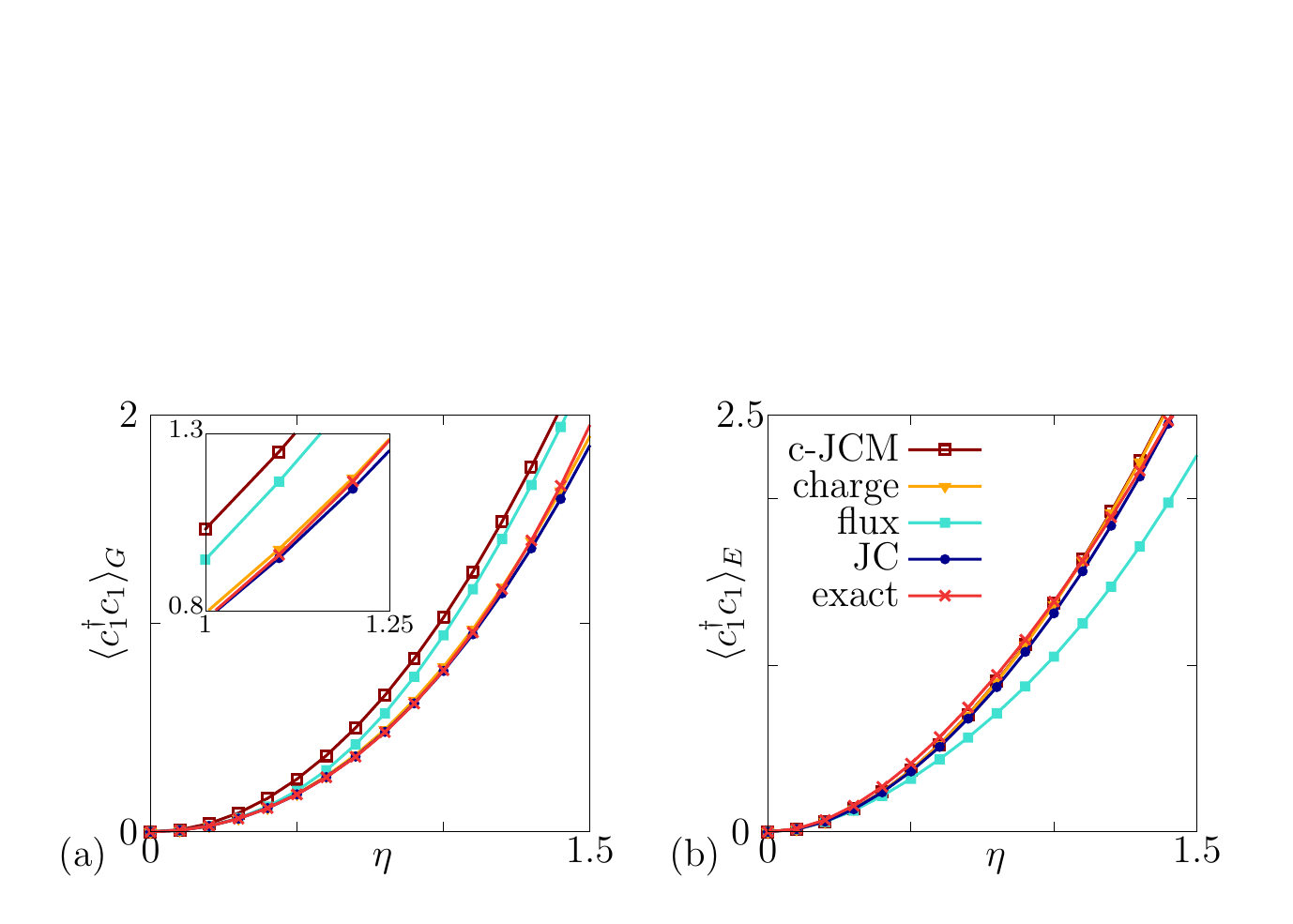}
\vspace*{-0.55cm}\caption{\textbf{Predictions of the number of flux-gauge photons as functions of coupling strength}. In all plots $E_{\rm l}=0.33\mu$eV, $E_{\rm J}=10E_{\rm l}=E_{\rm c}$, $\delta=5$ and $\phi_{\rm ext}=\uppi/2e$. \textbf{(a)}. The ground state average flux-gauge photon number is plotted with coupling $\eta$, for the flux-gauge and charge-gauge QRMs, for the JC-gauge two-level model, the exact theory, and the charge-gauge JCM (c-JCM). The flux-gauge JCM is inaccurate in the regime considered and is not shown. \textbf{(b)} Same as (a) for the first excited state average. In both graphs the plots corresponding to two-level models in gauges other than the flux gauge are found using the definition on the right-hand-side of inequality~(\ref{ineq}). The charge-gauge QRM, JC-gauge two-level model, and the exact number prediction are very close together, and in the case of the excited state the charge-gauge JCM is also accurate in this regime.}\label{g}
\end{center}
\end{minipage}
\end{figure}

Supplementary Fig.~\ref{g} shows the averages $\langle c_1^\dagger c_1\rangle_G$ and $\langle c_1^\dagger c_1\rangle_E$ in the ground and first energy states $\ket{G}$ and $\ket{E}$ found using various two-level models and the exact model. The predictions of two-level models in gauges other than the flux-gauge are found using the definition given by the right-hand-side of the inequality~(\ref{ineq}). The JC-gauge two-level model is more accurate than the flux-gauge QRM, the flux and charge-gauge JCMs, and is comparable to the charge-gauge QRM.

\begin{figure}[H]
\begin{minipage}{\columnwidth}
\begin{center}
\vspace*{-4.3cm}
\hspace*{-0.6cm}\includegraphics[scale=1.015]{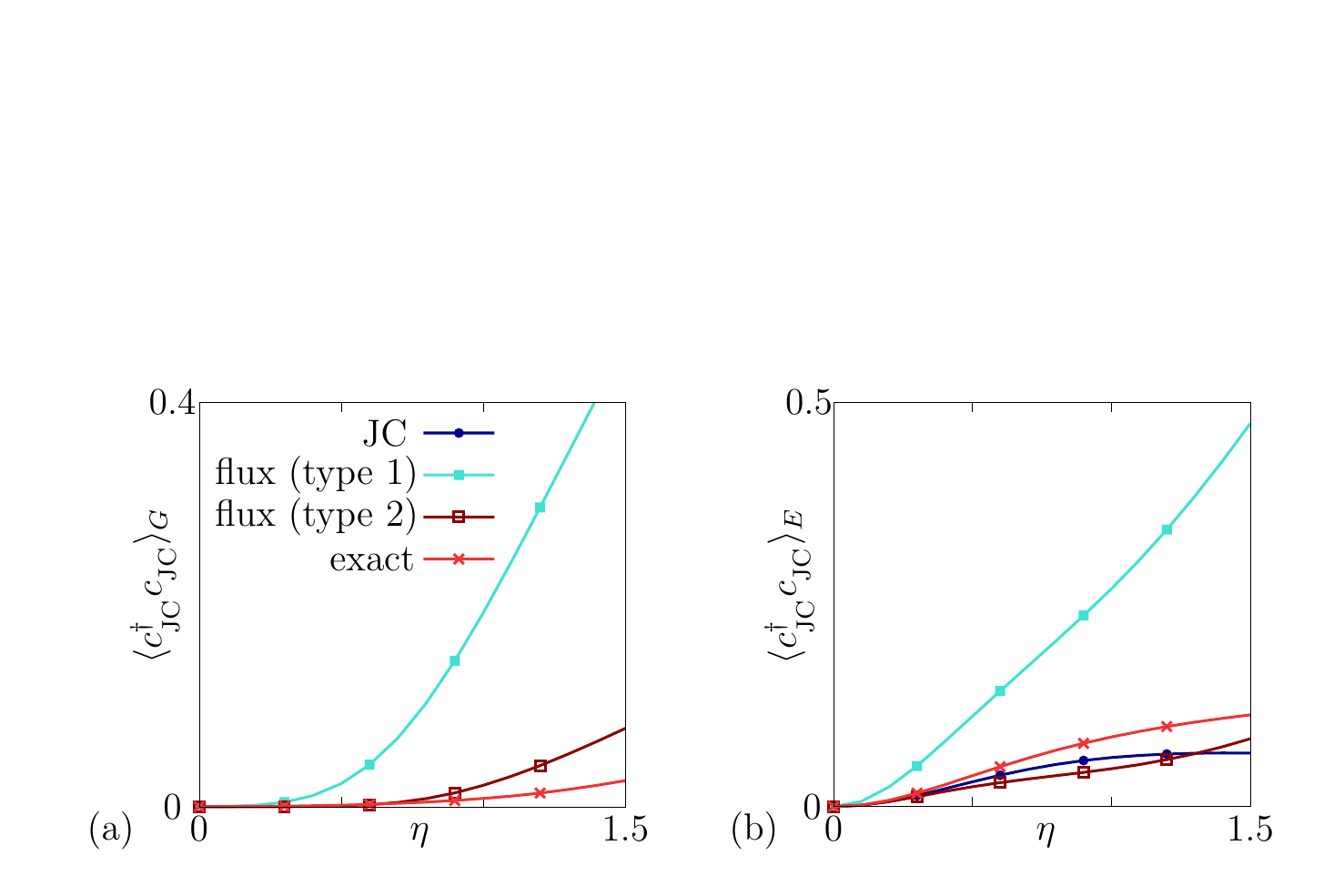}
\vspace*{-0.4cm}\caption{\textbf{Predictions of the number of JC-gauge photons as functions of coupling strength}. In all plots $E_{\rm l}=0.33\mu$eV, $E_{\rm J}=10E_{\rm l}=E_{\rm c}$, $\delta=5$ and $\phi_{\rm ext}=\uppi/2e$. \textbf{(a)} The ground state average JC-gauge photon number is plotted with coupling $\eta$, for the flux-gauge and the exact theory. The exact result remains approximately zero well into the ultrastrong regime indicating that the JC-gauge two-level model remains accurate. \textbf{(b)} Same as (a) for the first excited state average. The JC-gauge two-level model prediction is not identically zero in the excited state and has been included. In both graphs the flux-gauge QRM plots are given for definitions provided by both the right and left-hand-sides of the inequality (\ref{ineq}), which are labelled type 1 and type 2 respectively. In both cases the flux-gauge QRM is less accurate than the JC-gauge two-level model in the ultrastrong-coupling regime.}\label{g2}
\end{center}
\end{minipage}
\end{figure}

As a second example we consider the JC-gauge photon number operator $c_{\rm JC}^\dagger c_{\rm JC}$. In the ground state of the JC-gauge two-level model the average of this observable is zero for all coupling strengths. Thus, the exact average should be approximately zero for coupling strengths for which higher material levels can be neglected within the JC-gauge. Supplementary Fig.~\ref{g2} shows that the exact average is zero for sufficiently small coupling, but begins to grow for larger coupling strengths, which gives an indication of the relative validity of the JC-gauge two-level truncation. The flux-gauge QRM prediction for this observable is given for definitions provided by both the right and left-hand-sides of the inequality (\ref{ineq}). These definitions are labelled type 1 and type 2 respectively. In both cases the flux-gauge QRM overestimates the average and is less accurate than the JC-gauge two-level model. 
%%%%%%%%%%%%%%%%%%%%%%%%%%%%%%%%%%%%%%%%%%%%%%%%%%%%%%%%%%%%%%%%%%%%%%%%%%%%%%%%%%%%%%%%%%%%%%%
%%	F I G U R E S  S T A R T
%%
%%%%%%%%%%%%%%%%%%%%%%%%%%%%%%%%%%%%%%%%%%%%%%%%%%%%%%%%%%%%%%%%%%%%%%%%%%%%%%%%%%%%%%%%%%%%%
%%%%%%%%%%%%%%%%%%%%%%%%%%%%%%%%%%%%%%%%%%%%%%%%%%%%%%%%%%%%%%%%%%%%%%%%%%%%%%%%%%%%%%%%%%%%%
%%%%%%%%%%%%%%%%%%%%%%%%%%%%%%%%%%%%%%%%%%%%%%%%%%%%%%%%%%%%%%%%%%%%%%%%%%%%%%%%%%%%%%%%%%%%
%%
%%	F I G U R E S  E N D
%%
%%%%%%%%%%%%%%%%%%%%%%%%%%%%%%%%%%%%%%%%%%%%%%%%%%%%%%%%%%%%%%%%%%%%%%%%%%%%%%%%%%%%%%%%%%%%

\end{document}